\documentclass{LMCS}

\def\dOi{9(4:11)2013}
\lmcsheading%
{\dOi}
{1--34}
{}
{}
{Jan.~15, 2013}
{Nov.~12, 2013}
{}

\ACMCCS{[{\bf Theory of computation}]:  Models of computation---Concurrency}
\subjclass{F.1.2 Theory of computation (Concurrency)}

\usepackage{hyperref}
\usepackage{cite}
\usepackage{url}
\usepackage{float}
\newtheorem{theorem}{Theorem}
\newtheorem{lemma}[theorem]{Lemma}

\usepackage{pstricks}
\psset{xunit=1pt,yunit=1pt,runit=1pt,linewidth=0.5pt}
\newcommand{\GRAY}[1]{{\gray\psset{linecolor=gray}#1}}
\newcommand{\psst}{\pscircle[fillcolor=white]}
\newcommand{\arc}{\psline[linearc=2]{->}}
\newcommand{\aarc}[1]{\psline[linearc=#1]{->}}
\newcommand{\ainit}[1]{\rput[lb](#1){\arc(-4,8)(-0.9,1.8)}}
\newcommand{\aloopru}[1]{\rput[lb](#1){\arc(2,0)(10,0)(7,7)(1.4,1.4)}}
\newcommand{\aloopul}[1]{\rput[lb](#1){\arc(0,2)(0,10)(-7,7)(-1.4,1.4)}}
\newcommand{\aloopld}[1]{\rput[lb](#1){\arc(-2,0)(-10,0)(-7,-7)(-1.4,-1.4)}}
\newcommand{\aloopdr}[1]{\rput[lb](#1){\arc(0,-2)(0,-10)(7,-7)(1.4,-1.4)}}
\newcommand{\aloopur}[1]{\rput[lb](#1){\arc(0,2)(0,10)(7,7)(1.4,1.4)}}
\newcommand{\aloopdl}[1]{\rput[lb](#1){\arc(0,-2)(0,-10)(-7,-7)(-1.4,-1.4)}}
\newcommand{\aloopvd}[1]{\rput[lb](#1){\arc(-1.4,1.4)(-7,7)(-10,0)(-2,0)}}
\newcommand{\aloopmr}[1]{\rput[lb](#1){\arc(-1.4,-1.4)(-7,-7)(0,-10)(0,-2)}}
\newcommand{\arcud}[2]{\rput[bl](#1){%
\aarc{10}(1.4,1.4)(7,7)(21,7)(26.6,1.4)\rput[b](14,8.5){#2}}
}
\newcommand{\arcdu}[2]{\rput[bl](#1){%
\aarc{10}(1.4,-1.4)(7,-7)(21,-7)(26.6,-1.4)\rput[t](14,-8.5){#2}}
}
\newcommand{\acrud}[2]{\rput[bl](#1){%
\aarc{10}(-1.4,1.4)(-7,7)(-21,7)(-26.6,1.4)\rput[b](-14,8.5){#2}}
}
\newcommand{\acrdu}[2]{\rput[bl](#1){%
\aarc{10}(-1.4,-1.4)(-7,-7)(-21,-7)(-26.6,-1.4)\rput[t](-14,-8.5){#2}}
}

\newcommand{\DLG}{\begin{pspicture}(6,10)(0,2)
\ainit{4,2}\psst(4,2){2}
\end{pspicture}}

\newcommand{\LLG}{\begin{pspicture}(20,10)(0,2)
\ainit{4,2}\aloopru{4,2}\rput[l](15,5){$\tau$}\psst(4,2){2}
\end{pspicture}}

\newcommand{\BLG}{\begin{pspicture}(33,10)(0,2)
\arc(15,10)(15,4)
\aloopvd{15,2}\rput[r](5,5){$\tau$}
\arc(17,2)(29,2)\rput[B](23,4){$\tau$}
\psst(15,2){2}\psst(31,2){2}
\end{pspicture}}

\newcommand{\LTSa}[1]{\begin{pspicture}(22,10)(0,2)
\ainit{4,2}\arc(6,2)(18,2)\rput[B](12,4){#1}\psst(4,2){2}\psst(20,2){2}
\end{pspicture}}

\newcommand{\LTSsigma}{\begin{pspicture}(70,10)(0,2)
\ainit{4,2}
\arc(6,2)(18,2)\rput[B](12,5.5){$a_1$}
\arc(22,2)(34,2)\rput[B](28,5.5){$a_2$}
\rput(44,2){$\cdots$}
\arc(54,2)(66,2)\rput[B](60,5.5){$a_n$}
\psst(4,2){2}\psst(20,2){2}\psst(68,2){2}
\end{pspicture}}

\newcommand{\LTSsigmaloop}[1]{\begin{pspicture}(83,10)(0,2)
\ainit{4,2}
\arc(6,2)(18,2)\rput[B](12,5.5){$a_1$}
\arc(22,2)(34,2)\rput[B](28,5.5){$a_2$}
\rput(44,2){$\cdots$}
\arc(54,2)(66,2)\rput[B](60,5.5){$a_n$}
\aloopru{68,2}\rput[r](83,5){#1}
\psst(4,2){2}\psst(20,2){2}\psst(68,2){2}
\end{pspicture}}

\newcommand{\LTSxi}{\begin{pspicture}(49,10)(0,2)
\ainit{4,2}
\arc(6,2)(18,2)\rput[B](12,5.5){$a_1$}
\arc(22,2)(34,2)\rput[B](28,5.5){$a_2$}
\rput(44,2){$\cdots$}
\psst(4,2){2}\psst(20,2){2}
\end{pspicture}}

\newcommand{\LTSsbtloop}{
\begin{pspicture}(99,10)(0,2)
\ainit{4,2}
\arc(6,2)(18,2)\rput[B](12,5.5){$a_1$}
\arc(22,2)(34,2)\rput[B](28,5.5){$a_2$}
\rput(44,2){$\cdots$}
\arc(54,2)(66,2)\rput[B](60,5.5){$a_n$}
\arc(70,2)(82,2)\rput[B](76,4){$b$}
\aloopru{84,2}\rput[r](99,5){$\tau$}
\psst(4,2){2}\psst(20,2){2}\psst(68,2){2}\psst(84,2){2}
\end{pspicture}}

\newcommand{\LTSlr}[2]{
\begin{pspicture}(36,10)(0,2)
\ainit{18,2}
\arc(16,2)(4,2)\rput[B](10,4){#1}\arc(20,2)(32,2)\rput[B](26,4){#2}
\psst(2,2){2}\psst(18,2){2}\psst(34,2){2}
\end{pspicture}}

\newcommand{\ar}[1]{\,{-}#1{\rightarrow}}
\newcommand{\Ar}[1]{\,{=}#1{\Rightarrow}}
\newcommand{\Arr}[1]{\,{=}#1{\Rightarrow}\,}

\newcommand{\new}[2]{#1^{[#2]}}
\newcommand{\newup}[2]{\lceil#1\rceil^{[#2]}}
\newcommand{\newdn}[2]{\lfloor#1\rfloor_{[#2]}}

\newcommand{\mc}{\mathcal}
\newcommand{\ms}{\mathsf}
\newcommand{\mi}{\mathit}

\newcommand{\pp}{\:||\:}

\newcommand{\Stop}{\ms{Stop}}
\newcommand{\Run}[1]{\ms{Run}(#1)}
\newcommand{\RD}[1]{\ms{RD}(#1)}
\newcommand{\RDL}[1]{\ms{RDL}(#1)}

\newcommand{\bs}{\equiv}
\newcommand{\cgr}{\cong}
\newcommand{\CFFD}{\doteq}
\newcommand{\quo}[1]{``\,#1''}

\begin{document}

\title{All Linear-Time Congruences for Familiar Operators}

\author{Antti Valmari}
\address{Department of Mathematics\\
Tampere University of Technology\\
Tampere, Finland}
\email{Antti.Valmari@tut.fi}

\keywords{process algebra; semantics; compositionality; verification}

\begin{abstract}
The detailed behaviour of a system is often represented as a labelled
transition system (LTS) and the abstract behaviour as a
stuttering-insensitive semantic congruence.
Numerous congruences have been presented in the literature.
On the other hand, there have not been many results proving the absence of
more congruences.
This publication fully analyses the linear-time (in a well-defined sense)
region with respect to action prefix, hiding, relational renaming, and
parallel composition.
It contains 40 congruences.
They are built from the alphabet, two kinds of traces, two kinds of divergence
traces, five kinds of failures, and four kinds of infinite traces.
In the case of finite LTSs, infinite traces lose their role and the number of
congruences drops to 20.
The publication concentrates on the hardest and most novel part of the result,
that is, proving the absence of more congruences.
\end{abstract}

\maketitle

\section{Introduction}

A sequential program can usually be thought of as computing a partial function
from the set of possible inputs to the set of possible outputs.
Sometimes the program is not assumed to be deterministic, in which case its
meaning is not a partial function but a more general relation.
It is widely agreed that relations from inputs to outputs are usually the most
appropriate class of mathematical objects for modelling the semantics of
sequential programs at the abstract level.
Two programs are equivalent if and only if they compute the same relation.

The situation is entirely different with concurrent systems.
Process algebra researchers have introduced numerous abstract equivalence
notions for comparing the behaviours of systems or subsystems.
Many are surveyed in~\cite{vGl93}.
It is desirable that an equivalence is a congruence, that is, if a subsystem
is replaced by an equivalent subsystem, then the system as a whole remains
equivalent.
Whether or not an equivalence is a congruence depends on the set of operators
used in building systems from subsystems.
Although the congruence requirement narrows the range down, there is no
consensus about which abstract congruence is the most appropriate.
Indeed, the abstract congruence that is best for some purpose is not
necessarily the best for another purpose.

Behaviours of (sub)systems are often represented as \emph{labelled transition
systems}, abbreviated \emph{LTS}.
The congruence property makes it possible to apply reductions to subsystems or
their LTSs, and thus construct a reduced LTS of the system as a whole that is
equivalent to the full LTS of the system but often much smaller.
This \emph{compositional} approach is a key ingredient in many advanced
process-algebraic verification methods, see,
e.g.,~\cite{GSL96,MaV90,Val01}.

We say that ``$\cgr_1$'' \emph{implies} ``$\cgr_2$'', if and only if $L \cgr_1
L'$ implies $L \cgr_2 L'$ for every $L$ and $L'$.
We say that ``$\cgr_1$'' is \emph{weaker} (or coarser) than ``$\cgr_2$'', if
and only if ``$\cgr_2$'' implies ``$\cgr_1$'' but not vice versa.
We say that ``$\cgr$'' \emph{preserves} a property, if and only if $L \cgr L'$
implies that either none or both of $L$ and $L'$ have the property.
If, for instance, ``$\cgr$'' preserves deadlocks, $L$ is complicated, $L'$ is
simple, and we can reason that $L \cgr L'$, then we can analyse the deadlocks
of $L$ by analysing the deadlocks of $L'$.
On the other hand, if ``$\cgr$'' also preserves some other information (say,
livelocks) about which $L$ and $L'$ disagree, then $L \not\cgr L'$.
In that case, we cannot use $L'$ to analyse the deadlocks of $L$ because we
cannot reason that $L \cgr L'$.
Therefore, we would ideally like to use the weakest possible
deadlock-preserving congruence in this analysis task.

Finding the weakest congruence that preserves a given property has been
tedious.
A handful of such results has been published
(e.g.,~\cite{DeV95,GaF10,vGl10,KaV92,Puh01,PuV99,ReV07,Val95}), but if none of
them directly matches, then the user is more or less left with empty hands.
Furthermore, to fully exploit the weakest congruence, reduction algorithms
have to be adapted to it.
The prospect of rewriting the reduction tools for each property is not
attractive.

This publication shows that for a significant set of properties and widely
accepted set of process operators, the situation is not that bad.
This publication simplifies the selection of the abstract congruence that is
most appropriate for a task, by listing \emph{all} abstract congruences within
a reasonably wide region with respect to a reasonable set of operators.
The operators are parallel composition, hiding, relational renaming, and
action prefix.
The list will make it easy to answer such questions as ``what is the weakest
congruence that distinguishes $\LTSa{$a$}$ from $\LTSlr{$\tau$}{$a$}$?''

By \emph{abstract} we mean that invisible actions are not directly observable,
although they may have indirect observable consequences.
In the vocabulary of linear temporal logic~\cite{MaP92}, we only consider
stuttering-insensitive properties.
It is generally accepted that this is a reasonable restriction in the case of
concurrent systems.
Basically all process-algebraic verification methods make it.

The region that we cover is abstract \emph{linear-time} congruences, in the
following sense.
A linear-time property holds or fails to hold on an individual complete
execution of the system.
The system has the property if and only if all its complete executions have
it.
We originally only consider the execution of visible actions, deadlock, and
livelock as directly observable.
Then the congruence requirement will bring so-called refusal sets into
consideration in the end, but not in the middle, of a sequence of visible
actions.
The modern version~\cite{Ros10} of Hoare's CSP- or failures-divergences
equivalence~\cite{Hoa85} is within our scope, while Milner's observation
equivalence or weak bisimilarity~\cite{Mil89} is not.
Our notion of linear-time is slightly more general than that of the famous
stuttering-insensitive linear temporal logic of~\cite{MaP92}.
This is because we do but the logic does not distinguish deadlock from
livelock.
The congruence that matches the logic precisely will be found in
Section~\ref{S:BLisLL}.
On the other hand, we will see in Section~\ref{S:CFFD} that our notion of
linear-time is less general than another line of thought yields.

Two results of this kind were discussed in Chapters~11 and~12 of~\cite{Ros10}.
With the CSP set of operators and a certain notion of finite linear-time
observations, there are only three congruences.
Therefore, if the given property meets that notion, to find the weakest
congruence that preserves it, it suffices to test the three congruences.
If also infinite behaviour is observable, another set of only three
congruences is obtained.
Our range covers 40 congruences.
Four of them are the same as in~\cite{Ros10} and two are trivial.
The remaining 34 are obtained because we cover a different set of properties
and use a smaller set of operators than~\cite{Ros10}.
The additional two congruences in~\cite{Ros10} assume the ability to also
observe refusal sets in the middle of a trace.

This publication is based on~\cite{Val12a,Val12b}.
The former solved the problem for finite LTSs, finding 20 congruences.
The case of infinite LTSs was analysed in~\cite{Val12b}.
Some of the earlier congruences were split into two and some into three, so
the number grew to 40.
In~\cite{Val12a,Val12b} and this publication, we concentrate on proving that
there are no other congruences than those that we discuss, and skip the proofs
that they indeed are congruences.

Section~\ref{S:def} presents the background definitions.
Section~\ref{S:CFFD} introduces the strongest abstract linear-time congruence
(in our sense).
Congruences that are weaker than it are found in Sections~\ref{S:DLisLL}
to~\ref{S:BLisLL}.
Finally Section~\ref{S:conclusion} summarizes the publication.

\section{Basic Definitions}\label{S:def}

In this publication, systems are composed of labelled transition systems using
the action prefix, hiding, relational renaming, and parallel composition
operators.
In this section we define these and some related concepts, including
bisimilarity.

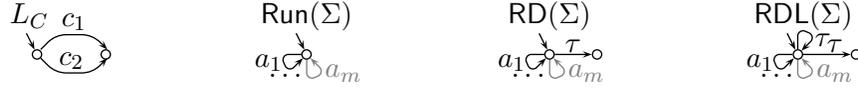
\begin{figure}[t]
\mbox{}\hfill
\begin{pspicture}(38,30)
\ainit{10,10}\rput[b](7,20){$L_C$}
\arcud{10,10.5}{$c_1$}
\arcdu{10,10}{}\rput[b](23,4.5){$c_2$}
\psst(10,10){2}
\psst(36,10){2}
\end{pspicture}\hfill
\begin{pspicture}(36,30)
\ainit{18,10}\rput[b](18,19){$\Run{\Sigma}$}
\aloopld{18,10}\rput[r](9,7){$a_1$}
\GRAY{\aloopdr{18,10}\rput[l](25,3){$a_m$}}
\rput(10,2){$\ldots$}
\psst(18,10){2}
\end{pspicture}\hfill
\begin{pspicture}(38,30)
\ainit{18,10}\rput[b](18,19){$\RD{\Sigma}$}
\aloopld{18,10}\rput[r](9,7){$a_1$}
\GRAY{\aloopdr{18,10}\rput[l](25,3){$a_m$}}
\rput(10,2){$\ldots$}
\arc(20,10)(34,10)\rput[b](27,11){$\tau$}
\psst(18,10){2}\psst(36,10){2}
\end{pspicture}\hfill
\begin{pspicture}(42,30)
\ainit{18,10}\rput[b](20,19){$\RDL{\Sigma}$}
\aloopur{18,10}\rput[tl](24.5,18){$\tau$}
\aloopld{18,10}\rput[r](9,7){$a_1$}
\GRAY{\aloopdr{18,10}\rput[l](25,3){$a_m$}}
\rput(10,2){$\ldots$}
\arc(20,10)(38,10)\rput[b](32,11){$\tau$}
\psst(18,10){2}\psst(40,10){2}
\end{pspicture}\hfill
\mbox{}
\caption{Some simple LTSs.
The alphabet of $L_C$ is $\{c_1, c_2\}$.
The alphabet of the others is $\Sigma = \{a_1, a_2, \ldots{\gray, a_m}\}$,
where {\gray grey notation} indicates that, despite the drawing, $\Sigma$ (and
consequently $\Delta$) may be infinite.}\label{F:simpleLTSs}
\end{figure}

We reserve the symbol $\tau$ to denote so-called invisible actions.
A \emph{labelled transition system} or \emph{LTS} is the tuple $(S, \Sigma,
\Delta, \hat s)$, where $\tau \notin \Sigma$, $\Delta \subseteq S \times
(\Sigma \cup \{\tau\}) \times S$, and $\hat s \in S$.
We call $S$ the set of \emph{states}, $\Sigma$ the \emph{alphabet}, $\Delta$
the set of \emph{transitions}, and $\hat s$ the \emph{initial state}.
An LTS is \emph{finite} if and only if its $S$ and $\Sigma$ (and thus also
$\Delta$) are finite.
Unless otherwise stated, $L_1$ denotes the LTS $(S_1, \Sigma_1, \Delta_1, \hat
s_1)$, and similarly with $L$, $L'$, $L_2$, and so on.
When we show an LTS as a drawing, \emph{unless otherwise stated, its alphabet
is precisely the labels in the drawing excluding~$\tau$}.
Fig.~\ref{F:simpleLTSs} shows as examples some simple LTSs that are needed
later.

\label{D:bisim}%
LTSs $L_1$ and $L_2$ are \emph{bisimilar}, denoted with $L_1 \bs L_2$, if and
only if there is a relation ``$\sim$''~$\subseteq S_1 \times S_2$ such that
\begin{enumerate}

\item
$\Sigma_1 = \Sigma_2$,

\item
$\hat s_1 \sim \hat s_2$, and

\item
for every $s_1 \in S_1$, $s_2 \in S_2$, $s'_1 \in S_1$, $s'_2 \in S_2$, and $a
\in \Sigma \cup \{\tau\}$ such that $s_1 \sim s_2$,
\begin{enumerate}
\item
if $(s_1,a,s'_1) \in \Delta_1$, then there is an $s'$ such that $s'_1 \sim s'$
and $(s_2,a,s') \in \Delta_2$, and
\item
if $(s_2,a,s'_2) \in \Delta_2$, then there is an $s'$ such that $s' \sim s'_2$
and $(s_1,a,s') \in \Delta_1$.
\end{enumerate}

\end{enumerate}
The relation ``$\sim$'' is a \emph{bisimulation}.

It is well known that bisimilarity is a very strong equivalence.
For the purposes of this publication (and, indeed, almost everywhere in
concurrency theory), bisimilar LTSs can be informally thought of as identical.
Formal justification for this comes from the fact (whose proof we skip) that
replacing an LTS by a bisimilar one in any of our definitions may change the
resulting LTS to a bisimilar one but cannot cause any other difference.
For instance, if an LTS deadlocks, then also all its bisimilar LTSs deadlock.

Because the purpose of an LTS is to represent the behaviour of a system, it
seems intuitively that only the part of the LTS that is reachable from the
initial state is significant.
Indeed, if $L'$ is the reachable part of $L$, by letting $s \sim s'$ if and
only if $s = s' \in S'$ we see that $L \bs L'$.
So also in our theory, only the reachable part matters.

If $\Phi$ is any set of pairs, we define $\mc{D}(\Phi) := \{ a \mid \exists b:
(a,b) \in \Phi \}$ (the \emph{domain}) and $\mc{R}(\Phi) := \{ b \mid \exists
a: (a,b) \in \Phi \}$ (the \emph{range}).
We also define $\Phi(a,b) :\Leftrightarrow (a,b) \in \Phi \vee a = b \notin
\mc{D}(\Phi)$.
This definition makes $\Phi(a,a)$ hold whenever $a$ is not in the domain of
$\Phi$.

The operators that we use for building systems are defined as follows:

\begin{description}

\item[Action prefix]
Let $a \neq \tau$.
The LTS $L' = a.L$ is defined as $S' = S \cup \{\hat s'\}$, where $\hat s'
\notin S$, $\Sigma' = \Sigma \cup \{a\}$, and $\Delta' = \Delta \cup \{(\hat
s',a,\hat s)\}$.
That is, $a.L$ executes $a$ and then behaves like $L$.
We do not define $\tau.L$ as we will not need it, but it is clear that it can
be built from $a.L$ and the next operator by choosing an $a$ that is not in
$\Sigma$.

\item[Hiding]
Let $A$ be a set.
The LTS $L' = L \setminus A$ is defined as $S' = S$, $\Sigma' = \Sigma
\setminus A$, $\Delta' = \{ (s,a,s') \mid \exists b: (s,b,s') \in \Delta$
$\wedge$ $(a = b \notin A \vee a = \tau \wedge b \in A) \}$, and $\hat s' =
\hat s$.
That is, $L \setminus A$ behaves otherwise like $L$, but all actions in $A$
are replaced by $\tau$.

\item[Relational renaming]
Let $\Phi$ be a set of pairs such that $\tau \notin \mc{D}(\Phi) \cup
\mc{R}(\Phi)$.
The LTS $L' = L\Phi$ is defined as $S' = S$, $\hat s' = \hat s$, $\Sigma' = \{
b \mid \exists a \in \Sigma: \Phi(a,b) \}$, and $\Delta' = \{ (s,b,s') \mid
\exists a: (s,a,s') \in \Delta \wedge \Phi(a,b) \}$.
That is, $L\Phi$ behaves otherwise like $L$, but the labels of transitions are
changed.
A label may be replaced by more than one label, resulting in more than one
copy of the original transition.
If $\Phi$ does not specify any new label for a transition, then it keeps its
original label.
This is in particular the case with $\tau$-transitions.

\item[Parallel composition]
The LTS $L = L_1 \pp L_2$ is defined as $S = S_1 \times S_2$, $\Sigma =
\Sigma_1 \cup \Sigma_2$, $\hat s = (\hat s_1, \hat s_2)$, and $((s_1,s_2), a,
(s'_1,s'_2)) \in \Delta$ if and only if
\begin{enumerate}
\item
$a \notin \Sigma_2$, $(s_1,a,s'_1) \in \Delta_1$, and $s'_2 = s_2$,
\item
$a \notin \Sigma_1$, $(s_2,a,s'_2) \in \Delta_2$, and $s'_1 = s_1$, or
\item
$a \in \Sigma_1 \cap \Sigma_2$, $(s_1,a,s'_1) \in \Delta_1$, and $(s_2,a,s'_2)
\in \Delta_2$.
\end{enumerate}
That is, if $a$ belongs to the alphabets of both $L_1$ and $L_2$, it is
executed simultaneously by both.
If $a = \tau$ or $a$ belongs to the alphabet of precisely one of $L_1$ and
$L_2$, then it is executed by one of $L_1$ and $L_2$ while the other stays in
the state where it is.
Clearly $L_2 \pp L_1 \bs L_1 \pp L_2$ and $L_1 \pp (L_2 \pp L_3) \bs (L_1 \pp
L_2) \pp L_3$, so we may write $L_1 \pp \cdots \pp L_n$ without confusion.

\end{description}
The CSP language~\cite{Ros10} has these operators (and many more), and every
major process-algebraic language has at least something similar.
Therefore, requiring the congruence property with respect to these operators
is justified.
One has to keep in mind, however, that if the language does not have all these
operators, then it may have more abstract linear-time congruences than the
ones in this publication.
Indeed, we will see after Theorem~\ref{T:noSigma} that the ability of the
renaming operator to convert a single action into many actions is important,
and so is the availability of the action prefix operator.

Because the notion of congruence depends on the set of operators and because
listing the set in theorems is clumsy, we state the following:

\begin{quote}
\emph{In the theorems of this publication, \quo{$\cgr$} is a congruence means
that it is an equivalence and for all LTSs $L$ and $L'$, if $L \cgr L'$, then
$a.L \cgr a.L'$, $L \setminus A \cgr L' \setminus A$, $L\Phi \cgr L'\Phi$, $L
\pp L'' \cgr L' \pp L''$, and $L'' \pp L \cgr L'' \pp L'$}.
\end{quote}
It follows by structural induction that if $f(L_1, \ldots, L_n)$ is any
expression only made of these four operators, and if $L_i \cgr L'_i$ for $1
\leq i \leq n$, then $f(L_1, \ldots, L_n) \cgr f(L'_1, \ldots, L'_n)$.

\section{The Strongest Abstract Linear-time Congruence}\label{S:CFFD}

In this section, we first define some concepts and notation that are useful
for discussing abstract linear-time equivalences.
Then we transform the notion of ``linear-time'' of~\cite{MaP92} to the
vocabulary of this publication.
(Unlike~\cite{MaP92}, we distinguish between deadlock and livelock.)
The resulting abstract equivalence is not a congruence.
We analyse what has to be added to make it a congruence.
Thanks to the additions, some original information becomes redundant.
So we throw it away.
We call the result an abstract linear-time congruence, because it does not
preserve more information than is necessary to cover linear temporal logic in
the sense described above.
It is the strongest such congruence, because it does not preserve less
information than that.
Finally we set the target for the rest of this publication.

For discussing abstract equivalences, it is handy to have notation for talking
about paths between states such that only the non-$\tau$ labels along the path
are shown.
Let $\Sigma^*$ and $\Sigma^\omega$ denote the sets of all finite and infinite
sequences of elements of $\Sigma$.
By $s \Arr{\varepsilon} s'$ we mean that there are $s_0$, \ldots, $s_n$ such
that $s = s_0$, $s_n = s'$, and $(s_{i-1},\tau,s_i) \in \Delta$ for $1 \leq i
\leq n$.
By $s \Arr{a_1 a_2 \cdots a_n} s'$, where $a_1 a_2 \cdots a_n \in \Sigma^*$,
we mean that there are $s_0$, $s'_0$, \ldots, $s_n$, $s'_n$ such that $s_0 =
s$, $s'_n = s'$, $s_i \Arr{\varepsilon} s'_i$ for $0 \leq i \leq n$, and
$(s'_{i-1},a_i,s_i) \in \Delta$ for $1 \leq i \leq n$.
If we do not want to mention $s'$, we write $s \Ar{a_1 a_2 \cdots a_n}$, and
$s \Ar{a_1 a_2 \cdots}$ denotes the similar notion for infinite sequences
$a_1 a_2\cdots$.
An infinite path can also consist of an uninterrupted infinite sequence of
invisible transitions.
This is denoted with $s \ar{\tau^\omega}$.

Let $s \in S$.
We say that $s$ is \emph{a deadlock} or \emph{deadlocked} if and only if
$\forall a: \forall s': (s,a,s') \notin \Delta$.
We say that $s$ is \emph{stable} if and only if $\forall s': (s,\tau,s')
\notin \Delta$.

\newcommand{\Tr}{\mi{Tr}}
\newcommand{\Dl}{\mi{D}\ell}
\newcommand{\Div}{\mi{Div}}
\newcommand{\Inf}{\mi{Inf}}
\newcommand{\Sf}{\mi{Sf}}
An \emph{execution} of $L$ is any path that starts at $\hat s$.
An execution is \emph{complete} if and only if it is infinite or leads to a
deadlock.
If an infinite execution only has a finite number of visible actions, then it
consists of a finite prefix and a \emph{livelock}, that is, an infinite path
only consisting of $\tau$-transitions.

In the linear temporal logic of~\cite{MaP92}, ``linear-time'' means that the
models of logical formulae are certain kind of abstractions of individual
complete executions, and a system satisfies a formula if and only if all its
complete executions satisfy it.
Analogously, we say that the linear-time semantics of $L$ consists of the
complete executions of $L$.
There is, however, one difference: in~\cite{MaP92}, deadlocking executions
are extended to infinite by repeating the last state forever, that is,
deadlocks are unified with livelocks.
We will not do so, because not unifying them gives a more natural and richer
theory, from which the theory with the unification is trivially obtained as a
corollary.

The abstract linear-time semantics of $L$ consists of the abstractions of the
complete executions of $L$, that is, \emph{deadlocking traces},
\emph{divergence traces}, and \emph{infinite traces}, defined as follows:
\begin{eqnarray}
\Dl(L) & := & \{ \sigma \in \Sigma^* \mid \exists s: \hat s \Arr{\sigma} s
\wedge \forall a: \forall s': (s,a,s') \notin \Delta \}\nonumber\\
\Div(L) & := & \{ \sigma \in \Sigma^* \mid \exists s: \hat s \Arr{\sigma} s
\wedge s \ar{\tau^\omega} \}\nonumber\\
\Inf(L) & := & \{ \xi \in \Sigma^\omega \mid \hat s \Ar{\xi} \}\nonumber
\end{eqnarray}
For uniformity, from now on $\Sigma(L)$ denotes the alphabet of $L$.

We say that \emph{the equivalence induced by $\Sigma$, $\Dl$, $\Div$, and
$\Inf$} is the one defined by $\Sigma(L) = \Sigma(L') \wedge \Dl(L) = \Dl(L')
\wedge \Div(L) = \Div(L') \wedge \Inf(L) = \Inf(L')$.
Unfortunately, it is not a congruence.
To fix this, we define \emph{stable failures}:
\begin{eqnarray}
\Sf(L) & := & \{ (\sigma,A) \in \Sigma^* \times 2^\Sigma \mid \exists s: \hat
s \Arr{\sigma} s \wedge \forall a \in A \cup \{\tau\}: \forall s': (s,a,s')
\notin \Delta \}\nonumber
\end{eqnarray}

\begin{figure}
\mbox{}\hfill\begin{pspicture}(201,33)
\ainit{6,13}\aloopdr{6,13}\rput[r](5,7){$\tau$}
\arc(8,13)(32,13)\rput[B](20,16){$b_1$}
\aloopdr{34,13}\rput[r](33,7){$\tau$}
\arc(36,13)(60,13)\rput[B](48,16){$b_2$}
\aloopdr{62,13}\rput[r](61,7){$\tau$}
\arc(64,13)(88,13)\rput[B](76,16){$b_3$}
\rput(98,13){$\cdots$}
\rput(98,6){$\cdots$}
\arc(108,13)(132,13)\rput[B](120,16){$b_n$}
\aloopdr{134,13}\rput[r](133,7){$\tau$}
\arc(136,13)(160,13)\rput[B](148,16){$\tau$}
\arcud{162,13}{$a_1$}
\GRAY{\arcdu{162,13}{$a_m$}}
\rput(176,15){$\vdots$}
\aloopdr{190,13}\rput[l](197,7){$\tau$}
\psst(6,13){2}\psst(34,13){2}\psst(62,13){2}\psst(134,13){2}\psst(162,13){2}
\psst(190,13){2}
\end{pspicture}\hfill\mbox{}
\caption{An LTS for detecting the stable failure $(b_1 \cdots b_n, \{a_1,
\ldots{\gray, a_m}\})$.
}\label{F:testDl}
\end{figure}
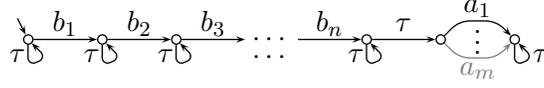

It was proven in~\cite{Val95} that any congruence ``$\cgr$'' that preserves
$\Sigma$ and $\Dl$ also preserves $\Sf$.
We repeat the proof here to get familiar with the proof technique.
To talk about a finite set $\{a_1, \ldots, a_m\}$ or the infinite set $\{a_1,
a_2, \ldots\}$, we use the notation $\{a_1, \ldots{\gray, a_m}\}$ where ``$,
a_m$'' is grey.

\proof
Assume that $(b_1 \cdots b_n, \{a_1, \ldots{\gray, a_m}\}) \in \Sf(L)$.
Let $T$ be the LTS in Fig.~\ref{F:testDl} with $\Sigma(T) = \Sigma(L)$.
By letting $L$ execute $b_1 \cdots b_n$ so that it then refuses $a_1$,
\ldots{\gray, $a_m$}, and $\tau$, we see that $b_1 \cdots b_n \in \Dl(L \pp
T)$.
Let $L \cgr L'$.
We have $\Sigma(L') = \Sigma(L)$ because ``$\cgr$'' preserves $\Sigma$.
By the congruence property $L \pp T \cgr L' \pp T$.
That ``$\cgr$'' preserves $\Dl$ yields $b_1 \cdots b_n \in \Dl(L' \pp T)$.
That is only possible if $L'$ can execute $b_1 \cdots b_n$ such that it then
refuses $a_1$, \ldots{\gray, $a_m$}, and $\tau$.
That is, $(b_1 \cdots b_n, \{a_1, \ldots{\gray, a_m}\}) \in \Sf(L')$.
We have proven that $\Sf(L) \subseteq \Sf(L')$.
By symmetry, $\Sf(L') \subseteq \Sf(L)$.\qed

Therefore, we must add $\Sf$ to the semantics.
We have $\Dl(L) = \{ \sigma \mid (\sigma,\Sigma) \in \Sf(L) \}$.
This implies that if $\Sigma(L) = \Sigma(L')$ and $\Sf(L) = \Sf(L')$, then
$\Dl(L) = \Dl(L')$.
As a consequence, the equivalence induced by $\Sigma$, $\Dl$, $\Sf$, $\Div$,
and $\Inf$ is the same as the equivalence induced by $\Sigma$, $\Sf$, $\Div$,
and $\Inf$.
That is, we no longer need $\Dl$ as such in the semantics.

The equivalence induced by $\Sigma$, $\Sf$, $\Div$, and $\Inf$ is a
congruence~\cite{VaT95}.
It is implied by ``$\bs$''.
It has traditionally been called \emph{chaos-free failures divergences
equivalence} or \emph{CFFD-equivalence} for the reason explained in
Section~\ref{S:BLisLL}.
We will denote it with ``$\CFFD$''.

Finite, not necessarily complete executions induce \emph{traces}:
\begin{eqnarray}
\Tr(L) & := & \{ \sigma \in \Sigma^* \mid \hat s \Ar{\sigma} \}\nonumber
\end{eqnarray}
If $(\sigma, A) \in \Sf(L)$, then clearly $(\sigma, \emptyset) \in \Sf(L)$ and
$\sigma \in \Tr(L)$.
We will later define also other subsets of $\Sigma^* \times 2^\Sigma$ that
have the similar property.
With $\Sf$ and them, the following notation will be handy:
\begin{eqnarray}
X^\Tr(L) & := & \{ \sigma \mid (\sigma, \emptyset) \in X(L) \}\nonumber
\end{eqnarray}

CFFD-equivalence contains full information on traces even without explicitly
mentioning them, because of the following easily proven fact:
\begin{eqnarray}\label{E:Tr=DivSf}
\Tr(L) & = & \Div(L) \cup \Sf^\Tr(L)
\end{eqnarray}
We will also need the following fact.
\begin{eqnarray}\label{E:Inf}
\Inf(L) & \subseteq & \{ a_1 a_2 \cdots \in \Sigma^\omega \mid \forall i: a_1
a_2 \cdots a_i \in \Tr(L) \}
\end{eqnarray}
In the case of finite LTSs, even $\Inf$ is unnecessary because of the
following (see, e.g.,~\cite{VaT91,VaT95}):
\begin{eqnarray}\label{E:Inf=}
\Inf(L) & = & \{ a_1 a_2 \cdots \in \Sigma^\omega \mid \forall i: a_1 a_2
\cdots a_i \in \Tr(L) \}\textrm{ , if $L$ is finite.}
\end{eqnarray}

To summarize, if we define ``$\CFFD$'' as the equivalence induced by $\Sigma$,
$\Sf$, $\Div$, and $\Inf$, then $L \CFFD L'$ also implies $\Tr(L) = \Tr(L')$
and $\Dl(L) = \Dl(L')$.
For finite LTSs, the assumption $\Inf(L) = \Inf(L')$ is not needed.
Because we derived ``$\CFFD$'' by starting with the abstract linear-time
semantics and strengthening it only as much as was necessary to make it a
congruence, it is reasonable to call it the strongest abstract linear-time
congruence.

This is not the only possible use of the phrase ``linear-time'', however.
For instance, one could classify as linear-time everything that can be defined
in terms of individual executions and the next-label sets $N(s)$ of the states
$s$ along each execution, where $N(s) := \{ a \mid \exists s': (s,a,s') \in
\Delta \}$.
Now $\Sf(L)$ can be rephrased as the set of pairs $(\sigma, A) \in \Sigma^*
\times 2^\Sigma$ such that $\sigma$ leads to a stable state $s$ such that $A
\subseteq \Sigma \setminus N(s)$.
So ``$\CFFD$'' is linear-time also in this sense.
However, so is also the equivalence obtained otherwise similarly, but letting
$A = \Sigma \setminus N(s)$.
This equivalence is a congruence.
It is trivially strictly stronger than ``$\CFFD$'', so it is outside our
notion of linear-time.

Our goal is to find all congruences that are implied by ``$\CFFD$''.
For any stuttering-insensitive linear-time property in the sense
of~\cite{MaP92}, its optimal congruence is among them.

To break our task into smaller parts, let us consider all possibilities when
$\Sigma = \emptyset$.
Then $\Tr(L) = \{\varepsilon\}$, $\Inf(L) = \emptyset$, $\Sf(L)$ is either
$\emptyset$ or $\{(\varepsilon, \emptyset)\}$, and $\Div(L)$ is either
$\emptyset$ or $\{\varepsilon\}$.
By~(\ref{E:Tr=DivSf}) they cannot both be empty.
This leaves three possibilities.
They can be drawn as follows.

\begin{center}\DLG\hspace{2cm}\LLG\hspace{2cm}\BLG\end{center}
We will study each of the cases $\DLG \cgr \LLG$, $\DLG \cgr \BLG \not\cgr
\LLG$, $\DLG \not\cgr \LLG \not\cgr \BLG \not\cgr \DLG$, and $\DLG\not\cgr
\LLG \cgr \BLG$ in turn.

\section{When Deadlock Is Livelock}\label{S:DLisLL}

In this section we find all congruences that are implied by ``$\CFFD$'' and
unify deadlock with livelock, that is, have $\DLG \cgr \LLG$.
Theorems~\ref{T:noSigma}, \ref{T:Sigma}, \ref{T:Trace}, and~\ref{T:Tr} say
that if ``$\cgr$'' preserves any information whatsoever, then it preserves at
least $\Sigma$; if it preserves more than that, then it also preserves $\Tr$;
if it preserves more than that, then it also preserves $\Inf$; and that is
all.
The technique used in all but one such proofs in this publication is developed
and illustrated.
It is based on Lemma~\ref{L:nocongr}.
The section also presents two lemmas related to preserving or not preserving
$\Tr$.
Theorem~\ref{T:noSigma} is different from others in this section in that it
uses a different proof technique and does not make the assumptions mentioned
above.
So it also applies to bisimulation-based semantics.
However, perhaps surprisingly, it depends on the presence of both action
prefix and relational renaming in our set of operators.

We define \emph{the dullest congruence} by $L \cgr L'$ holds for every $L$
and $L'$.
It is obviously the weakest of all congruences.
The next theorem implies that it is the only congruence that does not imply
$\Sigma(L) = \Sigma(L')$, that is, preserve $\Sigma$.
We define $\Stop(A)$ as the 1-state LTS whose alphabet is $A$ and which has no
transitions.
(So $\Stop(\emptyset) = \DLG$.)

\begin{theorem}\label{T:noSigma}
If ``$\cgr$'' is implied by ``$\bs$'', is a congruence, and does not preserve
$\Sigma$, then ``$\cgr$'' is the dullest congruence.
\end{theorem}
\proof
Because ``$\cgr$'' does not preserve $\Sigma$, there are LTSs $M_1$ and $M_2$
and an $a$ such that $M_1 \cgr M_2$ and $a \in \Sigma_1 \setminus \Sigma_2$.
Let $C = (\{c\} \cup \Sigma_1 \cup \Sigma_2) \setminus \{a\}$, where $c \neq
a$ and $c \neq \tau$.
When $i \in \{1,2\}$, let $f(M_i) = (c.M_i \pp \Stop(\{c\})) \setminus C$.
Because $c.M_i$ initially commits to $c$ and $\Stop(\{c\})$ blocks all
$c$-transitions, $f(M_i)$ has no reachable transitions and only one reachable
state.
$C$ contains all visible actions of $f(M_i)$ except $a$.
So $f(M_1) \bs \Stop(\{a\})$ and $f(M_2) \bs \Stop(\emptyset) \bs \DLG$.
Because $M_1 \cgr M_2$, we have $\Stop(\{a\}) \bs f(M_1) \cgr f(M_2) \bs
\DLG$.
This yields $\DLG \cgr \Stop(\{a\})$, because ``$\bs$'' implies ``$\cgr$'' and
``$\cgr$'' is an equivalence.

We prove next that each LTS with the empty alphabet is equivalent to $\DLG$.
Let $L' = (S', \emptyset, \Delta', \hat s')$ be an LTS.
Let $L'_a = (S', \{a\}, \Delta'_a, \hat s')$, where $\Delta'_a =
\{ (s,a,s') \mid (s,\tau,s') \in \Delta' \}$.
By the definition of ``$\setminus$'', $L' \bs L'_a \setminus \{a\} \bs (L'_a
\pp \DLG) \setminus \{a\} \cgr (L'_a \pp \Stop(\{a\})) \setminus \{a\} \bs
\DLG$.

Then we prove that each LTS is equivalent to an LTS with the empty alphabet.
Let $L = (S, \Sigma, \Delta, \hat s)$, $\Phi_a^\Sigma = \{a\} \times \Sigma$,
$\Delta' = \Delta \cap (S \times \{\tau\} \times S)$, and $L' = (S, \emptyset,
\Delta', \hat s)$.
By the definition of ``$\Phi$'', $\DLG \bs \DLG\Phi_a^\Sigma \cgr
\Stop(\{a\})\Phi_a^\Sigma \bs \Stop(\Sigma)$.
Therefore, $L \bs L \pp \DLG \cgr L \pp \Stop(\Sigma) \bs (S, \Sigma, \Delta',
\hat s) \bs L' \pp \Stop(\Sigma) \cgr L' \pp \DLG \bs L'$.

As a conclusion, every LTS is equivalent to $\DLG$ and thus to any other.\qed

This theorem relies on the ability of $\Phi$ to convert a single action to an
infinite set of actions.
Without that ability, the following would be a congruence: $L \cgr L'$ if and
only if $(\Sigma(L) \setminus \Sigma(L')) \cup (\Sigma(L') \setminus
\Sigma(L))$ is finite.
Also action prefix is necessary for this theorem.
Without it, the following would be a congruence: $L \cgr L'$ if and only if $L
\bs L'$ or both $\hat s \ar{\tau^\omega}$ and $\hat s' \ar{\tau^\omega}$.
That is, initially diverging LTSs could be declared equivalent, even if they
had different alphabets.

Theorem~\ref{T:noSigma} says that if a congruence makes any distinctions
between LTSs at all, then it preserves at least $\Sigma$.
On the other hand, it is easy to check from the definitions that the
equivalence induced by $\Sigma$ is a congruence.
So it is the second weakest congruence.
We have now two congruences that are both trivial.

The next lemma will be needed soon.

\begin{lemma}\label{L:Inf=>Tr}
Any congruence that preserves $\Inf$ also preserves $\Sigma$ and $\Tr$.
\end{lemma}
\proof
Let ``$\cgr$'' be a congruence that preserves $\Inf$.
Then $
\begin{pspicture}(20,10)(0,2)
\ainit{4,2}\aloopru{4,2}\rput[l](15,5){$a$}\psst(4,2){2}
\end{pspicture}
\not\cgr \LTSa{$a$}$, so ``$\cgr$'' preserves $\Sigma$ by
Theorem~\ref{T:noSigma}.
Let $L \cgr L'$, $\Sigma = \Sigma(L) = \Sigma(L')$, and $b \notin \Sigma \cup
\{\tau\}$.
If $\sigma = a_1 a_2 \cdots a_n \in \Tr(L)$, then let $T_\sigma^b$ be
$\LTSsigmaloop{$b$}$ with the alphabet $\Sigma \cup \{b\}$.
We have $\sigma b^\omega \in \Inf(L \pp T_\sigma^b) = \Inf(L' \pp
T_\sigma^b)$, yielding $\sigma \in \Tr(L')$.
So $\Tr(L) \subseteq \Tr(L')$.
By symmetry, $\Tr(L') \subseteq \Tr(L)$.\qed

The following lemma is central.
Many of the subsequent proofs use it.
In it, $X_1$, \ldots, $X_k$ are functions from LTSs to sets, like $\Tr$ and
$\Sf$.

\begin{lemma}\label{L:nocongr}
Assume that ``$\cgr$'' is an equivalence, is implied by ``$\CFFD$'', and
preserves $\Sigma$ and $X_1$, \ldots, $X_k$.
Assume that there is a function $f$ such that for every LTS $L$ we have $L
\cgr f(L)$, and $\Sf(f(L))$, $\Div(f(L))$, and $\Inf(f(L))$ can be represented
as functions of $\Sigma(L)$ and $X_1(L)$, \ldots, $X_k(L)$.
Then ``$\cgr$'' is the equivalence induced by $\Sigma$ and $X_1$, \ldots,
$X_k$.
\end{lemma}
\proof
Obviously ``$\cgr$'' implies the equivalence induced by $\Sigma$ and $X_1$,
\ldots, $X_k$.

To prove the implication in the opposite direction, let $\Sigma(L) =
\Sigma(L')$ and $X_i(L) = X_i(L')$ for $1 \leq i \leq k$.
We need to prove that $L \cgr L'$.
We have $\Sigma(f(L)) = \Sigma(L) = \Sigma(L') = \Sigma(f(L'))$, because $L
\cgr f(L)$ and ``$\cgr$'' preserves $\Sigma$.
When $X \in \{\Sf, \Div, \Inf\}$, let $\lambda_X$ be the function that
represents $X(f(L))$ as was promised.
Then $X(f(L)) = \lambda_X( \Sigma(L), X_1(L), \ldots, X_k(L))$
$=$ $\lambda_X( \Sigma(L'), X_1(L'), \ldots, X_k(L'))$ $=$ $X(f(L'))$.
We get $f(L) \CFFD f(L')$.
So $L \cgr f(L) \CFFD f(L') \cgr L'$ and $L \cgr L'$.\qed

The following proof illustrates, in a simple context, the use of
Lemma~\ref{L:nocongr}.
The $f$ in the proof preserves the congruence and consequently also $\Sigma$,
$\Tr$, and $\Inf$.
It throws away all information on $\Sf$ and $\Div$, except what can be derived
from $\Tr$ and $\Inf$ via such facts as $\Div(L) \subseteq \Tr(L)$.
Throwing information away is possible because of the assumption $\DLG \cgr
\LLG$.
Although $\Div(f(L))$ is neither $\emptyset$ nor $\Sigma(L)^*$, it contains no
genuine information, because it is fully determined by $\Tr(L)$.

\begin{theorem}\label{T:Tr}
If ``$\cgr$'' is a congruence, ``$\CFFD$'' implies ``$\cgr$'', ``$\cgr$''
preserves $\Inf$, and $\DLG \cgr \LLG$, then ``$\cgr$'' is the equivalence
induced by $\Sigma$, $\Tr$, and $\Inf$.
\end{theorem}
\proof
By Lemma~\ref{L:Inf=>Tr}, ``$\cgr$'' preserves $\Sigma$ and $\Tr$.
Let $f(L) = L \pp \LLG$.
We have $L \bs L \pp \DLG \cgr L \pp \LLG = f(L)$.
Clearly $\Sf(f(L)) = \emptyset$, $\Div(f(L)) = \Tr(L)$, and $\Inf(f(L)) =
\Inf(L)$.
Lemma~\ref{L:nocongr} gives the claim if we choose $k = 2$, $X_1 = \Tr$, and
$X_2 = \Inf$.\qed

In forthcoming proofs, we will employ renaming and hiding such that precisely
those actions synchronize which we want to synchronize.
To facilitate that, we introduce the following notation for temporarily
attaching an integer $i$ to symbols other than $\tau$.
In the notation, $a \neq \tau \notin A$ and $a_j \neq \tau$ for $1 \leq j$.
\begin{center}
\begin{tabular}{rcl}
$\new{a}{i}$ & $:=$ & $(a,i)$\\
$\new{(a_1 a_2 \cdots a_n)}{i}$ & $:=$ & $\new{a_1}{i} \new{a_2}{i} \cdots
\new{a_n}{i}$\\
$\new{(a_1 a_2 \cdots)}{i}$ & $:=$ & $\new{a_1}{i} \new{a_2}{i} \cdots$\\
$\new{A}{i}$ & $:=$ & $\{ \new{a}{i} \mid a \in A\}$\\
$\newup{L}{i}$ & $:=$ & $L\Phi$, where $\Phi = \{ (a,\new{a}{i}) \mid a \in
\Sigma \}$\\
$\newdn{L}{i}$ & $:=$ & $L\Phi$, where $\Phi = \{ (\new{a}{i},a) \mid
\new{a}{i} \in \Sigma \}$
\end{tabular}
\end{center}
We will use this notation in the proof of the following lemma, to ensure that
certain sets are disjoint.

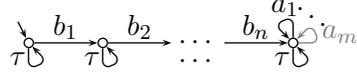
\begin{figure}[t]
\mbox{}\hfill\begin{pspicture}(128,35)
\ainit{6,13}
\aloopdr{6,13}\rput[r](5,7){$\tau$}
\arc(8,13)(32,13)\rput[B](20,16){$b_1$}
\aloopdr{34,13}\rput[r](33,7){$\tau$}
\arc(36,13)(60,13)\rput[B](48,16){$b_2$}
\rput(70,13){$\cdots$}
\rput(70,6){$\cdots$}
\arc(80,13)(104,13)\rput[B](92,16){$b_n$}
\aloopdr{106,13}\rput[r](105,7){$\tau$}
\aloopul{106,13}\rput[b](103,23){$a_1$}
\GRAY{\aloopru{106,13}\rput[lb](117,13){$a_m$}}
\rput(113,26){$\ddots$}
\psst(6,13){2}\psst(34,13){2}\psst(106,13){2}
\end{pspicture}\hfill\mbox{}
\caption{An LTS for detecting the trace $b_1 b_2 \cdots b_n$.}\label{F:testTr}
\end{figure}

\begin{lemma}\label{L:trdiff}
If ``$\cgr$'' is a congruence and preserves $\Sigma$ but not $\Tr$, then for
any set $A$ such that $\tau \notin A$ there are LTSs $M_1^A$ and $M_2^A$ such
that $M_1^A \cgr M_2^A$, $\Sigma(M_1^A) = \Sigma(M_2^A) = A$, $\Sf(M_1^A) =
\Sf(M_2^A) = \emptyset$, $\Tr(M_1^A) = \Div(M_1^A) = A^*$, $\Tr(M_2^A) =
\Div(M_2^A) = \{\varepsilon\}$, $\Inf(M_1^A) = A^\omega$, and $\Inf(M_2^A) =
\emptyset$.
\end{lemma}
\proof
There are $M_1$, $M_2$, and $\sigma$ such that $M_1 \cgr M_2$ and $\sigma \in
\Tr(M_1) \setminus \Tr(M_2)$.
Let $\Sigma_M = \Sigma(M_1) = \Sigma(M_2)$, $b_1 \cdots b_n =
\new{\sigma}{1}$, and $\{a_1, \ldots{\gray, a_m}\} = \new{A}{2}$.
When $i \in \{1,2\}$, let
$$M_i^A \ =\ \newdn{\,(T_\sigma \pp \newup{M_i}{1}) \setminus
\new{\Sigma_M}{1}\,}{2}\textrm{ ,}$$
where $\Sigma(T_\sigma) = \new{\Sigma_M}{1} \cup \new{A}{2}$ and otherwise
$T_\sigma$ is like in Fig.~\ref{F:testTr}.
In the rightmost state of $T_\sigma$, there is an $a$-loop for every $a \in
\new{A}{2}$.
We have $M_1^A \cgr M_2^A$ because of the congruence property of ``$\cgr$''.
Because $\new{X}{1}$ and $\new{Y}{2}$ are disjoint for any $X$ and $Y$, we
have $\Sigma(\,(T_\sigma \pp \newup{M_i}{1}) \setminus \new{\Sigma_M}{1}\,)$
$=$ $(\new{\Sigma_M}{1} \cup \new{A}{2} \cup \new{\Sigma_M}{1}) \setminus
\new{\Sigma_M}{1} = \new{A}{2}$.
This yields $\Sigma(M_1^A) = \Sigma(M_2^A) = A$.
Because $T_\sigma$ does not have stable states, we get $\Sf(M_1^A) =
\Sf(M_2^A) = \emptyset$.

Thanks to how renaming and hiding are used, $T_\sigma$ executes its
$a$-transitions without $M_i$, while it executes its $b$-transitions
synchronously with $M_i$ and invisibly from the environment.
The environment sees the $a$-transitions with their $A$-names (instead of
$\new{A}{2}$-names).
Because of the synchronization with $T_\sigma$, $M_i$ can only execute
$\tau$-transitions and some prefix of $\sigma$.
Because $M_1$ can but $M_2$ cannot execute $\sigma$ completely, $T_\sigma$ can
reach its rightmost state when in $M_1^A$ but not when in $M_2^A$.
Therefore, $\Tr(M_1^A) = \Div(M_1^A) = A^*$, $\Inf(M_1^A) = A^\omega$,
$\Tr(M_2^A) = \Div(M_2^A) = \{\varepsilon\}$, and $\Inf(M_2^A) =
\emptyset$.\qed

Let $\Run{A}$ denote the LTS whose alphabet is $A$, which has one state, and
whose transitions are $\{ (\hat s, a, \hat s) \mid a \in A \}$ (please see
Fig.~\ref{F:simpleLTSs}).
The following theorem tells that all remaining congruences in this section
preserve $\Sigma$ and $\Tr$.

\begin{theorem}\label{T:Sigma}
If ``$\cgr$'' is a congruence, ``$\CFFD$'' implies ``$\cgr$'', ``$\cgr$''
preserves $\Sigma$ but not $\Tr$, and $\DLG \cgr \LLG$, then ``$\cgr$'' is the
equivalence induced by $\Sigma$.
\end{theorem}
\proof
Let $L$ be any LTS and $A = \Sigma(L)$.
We can reason $\Run{A} \bs \Run{A} \pp \DLG \cgr \Run{A} \pp \LLG \CFFD M_1^A
\cgr M_2^A$, where $M_1^A$ and $M_2^A$ are the LTSs in Lemma~\ref{L:trdiff}.
By choosing $f(L) = L \pp M_2^A$ we get $L \bs L \pp \Run{A} \cgr L \pp M_2^A
= f(L)$, so $L \cgr f(L)$.
Because $M_2^A$ lacks stable failures and blocks all visible actions of $L$ in
$L \pp M_2^A$, we have $\Sf(f(L)) = \emptyset$, $\Div(f(L)) =
\{\varepsilon\}$, and $\Inf(f(L)) = \emptyset$.
They are constants, so Lemma~\ref{L:nocongr} yields the claim if we choose $k
= 0$ in it.\qed
It is widely known that the equivalence induced by $\Sigma$ and $\Tr$ is a
congruence.
The next theorem says that climbing up the ladder, $\Inf$ has to be preserved.

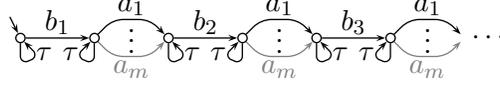
\begin{figure}[t]
\mbox{}\hfill\begin{pspicture}(187,26)
\ainit{4,13}
\aloopdr{4,13}\rput[l](10,7){$\tau$}
\arc(6,13)(30,13)\rput[B](18,16){$b_1$}
\arcud{32,13}{$a_1$}
\GRAY{\arcdu{32,13}{$a_m$}}
\aloopdl{32,13}\rput[r](26,7){$\tau$}
\rput(46,15){$\vdots$}
\aloopdr{60,13}\rput[l](66,7){$\tau$}
\arc(62,13)(86,13)\rput[B](74,16){$b_2$}
\arcud{88,13}{$a_1$}
\GRAY{\arcdu{88,13}{$a_m$}}
\aloopdl{88,13}\rput[r](82,7){$\tau$}
\rput(102,15){$\vdots$}
\aloopdr{116,13}\rput[l](122,7){$\tau$}
\arc(118,13)(142,13)\rput[B](130,16){$b_3$}
\arcud{144,13}{$a_1$}
\GRAY{\arcdu{144,13}{$a_m$}}
\aloopdl{144,13}\rput[r](138,7){$\tau$}
\rput(158,15){$\vdots$}
\rput(182,13){$\cdots$}
\psst(4,13){2}\psst(32,13){2}\psst(60,13){2}\psst(88,13){2}\psst(116,13){2}
\psst(144,13){2}
\end{pspicture}\hfill\mbox{}
\caption{An LTS for detecting the infinite trace $b_1 b_2
\cdots$.}\label{F:testinftr}
\end{figure}

\begin{theorem}\label{T:Trace}
If ``$\cgr$'' is a congruence, ``$\CFFD$'' implies ``$\cgr$'', ``$\cgr$''
preserves $\Tr$ but not $\Inf$, and $\DLG \cgr \LLG$, then ``$\cgr$'' is the
equivalence induced by $\Sigma$ and $\Tr$.
\end{theorem}
\proof
There are $M_1$, $M_2$, and $\xi$ such that $M_1 \cgr M_2$ and $\xi \in
\Inf(M_1) \setminus \Inf(M_2)$.
Because ``$\cgr$'' preserves $\Tr$, Theorem~\ref{T:noSigma} implies that it
also preserves $\Sigma$, so we may let $\Sigma_M = \Sigma(M_1) = \Sigma(M_2)$.
Let $b_1 b_2 \cdots = \new{\xi}{1}$.
Let $A$ be any set such that $\tau \notin A$.
Let $\{a_1, a_2, \ldots{\gray, a_m}\} = \new{A}{2}$.
When $i \in \{1,2\}$, let
$$M_i^A \ =\ \newdn{\,(T_\xi \pp \newup{M_i}{1}) \setminus
\new{\Sigma_M}{1}\,}{2}\textrm{ ,}$$
where $\Sigma(T_\xi) = \new{\Sigma_M}{1} \cup \new{A}{2}$ and otherwise
$T_\xi$ is like in Fig.~\ref{F:testinftr}.
Because $\new{X}{i}$ and $\new{Y}{j}$ are disjoint whenever $i \neq j$, we
have $\Sigma(M_1^A) = \Sigma(M_2^A) = A$.
Thanks to the $\tau$-loops in Fig.~\ref{F:testinftr}, $\Sf(M_1^A) = \Sf(M_2^A)
= \emptyset$.
By~(\ref{E:Inf}), $M_1$ can execute any finite prefix of $\xi$.
This yields $\Tr(M_1^A) = \Div(M_1^A) = A^*$.
By the congruence property $M_1^A \cgr M_2^A$.
Because ``$\cgr$'' preserves $\Tr$, also $\Tr(M_2^A) = \Div(M_2^A) = A^*$.
Since $M_1$ can but $M_2$ cannot execute $\xi$ completely, we get $\Inf(M_1^A)
= A^\omega$ and $\Inf(M_2^A) = \emptyset$.

Let $L$ be any LTS and $A = \Sigma(L)$.
We can reason $\Run{A} \bs \Run{A} \pp \DLG \cgr \Run{A} \pp \LLG \CFFD M_1^A
\cgr M_2^A$, and $L \bs L \pp \Run{A} \cgr L \pp M_2^A$.
Lemma~\ref{L:nocongr} gives the claim if we choose $k = 1$, $X_1 = \Tr$, and
$f(L) = L \pp M_2^A$, because then $L \cgr f(L)$, $\Sf(f(L)) = \emptyset$,
$\Div(f(L)) = \Tr(L)$, and $\Inf(f(L)) = \emptyset$.\qed

The above proof constructed a function $f(L)$ that throws away all information
(modulo ``$\CFFD$'') except $\Sigma$ and $Tr$, while preserving ``$\cgr$''.
Information on $\Sf$ and $\Div$ was thrown away using the assumption that
$\DLG \cgr \LLG$.
Information on $\Inf$ was thrown away by starting with an arbitrary difference
on $\Inf$, and amplifying it to a function
$$f'(L,M) \ =\ L \pp \newdn{\,(T_\xi \pp \newup{M}{1}) \setminus
\new{\Sigma_M}{1}\,}{2}$$
so that $f'(L,M_1)$ preserves $\Inf(L)$ while $f'(L,M_2)$ wipes it out.
The permission to also throw away all information on $\Sf$ and $\Div$
simplified the design.
We have $L \cgr f'(L,M_1) \cgr f'(L,M_2) = f(L)$, where the first ``$\cgr$''
takes care of $\Sf$ and $\Div$, and the second of $\Inf$.
In the construction of $f$, despite the use of notation defined in this
section, ultimately only operators from Section~\ref{S:def} were used.

By Theorem~\ref{T:Tr}, there are no more congruences in this section.
In conclusion, altogether precisely four abstract linear-time congruences
satisfy $\DLG \cgr \LLG$: those induced by the first zero, one, two, or three
of $\Sigma$, $\Tr$, and $\Inf$.
That also the last one is a congruence is widely known and proven, e.g.,
in~\cite{VaT95}.

\section{When Deadlock Is Bothlock Is Not Livelock}\label{S:BLnotLL}

In this section we show that only three congruences that are implied by
``$\CFFD$'' satisfy $\DLG \cgr \BLG \not\cgr \LLG$.
We also introduce an ``internal choice'' operator $L \sqcap L'$ that will be
used in this and later sections.
It can be built from the four operators in Section~\ref{S:def}, so any
equivalence that is a congruence with respect to them also is a congruence
with respect to internal choice.

The next theorem tells that all congruences in this section preserve $\Sf$.
By Theorem~\ref{T:noSigma}, they also preserve $\Sigma$.

\begin{theorem}\label{T:Sf}
If ``$\cgr$'' is a congruence, ``$\CFFD$'' implies ``$\cgr$'', and $\BLG
\not\cgr \LLG$, then ``$\cgr$'' preserves $\Sf$.
\end{theorem}
\proof
If ``$\cgr$'' does not preserve $\Sf$, then there are $M_1$, $M_2$, $\sigma =
b_1 \cdots b_n$, and $A = \{a_1, \ldots{\gray, a_m}\}$ such that $M_1 \cgr
M_2$ and $(\sigma, A) \in \Sf(M_1) \setminus \Sf(M_2)$.
Let $\Sigma_M = \Sigma(M_1)$.
If $\Sigma(M_2) \neq \Sigma_M$, then Theorem~\ref{T:noSigma} yields $\LLG \cgr
\BLG$.
Otherwise, if $T_\sigma^A$ is the LTS in Fig.~\ref{F:testDl} with
$\Sigma(T_\sigma^A) = \Sigma_M$, we have $(M_2 \pp T_\sigma^A) \setminus
\Sigma_M \CFFD \LLG$ and $(M_1 \pp T_\sigma^A) \setminus \Sigma_M \CFFD \BLG$.
In both cases, $\LLG \cgr \BLG$, contrary to our assumption.
Thus ``$\cgr$'' preserves $\Sf$.\qed

The equivalence induced by $\Sigma$ and $\Sf$ is a congruence~\cite{VaT95}.
However, if the so-called interrupt operator found in CSP or
Lotos~\cite{BoB87} is employed, then it is no longer a
congruence~\cite{Val95}.

To prove the next result, the ``internal choice'' operator of CSP would be
handy.
It is equivalent to the CCS expression $\tau.P + \tau.Q$.
Fortunately, it can be built from our operators.
\begin{eqnarray}
L_1 \sqcap L_2 & := & \big((\,L_C \pp c_1.\newup{L_1}{1} \pp
c_2.\newup{L_2}{2}\,) \setminus \{c_1, c_2\}\big)\Phi \textrm{ ,}\nonumber
\end{eqnarray}
where $c_1 = \new{1}{0}$, $c_2 = \new{2}{0}$, $\Phi = \{(\new{a}{1},a) \mid a
\in \Sigma_1 \} \cup \{(\new{a}{2},a) \mid a \in \Sigma_2 \}$, and $L_C$ has
$S_C = \{\hat s_C, s_C\}$, $\Sigma_C = \{c_1, c_2\}$, $\Delta_C = \{ (\hat
s_C,c_1,s_C), (\hat s_C,c_2,s_C) \}$, and $\hat s_C \neq s_C$ (please see
Fig.~\ref{F:simpleLTSs}).
(Here $c_1$ and $c_2$ could be any distinct symbols that are not in
$\new{\Sigma_1}{1} \cup \new{\Sigma_2}{2}$.)

The CFFD-semantics of this operator is simple:
\begin{eqnarray}
\Sigma(L \sqcap L') & = & \Sigma(L) \cup \Sigma(L')\nonumber\\
\Sf(L \sqcap L') & = & \Sf(L) \cup \Sf(L')\nonumber\\
\Div(L \sqcap L') & = & \Div(L) \cup \Div(L')\nonumber\\
\Inf(L \sqcap L') & = & \Inf(L) \cup \Inf(L')\nonumber
\end{eqnarray}

The next congruence in this section also preserves $\Tr$.

\begin{theorem}\label{T:SfTr}
If ``$\cgr$'' is a congruence, ``$\CFFD$'' implies ``$\cgr$'', ``$\cgr$''
preserves $\Sf$ but not $\Tr$, and $\DLG \cgr \BLG$, then ``$\cgr$'' is the
equivalence induced by $\Sigma$ and $\Sf$.
\end{theorem}
\proof
Let $L$ be any LTS and $A = \Sigma(L)$.
By Theorem~\ref{T:noSigma}, ``$\cgr$'' preserves $\Sigma$.
The assumptions of Lemma~\ref{L:trdiff} hold, so we can use its $M_1^A$ and
$M_2^A$.
Let $f(L) = (L \pp \BLG) \sqcap M_1^A$.
We have $L \bs L \pp \DLG \cgr L \pp \BLG \CFFD (L \pp \BLG) \sqcap M_2^A$, so
$L \cgr f(L)$.
Furthermore, $\Sf(f(L)) = \Sf(L)$, $\Div(f(L)) = A^* = \Sigma(L)^*$, and
$\Inf(f(L)) = A^\omega = \Sigma(L)^\omega$.
With $k=1$ and $X_1 = \Sf$, Lemma~\ref{L:nocongr} gives the claim.\qed
The equivalence induced by $\Sigma$, $\Tr$, and $\Sf$ is a
congruence~\cite{VaT95}.

At the next level, also $\Inf$ has to be preserved.
To prove this, we need a more complicated construction than in the proof of
Theorem~\ref{T:Trace}, because this time $\Sf$ has to be preserved.

\begin{figure}[t]
\hfill
\begin{pspicture}(40,49)(-13,0)
\ainit{10,41}\rput[tr](0,49){$R_1^A$}
\aloopld{10,41}\rput[t](3,33){$\tau$}
\arc(10,39)(10,15)\rput[l](11,30.5){$\tau$}
\arcud{10,13}{$a_1$}
\GRAY{\arcdu{10,13}{$a_m$}}
\rput(24,15){$\vdots$}
\aarc{20}(38,15)(38,41)(12,41)\rput[tr](38,40){$\tau$}
\psst(10,41){2}\psst(10,13){2}\psst(38,13){2}
\end{pspicture}\hfill{}
\begin{pspicture}(137,49)(-10,0)
\ainit{12,41}\rput[tr](3,49){$R_2^A$}
\arc(14,41)(38,41)\rput[B](26,43){$\tau$}
\arc(42,41)(66,41)\rput[B](54,43){$\tau$}
\arc(40,39)(40,15)\rput[l](41,31){$\tau$}
\arc(70,41)(94,41)\rput[B](82,43){$\tau$}
\arc(68,39)(68,15)\rput[l](69,31){$\tau$}
\arc(98,41)(122,41)\rput[B](110,43){$\tau$}
\arc(96,39)(96,15)\rput[l](97,31){$\tau$}
\rput(132,41){$\cdots$}\rput(132,27){$\cdots$}

\aloopdl{12,13}\rput[r](6,7){$\tau$}
\acrud{40,13}{$a_1$}
\GRAY{\acrdu{40,13}{$a_m$}}
\rput(26,15){$\vdots$}
\acrud{68,13}{$a_1$}
\GRAY{\acrdu{68,13}{$a_m$}}
\rput(54,15){$\vdots$}
\acrud{96,13}{$a_1$}
\GRAY{\acrdu{96,13}{$a_m$}}
\rput(82,15){$\vdots$}
\acrud{124,13}{$a_1$}
\GRAY{\acrdu{124,13}{$a_m$}}
\rput(110,15){$\vdots$}
\rput(132,13){$\cdots$}

\psst(12,41){2}\psst(40,41){2}\psst(68,41){2}\psst(96,41){2}
\psst(12,13){2}\psst(40,13){2}\psst(68,13){2}\psst(96,13){2}
\end{pspicture}\hfill{}
\caption{$R_1^A$ has $\Sigma(R_1^A) = A = \{a_1, \ldots{\gray, a_m}\}$,
$\Sf(R_1^A) = A^* \times \{\emptyset\}$, $\Div(R_1^A) = A^*$, and $\Inf(R_1^A)
= A^\omega$.
$R_2^A$ has the same except $\Inf(R_2^A) = \emptyset$.}\label{F:Inf2}
\end{figure}
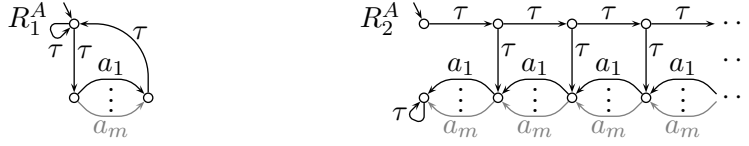

\begin{theorem}\label{T:TrSf}
If ``$\cgr$'' is a congruence, ``$\CFFD$'' implies ``$\cgr$'', ``$\cgr$''
preserves $\Tr$ and $\Sf$ but not $\Inf$, and $\DLG \cgr \BLG$, then
``$\cgr$'' is the equivalence induced by $\Sigma$, $\Tr$, and $\Sf$.
\end{theorem}
\proof
Let $M_1 \cgr M_2$, $\xi \in \Inf(M_1) \setminus \Inf(M_2)$, $b_1 b_2 \cdots =
\new{\xi}{1}$, and $A$ be any set such that $\tau \notin A$.
By Theorem~\ref{T:noSigma}, ``$\cgr$'' preserves $\Sigma$.
Let $\Sigma_M = \Sigma(M_1) = \Sigma(M_2)$.
When $i \in \{1,2\}$, let
$$M_i^A \ =\ \newdn{\,(T_\xi \pp \newup{M_i}{1}) \setminus
\new{\Sigma_M}{1}\,}{2}\textrm{ ,}$$
where $\Sigma(T_\xi) = \new{\Sigma_M}{1} \cup \new{A}{2}$ and otherwise
$T_\xi$ is like in Fig.~\ref{F:testinftr}.

Because $T_\xi$ does not have stable states, we have $\Sf(M_1^A) = \Sf(M_2^A)
= \emptyset$.
Because $\newup{M_2}{1}$ lacks the infinite trace $b_1 b_2 \cdots$, we have
$\Inf(M_2^A) = \emptyset$.
Let $R_1^A$ and $R_2^A$ be the LTSs in Fig.~\ref{F:Inf2}.
We have $\Div(R_2^A) = A^*$.
These imply $M_2^A \sqcap R_2^A \CFFD R_2^A$.
On the other hand, $\Inf(M_1^A) = \Inf(R_1^A) = A^\omega$, $\Sf(R_1^A) =
\Sf(R_2^A)$, and also $\Div(R_1^A) = A^*$, so $M_1^A \sqcap R_2^A \CFFD
R_1^A$.
As a consequence, $R_1^A \CFFD M_1^A \sqcap R_2^A \cgr M_2^A \sqcap R_2^A
\CFFD R_2^A$.

By choosing $A = \Sigma(L)$ and $f(L) = L \pp R_2^A$ we get $L \bs L \pp \DLG
\cgr L \pp \BLG \CFFD L \pp R_1^A \cgr L \pp R_2^A$, so $L \cgr f(L)$.
We have $\Sf(f(L)) = \Sf(L)$, $\Div(f(L)) = \Tr(f(L)) = \Tr(L)$, and
$\Inf(f(L)) = \emptyset$.
With $k = 2$, $X_1 = \Tr$, and $X_2 = \Sf$, Lemma~\ref{L:nocongr} gives the
claim.\qed

The equivalence induced by $\Sigma$, $\Tr$, $\Sf$, and $\Inf$ is the
intersection of the equivalences induced by $(\Sigma, \Tr, \Sf)$ and $(\Sigma,
\Tr, \Inf)$.
So it is the intersection of two congruences and thus a congruence.
We now show that it is the last one in this section.

\begin{theorem}\label{T:TrSfail}
If ``$\cgr$'' is a congruence, ``$\CFFD$'' implies ``$\cgr$'', ``$\cgr$''
preserves $\Sf$ and $\Inf$, and $\DLG \cgr \BLG$, then ``$\cgr$'' is the
equivalence induced by $\Sigma$, $\Tr$, $\Sf$, and $\Inf$.
\end{theorem}
\proof
By Lemma~\ref{L:Inf=>Tr}, ``$\cgr$'' preserves $\Sigma$ and $\Tr$.
Let $f(L) = L \pp \BLG$.
We have $L \bs L \pp \DLG \cgr L \pp \BLG = f(L)$.
Clearly $\Sf(f(L)) = \Sf(L)$, $\Div(f(L)) = \Tr(L)$, and $\Inf(f(L)) =
\Inf(L)$.
Letting $k = 3$, $X_1 = \Tr$, $X_2 = \Sf$, and $X_3 = \Inf$,
Lemma~\ref{L:nocongr} gives the claim.\qed

To summarize, precisely three abstract linear-time congruences satisfy $\DLG
\cgr \BLG \not\cgr\LLG$: those induced by $(\Sigma, \Sf)$, $(\Sigma, \Tr,
\Sf)$, and $(\Sigma, \Tr, \Sf, \Inf)$.

\section{When All Three Are Non-equivalent}\label{S:ALLnot}

\newcommand{\minD}{\mi{minD}}
\newcommand{\extT}{\mi{extT}}
\newcommand{\extI}{\mi{extI}}
\newcommand{\anT}{\mi{anT}}
\newcommand{\anI}{\mi{anI}}
\newcommand{\eanI}{\mi{eanI}}
\newcommand{\aenI}{\mi{aenI}}
\begin{figure}[t]
\mbox{}\hfill\begin{pspicture}(-12,-12)(301,221)
\newcommand{\blob}{\pspolygon[linewidth=3pt,linecolor=lightgray,linearc=3mm]}
\blob(-10,80)(-10,113)(120,40)(120,-10)(100,-10)
  \rput[l](123,16){\footnotesize Section~\ref{S:DLisLL}:
  $\DLG \cgr \LLG$}
\blob(22,110)(22,140)(152,67)(152,37)
  \rput[l](155,54){\footnotesize Section~\ref{S:BLnotLL}:
  $\DLG \cgr \BLG \not\cgr \LLG$}
\blob(22,154)(22,225)(152,152)(152,81)
  \rput[l](155,118){\footnotesize Section~\ref{S:ALLnot}:
  $\DLG \not\cgr \LLG \not\cgr \BLG \not\cgr \DLG$}
\psline[linewidth=3pt,linecolor=lightgray,linearc=3mm,linestyle=dashed]
(22,170)(86,134)(88,156)(152,120)

\newcommand{\Qx}{108}\newcommand{\Qy}{32}
\newcommand{\QT}[1]{\rput(-32,18){#1}}
\newcommand{\QA}[1]{\rput(-64,36){#1}}
\newcommand{\QI}[1]{\rput(-96,54){#1}}
\newcommand{\QF}[1]{\rput(30,30){#1}}
\newcommand{\QM}[1]{\rput(0,55){#1}}
\newcommand{\QE}[1]{\rput(0,85){#1}}
\newcommand{\Qc}[1]{\rput(\Qx,\Qy){\pscircle*(0,0){2}
\rput[r](-3,-2){\tiny{#1}}}}
\newcommand{\Qd}[1]{\rput(\Qx,\Qy){\rput[r](-3,-8){\tiny{#1}}}}
\newcommand{\Qe}[1]{\rput(\Qx,\Qy){\rput[r](-3,-14){\tiny{#1}}}}
\newcommand{\Qline}[2]{\psline[linewidth=2pt,linecolor=white](#1)(#2)
\psline(#1)(#2)}
\newcommand{\Qu}{\rput(\Qx,\Qy){\Qline{0,-55}{0,0}}}
\newcommand{\Quu}{\rput(\Qx,\Qy){\Qline{0,-30}{0,0}}}
\newcommand{\Ql}{\rput(\Qx,\Qy){\Qline{32,-18}{0,0}}}
\newcommand{\Qll}{\rput(\Qx,\Qy){\Qline{64,-36}{0,0}}}
\newcommand{\Qrrrr}{\rput(\Qx,\Qy){\Qline{-30,-30}{0,0}}}

\QF{\Qrrrr}\QF{\QT{\Qrrrr\Ql}}\QF{\QI{\Qrrrr\Qll}}
\QF{\QM{\Qu}}\QF{\QM{\QT{\Qu\Ql}}}\QF{\QM{\QI{\Qu\Qll}}}
\QF{\QE{\QT{\Quu}}}\QF{\QE{\QA{\Ql}}}\QF{\QE{\QI{\Quu\Ql}}}
\Quu\QT{\Ql}\QI{\Qll}

\rput(0,-30){\Qc{}}\Qc{$\Sigma$}\QT{\Qc{$\Tr$}}\QI{\Qc{$\Inf$}}
\QF{\Qc{$\Sf$\,}}\QF{\QT{\Qc{}}}\QF{\QI{\Qc{}}}
\QF{\QM{\Qc{$\minD$}\Qd{$\anT$}\Qe{$\anI$}}}
  \QF{\QM{\QT{\Qc{}}}}\QF{\QM{\QI{\Qc{}}}}
\QF{\QE{\QT{\Qc{$\Div$}\Qd{$\eanI$}}}}\QF{\QE{\QA{\Qc{$\aenI$}}}}
  \QF{\QE{\QI{\Qc{$\CFFD$\,}}}}
\end{pspicture}\hfill\mbox{}
\caption{The congruences in Sections~\ref{S:DLisLL}, \ref{S:BLnotLL},
and~\ref{S:ALLnot} as a Hasse diagram.
Names in $\mi{italics}$ indicate the new preserved set(s).}\label{F:ALLnot}
\end{figure}
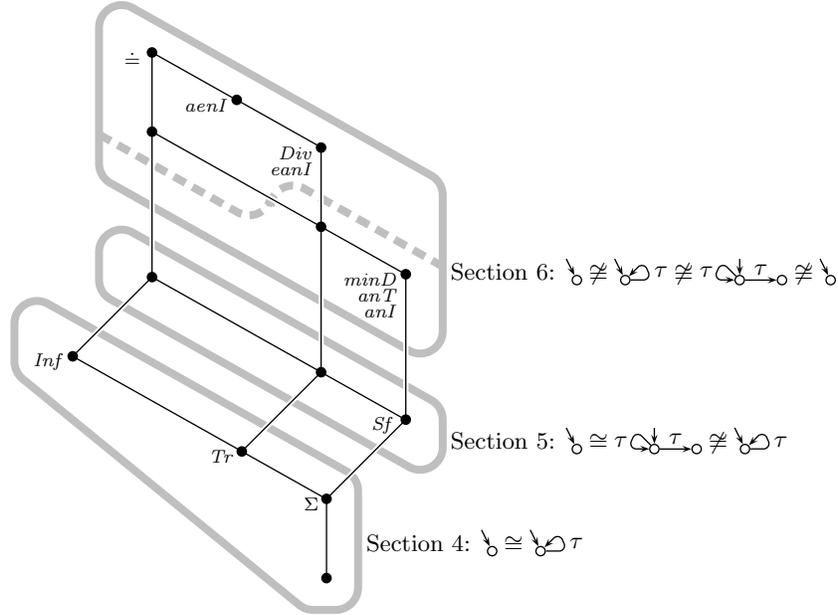

Figure~\ref{F:ALLnot} shows the results of the previous two sections and this
section.
In this section we survey the region where $\DLG \not\cgr \LLG \not\cgr \BLG
\not\cgr \DLG$.
We need new semantic sets.
They are defined in Subsection~\ref{S:minD}.
Many proofs in this and the next section treat end states of divergence traces
differently from end states of other traces.
For this to be possible, no state must be simultaneously the end state of both
a divergence trace and a nondivergent trace.
Subsection~\ref{S:Una} presents a construct with which LTSs can be transformed
into such a form, while preserving bisimilarity.
The theorems that there are no other congruences are presented in
Subsections~\ref{S:lower-ALLnot} and~\ref{S:upper-ALLnot}.

\subsection{New kinds of divergence and infinite traces}\label{S:minD}

In this subsection we define new semantic sets that are related to divergence
traces or infinite traces, and briefly study their properties.

\emph{Minimal divergence traces} $\minD$ are divergence traces whose proper
prefixes are not divergence traces.
Finite extensions of minimal divergence traces $\extT$ are an alternative
representation for the same information (assuming that $\Sigma$ is available).
Also infinite extensions $\extI$ can be derived from $\minD$.
\emph{Always-nondivergent traces} $\anT$ are traces which and whose proper
prefixes are not divergence traces, and similarly with
\emph{always-nondivergent infinite traces} $\anI$.
\emph{Eventually-always-nondivergent infinite traces} $\eanI$ may have a
finite number of divergence traces as prefixes.
\emph{Always-eventually-nondivergent infinite traces} $\aenI$ have an
infinite number of prefixes that are not divergence traces.
\begin{eqnarray}
\minD(L) & := & \{ a_1 \cdots a_n \in \Div(L) \mid \forall i; 0 \leq i < n:
a_1 \cdots a_i \notin \Div(L) \}\nonumber\\
\extT(L) & := & \{ a_1 \cdots a_n \in \Sigma(L)^* \mid \exists i; 0 \leq i
\leq n: a_1 \cdots a_i \in \minD(L) \}\nonumber\\
\extI(L) & := & \{ a_1 a_2 \cdots \in \Sigma(L)^\omega \mid \exists i; i \geq
0: a_1 \cdots a_i \in \minD(L) \}\nonumber\\
\anT(L) & := & \Tr(L) \setminus \extT(L)\nonumber\\
\anI(L) & := & \Inf(L) \setminus \extI(L)\nonumber\\
\eanI(L) & := & \{ a_1 a_2 \cdots \in \Inf(L) \mid \exists n; n \geq 0:
\forall i; i \geq n: a_1 \cdots a_i \notin \Div(L) \}\nonumber\\
\aenI(L) & := & \{ a_1 a_2 \cdots \in \Inf(L) \mid \forall n; n \geq 0:
\exists i; i \geq n: a_1 \cdots a_i \notin \Div(L) \}\nonumber
\end{eqnarray}
We have
$$
\begin{array}{rcll}
\minD(L) & = & \{ a_1 \cdots a_n \in \extT(L) \mid n=0 \vee a_1 \cdots a_{n-1}
\notin \extT(L) \} & \textrm{ ,}\\
\anT(L) & = & \Sf^\Tr(L) \setminus \extT(L) & \textrm{ , and}\\
\anI(L) & \subseteq &
\begin{array}{@{}ccccc@{}}
\eanI(L) & \subseteq & \aenI(L) & \subseteq & \Inf(L)
\end{array} &
\textrm{ .}
\end{array}
$$

\begin{lemma}\label{L:minD=>anT}
Any congruence that preserves $\minD$ also preserves $\Sigma$ and $\anT$.
\end{lemma}
\proof
By Theorem~\ref{T:noSigma}, it preserves $\Sigma$.
Let $L \cgr L'$ and $b \notin \Sigma(L) \cup \{\tau\}$.
For each $\sigma = a_1 \cdots a_n \in \Sigma(L)^*$, let $T_\sigma$ be the LTS
whose graph is $\LTSsbtloop$ and whose alphabet is $\Sigma(L) \cup \{b\}$.
We have $\sigma \in \Tr(L)$ if and only if $\sigma b \in \Div(L \pp
T_\sigma)$.
If $0 \leq i \leq n$, then $a_1 \cdots a_i \in \Div(L)$ if and only if $a_1
\cdots a_i \in \Div(L \pp T_\sigma)$.
Therefore, $\sigma \in \anT(L)$ if and only if $\sigma b \in \minD(L \pp
T_\sigma)$ if and only if $\sigma b \in \minD(L' \pp T_\sigma)$ if and only if
$\sigma \in \anT(L')$.\qed

\begin{lemma}\label{L:minD=>anI}
Any congruence that preserves $\minD$ also preserves $\Sigma$ and $\anI$.
\end{lemma}
\proof
By Theorem~\ref{T:noSigma}, it preserves $\Sigma$.
Let $L \cgr L'$, $a_1 a_2 \cdots \in \anI(L)$, and $T = \LTSxi$ with
$\Sigma(T) = \Sigma(L)$.
None of $a_1 \cdots a_i$ is in $\minD(L) = \minD(L')$, yielding $\minD(L' \pp
T) = \emptyset$.
On the other hand, $\varepsilon \in \minD(\,(L \pp T) \setminus \Sigma(L)\,) =
\minD(\,(L' \pp T) \setminus \Sigma(L')\,)$.
So $a_1 \cdots a_i \notin \Div(L' \pp T)$, $a_1 a_2 \cdots \in \Inf(L' \pp
T)$, $a_1 a_2 \cdots \in \Inf(L')$, and $a_1 a_2 \cdots \in \anI(L')$.\qed

\begin{lemma}\label{L:Div=>Tr}
Any congruence that preserves $\Div$ also preserves $\Tr$.
\end{lemma}
\proof
$\sigma \in \Tr(L) \Leftrightarrow \sigma \in \Div(L \pp \LLG)$.\qed

\begin{lemma}\label{L:Div=>eanI}
Any congruence that preserves $\Div$ also preserves $\Sigma$ and $\eanI$.
\end{lemma}
\proof
By Theorem~\ref{T:noSigma}, it preserves $\Sigma$.
Let $L \cgr L'$ and $\xi \in \eanI(L)$.
If no prefix of $\xi$ is in $\Div(L)$, then let $i = 1$, and otherwise let $i$
be $2$ plus the length of the longest prefix of $\xi$ that is in $\Div(L)$.
Let $a_i \notin \Sigma(L) \cup \{\tau\}$ and, when $1 \leq j \neq i$, let
$a_j$ be such that $\xi = a_1 \cdots a_{i-1} a_{i+1} \cdots$.
When $j \geq 0$, none of $a_1 \cdots a_{i-1} a_{i+1} \cdots a_{i+j}$ is in
$\Div(L) = \Div(L')$.
Let $T$ be the LTS whose alphabet is $\Sigma(L) \cup \{a_i\}$ and whose graph
is \LTSxi.
We have $a_1 \cdots a_{i+j} \notin \Div(L' \pp T)$ but $a_i \in \Div(\,(L \pp
T) \setminus \Sigma(L)\,) = \Div(\,(L' \pp T) \setminus \Sigma(L')\,)$.
As a consequence, $a_1 a_2 \cdots \in \Inf(L' \pp T)$, $\xi \in \Inf(L')$, and
$\xi \in \eanI(L')$.\qed

\subsection{Lower sub-region}\label{S:lower-ALLnot}

In this subsection we survey the part of the current region that is below the
dashed grey line in Fig.~\ref{F:ALLnot}.

Thanks to the next theorem, all congruences in this and the next section
preserve $\minD$.

\begin{theorem}\label{T:minD}
If ``$\cgr$'' is a congruence, ``$\CFFD$'' implies ``$\cgr$'', and $\DLG
\not\cgr \BLG$, then ``$\cgr$'' preserves $\minD$.
\end{theorem}
\proof
To derive a contradiction, let $L \cgr L'$ and $\sigma = a_1 \cdots a_n \in
\minD(L) \setminus \minD(L')$.
If there is an $i < n$ such that $a_1 \cdots a_i \in \minD(L')$, then swap the
roles of $L$ and $L'$, and use $a_1 \cdots a_i$ instead of $\sigma$.
Now no prefix of $\sigma$ is in $\minD(L')$.

If $\Sigma(L) \neq \Sigma(L')$, then Theorem~\ref{T:noSigma} yields $\DLG \cgr
\BLG$.
Otherwise, let $T_\sigma$ be the LTS whose graph is $\LTSsigma$ and alphabet
is $\Sigma(L)$.
We have $(L \pp T_\sigma) \setminus \Sigma(L) \CFFD \BLG$ or $(L \pp T_\sigma)
\setminus \Sigma(L) \CFFD \LLG$.
Furthermore, $(L' \pp T_\sigma) \setminus \Sigma(L) \CFFD \DLG$.
These imply $\DLG \cgr \BLG$ or $\DLG \cgr \LLG$.
If $\DLG \cgr \LLG$, then $\LLG \CFFD \BLG \pp \LLG \cgr \BLG \pp \DLG \bs
\BLG$, so $\DLG \cgr \BLG$.
All cases contradict the assumption $\DLG \not\cgr \BLG$.\qed

By Lemmas~\ref{L:minD=>anT} and~\ref{L:minD=>anI}, all congruences in this and
the next section also preserve $\Sigma$, $\anT$, and $\anI$.
Furthermore, in this section also Theorem~\ref{T:Sf} is applicable.
So $\Sf$ must be added to the semantics.
Doing so yields a congruence (proof skipped).
After adding $\Sf$, $\anT$ can be removed because $\anT(L) = \Sf^\Tr(L)
\setminus \extT(L)$.
Thus the weakest congruence in this section is induced by $\Sigma$, $\Sf$,
$\minD$, and $\anI$.

Adding $\Tr$ to this also yields a congruence.
The next theorem says that it is the next congruence.
We will need the construction in the proof of the theorem also in
Section~\ref{S:BLisLL}, so we isolate it in a lemma.

\begin{lemma}\label{L:noTr}
If ``$\cgr$'' is a congruence, ``$\CFFD$'' implies ``$\cgr$'', and ``$\cgr$''
preserves $\minD$ but not $\Tr$, then for every LTS $L$ there is an LTS $f(L)$
such that $L \cgr f(L)$, $\Tr(f(L)) = \anT(L) \cup \extT(L)$, $\Sf(f(L)) =
\Sf(L)$, $\Div(f(L)) = \extT(L)$, and $\Inf(f(L)) = \anI(L) \cup \extI(L)$.
\end{lemma}
\proof
By Theorem~\ref{T:noSigma}, ``$\cgr$'' preserves $\Sigma$.
Let $M_1 \cgr M_2$, $\sigma \in \Tr(M_1) \setminus \Tr(M_2)$, $b_1 \cdots b_n
= \new{\sigma}{1}$, and $c = \new{1}{0}$.
Let $\Sigma_M = \Sigma(M_1) = \Sigma(M_2)$.
For any LTS $L$, let $\Sigma_L = \Sigma(L)$, and let $g(L)$ be the LTS that is
obtained as follows: the label of every visible transition is transformed from
$a$ to $\new{a}{2}$, and a $c$-transition to the initial state of
Fig.~\ref{F:testTr} is added to every divergent state.
In Fig.~\ref{F:testTr}, $\{a_1, \ldots{\gray, a_m}\} = \new{\Sigma_L}{2}$.
The alphabet of $g(L)$ is $\{c\} \cup \new{\Sigma_M}{1} \cup
\new{\Sigma_L}{2}$.

When $i \in \{1,2\}$, let
$$f_i(L) \ =\ \newdn{\, (g(L) \pp c.\newup{M_i}{1}) \setminus (\{c\} \cup
\new{\Sigma_M}{1}) \,}{2} \textrm{ .}$$
By construction, $f_i(L)$ can do everything that $L$ can do, and also
try to hiddenly execute $c \new{\sigma}{1}$.
Attempts to execute $c \new{\sigma}{1}$ start at divergent states and, thanks
to the $\tau$-loops in Fig.~\ref{F:testTr}, do not lead to stable states.
Thus $\Sf(f_1(L)) = \Sf(f_2(L)) = \Sf(L)$.
Because $f_2(L)$ cannot execute $c \new{\sigma}{1}$ completely, $f_2(L) \CFFD
L$.
On the other hand, $f_1(L)$ can, so $\Tr(f_1(L)) = \anT(L) \cup \extT(L)$,
$\Div(f_1(L)) = \extT(L)$, and $\Inf(f_1(L)) = \anI(L) \cup \extI(L)$.
We have $L \CFFD f_2(L) \cgr f_1(L)$.
Therefore, $f_1$ qualifies as the $f$ of the claim.\qed

\begin{theorem}\label{T:minDTr}
If ``$\cgr$'' is a congruence, ``$\CFFD$'' implies ``$\cgr$'', and ``$\cgr$''
preserves $\Sf$ and $\minD$ but not $\Tr$, then ``$\cgr$'' is the equivalence
induced by $\Sigma$, $\Sf$, $\minD$, and $\anI$.
\end{theorem}
\proof
Lemma~\ref{L:minD=>anI} implies that ``$\cgr$'' preserves $\Sigma$ and $\anI$.
Because $\extT(L)$ and $\extI(L)$ are functions of $\Sigma(L)$ and $\minD(L)$,
the $f$ of Lemma~\ref{L:noTr} qualifies as the $f$ of Lemma~\ref{L:nocongr}
with $k = 3$, $X_1 = \Sf$, $X_2 = \minD$, and $X_3 = \anI$.\qed

\subsection{Unambiguation of LTSs}\label{S:Una}

\newcommand{\Det}{\ms{Det}}
\newcommand{\Una}{\ms{Una}}
\newcommand{\DD}{_\ms{D}}
\newcommand{\UU}{_\ms{U}}
\newcommand{\PPD}{\ms{PD}}
\newcommand{\PP}{_\ms{P}}
In this subsection we motivate and present two functions, called $\Una$ and
$\PPD$, that transform any LTS to a bisimilar LTS that has some useful
property.

\begin{figure}[t]
\mbox{}\hfill\begin{pspicture}(81,51)
\ainit{41,30}\rput[b](38,42){$L$}
\arc(39,30)(15,30)\rput[B](27,32){$b$}
\arcud{41,30}{$a$}\arcdu{41,30}{$b$}
\aloopul{13,30}\rput[r](6,33){$\tau$}
\aloopur{69,30}\rput[l](76,33){$a$}
\psst(13,30){2}\psst(41,30){2}\psst(69,30){2}
\end{pspicture}\hfill
\begin{pspicture}(12,0)(81,51)
\ainit{41,30}\rput[b](38,40){$\Det(L)$}
\arc(43,30)(67,30)\rput[B](55,32){$a$}
\aloopur{69,30}\rput[l](76,33){$a$}
\arc(41,28)(41,4)\rput[l](42,16){$b$}
\arc(42.4,3.4)(67.6,28.6)\rput(60,15){$a$}
\psst(41,30){2}\psst(69,30){2}\psst(41,2){2}
\end{pspicture}\hfill
\begin{pspicture}(81,51)
\ainit{41,30}\rput[b](38,40){$\Una(L)$}
\arc(39,30)(15,30)\rput[B](27,32){$b$}\arc(43,30)(67,30)\rput[B](55,32){$a$}
\aloopul{13,30}\rput[r](6,33){$\tau$}
\aloopur{69,30}\rput[l](76,33){$a$}
\arc(41,28)(41,4)\rput[l](42,16){$b$}
\arc(42.4,3.4)(67.6,28.6)\rput(60,15){$a$}
\psst(13,30){2}\psst(41,30){2}\psst(69,30){2}
\psst(41,2){2}
\end{pspicture}\hfill
\begin{pspicture}(81,51)
\ainit{41,30}\rput[b](38,40){$\PPD(L)$}
\arc(39,30)(15,30)\rput[B](27,32){$b$}\arc(43,30)(67,30)\rput[B](55,32){$a$}
\aloopul{13,30}\rput[r](6,33){$\tau$}
\aloopur{69,30}\rput[l](76,33){$a$}
\arc(41,28)(41,4)\rput[l](42,16){$b$}\arc(43,2)(67,2)\rput[B](55,4){$a$}
\aloopur{69,2}\rput[l](76,5){$a$}
\psst(13,30){2}\psst(41,30){2}\psst(69,30){2}
\psst(41,2){2}\psst(69,2){2}
\end{pspicture}\hfill\mbox{}
\caption{An example of $L$, $\Det(L)$, $\Una(L)$, and
$\PPD(L)$.}\label{F:DetUna}
\end{figure}
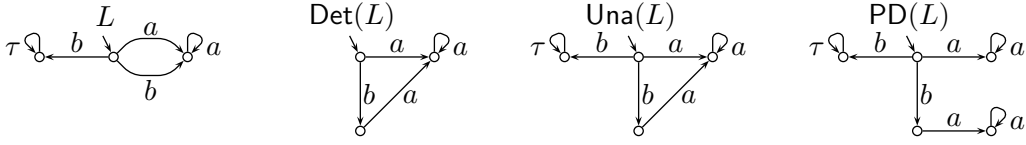

To continue the survey, we need a construction that preserves $\anI$ but not
$\Inf$.
It will block infinite traces after a minimal divergence trace, while not
affecting them before a minimal divergence trace.
Blocking does not have the desired effect unless \emph{all} executions of each
minimal divergence trace switch it on.
Forcing the execution of the switch at every divergent state does not suffice,
because the same trace may have two executions, one leading to a divergent and
the other to a nondivergent state.
This is exemplified by the trace $b$ of the $L$ in Fig.~\ref{F:DetUna}.
Even if we knew that this is the case with some nondivergent state, we could
not blindly implement the switch there, because it may also be reachable via
another, always-nondivergent trace.
An example is the trace $a$ in the figure.

To cope with this problem, we define a function $\Una$ that, given an LTS,
yields a bisimilar LTS where different traces lead to the same state only if
they have the same futures.
This is obtained by keeping track, in a new component of the state, of the set
of original states that can be reached via the trace that has been executed so
far.
To do that, we first define the \emph{determinization} of $L$ as the LTS
\begin{eqnarray}
\Det(L) & := & (S\DD, \Sigma, \Delta\DD, \hat s\DD)\textrm{, where}\nonumber\\
S_\sigma & = & \{ s \mid \hat s \Arr{\sigma} s \}\nonumber\\
S\DD & = & \{ S_\sigma \mid \sigma \in \Tr(L) \}\nonumber\\
\Delta\DD & = & \{ (S_\sigma, a, S_{\sigma a}) \mid a \neq \tau \wedge \sigma
a \in \Tr(L) \}\nonumber\\
\hat s\DD & = & S_\varepsilon\nonumber
\end{eqnarray}

\begin{lemma}\label{L:Det}
If $\sigma \in \Tr(L)$, then $\hat s\DD \Arr{\sigma} S_\sigma$.
If $\hat s\DD \Arr{\sigma} s\DD$, then $s\DD = S_\sigma$ and $\sigma \in
\Tr(L)$.
\end{lemma}
\proof
We prove the first claim by induction.
Clearly $\hat s\DD \Arr{\varepsilon} \hat s\DD = S_\varepsilon$.
If $\sigma a \in \Tr(L)$, then $a \neq \tau$ and $(S_\sigma, a, S_{\sigma a})
\in \Delta\DD$.
By the induction assumption $\hat s\DD \Arr{\sigma} S_\sigma$, yielding
$\hat s\DD \Arr{\sigma a} S_{\sigma a}$.

Also the second claim is proven by induction.
The definition of $\Delta\DD$ constructs no $\tau$-transitions, so if $\hat
s\DD \Arr{\varepsilon} s\DD$, then $s\DD = \hat s\DD = S_\varepsilon$.
Trivially $\varepsilon \in \Tr(L)$.
If $\hat s\DD \Arr{\sigma a} s\DD$, then consider the last transition along
the path.
By the definition of $\Delta\DD$, it is of the form $(S_\rho,a,S_{\rho a})$,
where $\rho a \in \Tr(L)$, $\hat s\DD \Arr{\sigma} S_\rho$, and $S_{\rho a} =
s\DD$.
By the induction assumption $S_\rho = S_\sigma$.
We get $s\DD = S_{\rho a} = \{ s \mid \exists s' \in S_\rho: s'
\Arr{a} s \} = \{ s \mid \exists s' \in S_\sigma: s' \Arr{a} s \} = S_{\sigma
a}$.
Because $\rho a \in \Tr(L)$, we have $\emptyset \neq S_{\rho a} = S_{\sigma
a}$, so $\sigma a \in \Tr(L)$.\qed

Then we define the \emph{unambiguation} of $L$ as
\begin{eqnarray}
\Una(L) & := & L \pp \Det(L)\textrm{ .}\nonumber
\end{eqnarray}

\begin{lemma}\label{L:Una-bs}
$\Una(L) \bs L$, that is, $\Una(L)$ is bisimilar with $L$.
\end{lemma}
\proof
Let $\Una(L) = (S\UU, \Sigma\UU, \Delta\UU, \hat s\UU)$.
The states of $\Una(L)$ are of the form $s\UU = (s,S_\sigma)$.
Let ``$\sim$'' $\subseteq$ $S \times S\UU$ be defined by $s \sim
(s',S_\sigma)$ if and only if $\hat s \Arr{\sigma} s = s'$.
We now show that ``$\sim$'' is a bisimulation.
``(1)'', etc., refer to the numbers in the definition on p.~\pageref{D:bisim}.
\begin{itemize}[label=(3a)]
\item[(1)] Clearly $\Sigma(\Una(L)) = \Sigma\UU = \Sigma \cup \Sigma = \Sigma =
\Sigma(L)$.

\item[(2)] We have $\hat s\UU = (\hat s, \hat s\DD) = (\hat s, S_\varepsilon)$ and
$\hat s \Arr{\varepsilon} \hat s$, so $\hat s \sim \hat s\UU$.

\item[(3)] Let $s \sim (s,S_\sigma)$, that is, $\hat s \Arr{\sigma} s$.

\item[(3a)] If $(s,\tau,s') \in \Delta$, then $((s,S_\sigma), \tau, (s',S_\sigma))
\in \Delta\UU$ and $\hat s \Arr{\sigma} s'$, yielding $s' \sim (s',S_\sigma)$.
If $(s,a,s') \in \Delta$ where $a \in \Sigma$, then $\hat s \Arr{\sigma a}
s'$.
The definition of $\Delta\DD$ yields $(S_\sigma, a, S_{\sigma a}) \in
\Delta\DD$, implying $((s,S_\sigma), a, (s',S_{\sigma a})) \in \Delta\UU$.
We have $s' \sim (s',S_{\sigma a})$.

\item[(3b)] If $((s,S_\sigma), \tau, (s',s'\DD)) \in \Delta\UU$, then by the
definitions of ``$||$'' and $\Delta\DD$ we have $(s,\tau,s') \in \Delta$ and
$s'\DD = S_\sigma$.
Furthermore, $\hat s \Arr{\sigma} s'$.
So $s' \sim (s',s'\DD)$.
If $((s,S_\sigma), a, (s',s'\DD)) \in \Delta\UU$ where $a \in \Sigma$, then
$(s,a,s') \in \Delta$.
It implies $\hat s \Arr{\sigma a} s'$.
We also have $(S_\sigma, a, s'\DD) \in \Delta\DD$, yielding $\sigma a \in
\Tr(\Det(L))$ and $s'\DD = S_{\sigma a}$ by Lemma~\ref{L:Det}.
Again $s' \sim (s',s'\DD)$.\qed
\end{itemize}

\noindent We say that a state of $\Una(L)$ is \emph{potentially divergent} if it can be
reached via a divergence trace, and \emph{certainly nondivergent} otherwise.
These phrases do not actually refer to the properties of the state but to the
properties of the traces that lead to it.
The essential useful property of $\Una(L)$ is stated in the following lemma.

\begin{lemma}\label{L:Una-div}
If state $s\UU$ of $\Una(L)$ is potentially divergent, then all traces that
lead to it belong to $\Div(L)$.
If state $s\UU$ of $\Una(L)$ is certainly nondivergent, then no trace that
leads to it belongs to $\Div(L)$.
\end{lemma}
\proof
If $\hat s\UU \Arr{\sigma} s\UU$, then $s\UU$ is of the form $(s, s\DD)$,
where $\hat s \Arr{\sigma} s$ and $\hat s\DD \Arr{\sigma} s\DD$.
By Lemma~\ref{L:Det}, $s\DD = S_\sigma$.
If also $\hat s\UU \Arr{\rho} s\UU$, then $S_\rho = s\DD = S_\sigma$.
If $\sigma \in \Div(L)$, then there is an $s' \in S_\sigma = S_\rho$ such that
$s' \ar{\tau^\omega}$, implying $\rho \in \Div(L)$.
Therefore, either none or all of the traces that lead to $s\UU$ are divergence
traces.\qed

In Fig.~\ref{F:DetUna}, the rightmost state of $L$ has been split to two
states in $\Una(L)$, a certainly nondivergent one led to by $a$ and a
potentially divergent one led to by $b$.

Then we define a function $\PPD$ that makes the following property hold while
preserving bisimilarity: for every state $s$, either no or all traces that
lead to $s$ have a divergence trace as a prefix.
This is obtained by adding a component to $\Una(L)$ that remembers if the
execution has gone through a divergence trace.
Formally, by $\PPD(L)$ we mean the LTS $(S\PP, \Sigma, \Delta\PP, \hat s\PP)$
that is obtained as follows.
Let $[\sigma] = \ms{pre}$ if $\sigma \in \anT(L)$ and $[\sigma] = \ms{post}$
otherwise.
Let $\sigma_\tau = \sigma$ and $\sigma_a = \sigma a$ if $a \in \Sigma$.
First $L$ is replaced by $\Una(L) = (S\UU, \Sigma, \Delta\UU, \hat s\UU)$.
Then let
\begin{eqnarray}
S\PP & = & \{ (s\UU, [\sigma]) \mid \hat s\UU \Arr{\sigma} s\UU \}\nonumber\\
\Delta\PP & = & \{ ((s\UU, [\sigma]), a, (s'\UU, [\sigma_a])) \mid \hat s\UU
\Arr{\sigma} s\UU \wedge (s\UU, a, s'\UU) \in \Delta\UU \}\nonumber\\
\hat s\PP & = & (\hat s\UU, [\varepsilon])\nonumber
\end{eqnarray}
We say that $(s\UU, x)$ is \emph{pre-divergent} if $x = \ms{pre}$ and
\emph{post-divergent} otherwise.

\begin{lemma}\label{L:PPD}
We have $\PPD(L) \bs L$.
If state $s\PP$ of $\PPD(L)$ is pre-divergent, then all traces that lead to it
belong to $\anT(L)$.
If state $s\PP$ of $\PPD(L)$ is post-divergent, then no trace that leads to it
belongs to $\anT(L)$.
\end{lemma}
\proof
We have $\PPD(L) \bs \Una(L) \bs L$, because the relation $(s\UU, [\sigma])
\sim s'\UU \Leftrightarrow s\UU = s'\UU$ is a bisimulation between $S\PP$ and
$S\UU$.

If $[\sigma_a] = \ms{pre}$, then $\sigma_a \in \anT(L)$, implying $\sigma \in
\anT(L)$ and $[\sigma] = \ms{pre}$.
Thus $\PPD(L)$ has no transitions from post-divergent to pre-divergent states.

Let $\hat s\PP \Arr{\rho} (s\UU,x)$ and $\rho \in \Div(L)$.
Because $(s\UU,x) \in S\PP$, there is a $\sigma$ such that $\hat s\UU
\Arr{\sigma} s\UU$ and $x = [\sigma]$.
Because $\rho \in \Div(L)$, $s\UU$ is potentially divergent.
By Lemma~\ref{L:Una-div}, all traces that lead to it are divergence traces.
That includes $\sigma$.
Thus $x = \ms{post}$.
As a consequence, each divergence trace only leads to post-divergent states.
By the first result in this proof, the same holds for each trace that has a
divergence trace as a prefix.

If an execution of $\PPD(L)$ leads to a post-divergent state, then $\hat s\PP$
is post-diver\-gent or the execution contains a transition of the form
$((s\UU, \ms{pre}), a, (s'\UU, \ms{post}))$.
In the first case, $[\varepsilon] = \ms{post}$, so $\varepsilon \in \Div(L)$.
In the second case, by the definition of $\Delta\PP$, there is a $\sigma$ such
that $\hat s\UU \Arr{\sigma} s\UU$, $\sigma \in \anT(L)$, and $\sigma_a \notin
\anT(L)$.
This implies $\sigma a \in \Div(L)$.
So $s'\UU$ is potentially divergent and all traces that lead to it are
divergence traces.
As a consequence, each post-divergent state has a divergence trace in each of
its histories.\qed

In Fig.~\ref{F:DetUna}, the rightmost state of $\Una(L)$ has been split to two
states in $\PPD(L)$, one such that all traces leading to it start with the
only divergence trace $b$, and another such that no trace leading to it starts
with $b$.

\subsection{Upper sub-region}\label{S:upper-ALLnot}

In this subsection we survey the rest of the current region.

Armed with $\PPD$, we can attack the case where $\Tr$, $\Sf$, and $\minD$ are
preserved, but $\Div$ and $\Inf$ are not.
This time there is no unique next congruence, but two.
Therefore, the proof consists of two parts, where the first throws away
information on divergence traces that are not minimal, and the second on
infinite traces that are not always-nondivergent.
Again, to reuse the construction in Section~\ref{S:BLisLL}, we present it as a
lemma that does not assume that $\Sf$ is preserved.

\begin{lemma}\label{L:noDiv}
Assume that ``$\cgr$'' is a congruence, ``$\CFFD$'' implies ``$\cgr$'', and
``$\cgr$'' preserves $\Tr$ and $\minD$ but not $\Div$.
\begin{enumerate}[label=(a)]

\item
For every LTS $L$ there is an LTS $f(L)$ such that $L \cgr f(L)$, $\Sf(f(L)) =
\Sf(L)$, $\Div(f(L)) = \Tr(L) \cap \extT(L)$, and $\Inf(f(L)) = \Inf(L)$.

\item
If ``$\cgr$'' does not preserve $\Inf$, then for every LTS $L$ there is an LTS
$f(L)$ such that $L \cgr f(L)$, $\Sf(f(L)) = \Sf(L)$, $\Div(f(L)) = \Tr(L)
\cap \extT(L)$, and $\Inf(f(L)) = \anI(L)$.
\end{enumerate}
\end{lemma}
\proof
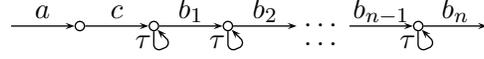
\begin{figure}[t]
\mbox{}\hfill\begin{pspicture}(180,20)
\arc(0,10)(24,10)\rput[B](12,13){$a$}
\arc(28,10)(52,10)\rput[B](40,13){$c$}
\arc(56,10)(80,10)\rput[B](68,13){$b_1$}
\aloopdr{54,10}\rput[r](53,4){$\tau$}
\arc(84,10)(108,10)\rput[B](96,13){$b_2$}
\rput(118,10){$\cdots$}
\rput(118,3){$\cdots$}
\aloopdr{82,10}\rput[r](81,4){$\tau$}
\arc(128,10)(152,10)\rput[B](140,13){$b_{n-1}$}
\aloopdr{154,10}\rput[r](153,4){$\tau$}
\arc(156,10)(180,10)\rput[B](168,13){$b_n$}
\psst(26,10){2}\psst(54,10){2}\psst(82,10){2}\psst(154,10){2}
\end{pspicture}\hfill\mbox{}
\caption{An LTS fragment for detecting the divergence trace $b_1 b_2 \cdots
b_n$.}\label{F:testdiv}
\end{figure}
Let $M_1 \cgr M_2$, $\sigma \in \Div(M_1) \setminus \Div(M_2)$, $b_1 \cdots
b_n = \new{\sigma}{1}$, $c = \new{1}{0}$, and $d = \new{2}{0}$.
By Theorem~\ref{T:noSigma}, ``$\cgr$'' preserves $\Sigma$, so we may let
$\Sigma_M = \Sigma(M_1) = \Sigma(M_2)$.
For any LTS $L$, let $\Sigma_L = \Sigma(L)$ and let $g(L)$ be the LTS that is
obtained as follows.
First $L$ is replaced by $\PPD(\newup{L}{2})$.
If $\hat s\PP$ is pre-divergent, then it is the new initial state, and each
transition $(s,a,s')$ where $s$ is pre-divergent and $s'$ is post-divergent is
replaced by a copy of the LTS fragment shown in Fig.~\ref{F:testdiv}.
Otherwise a copy of Fig.~\ref{F:testdiv} is added such that its $a$-transition
is left out, the start state of the $c$-transition is the new initial state,
and the LTS fragment leads to $\hat s\PP$.
The alphabet of $g(L)$ is $\{c\} \cup \new{\Sigma_M}{1} \cup
\new{\Sigma_L}{2}$.
When completing a minimal divergence trace of $\newup{L}{2}$, $g(L)$ executes
$c \new{\sigma}{1}$ before continuing, but otherwise it behaves like
$\newup{L}{2}$.

Later, in the proof of claim (b), we will introduce $\Sigma_N$, $N'_1$, and
$N'_2$.
To have a place for them in our construction, we now let $N'_0 = \DLG$ (with
$\Sigma(N'_0) = \emptyset$).
When $i \in \{1,2\}$ and $j \in \{0,1,2\}$, let $M'_i = c.\newup{M_i \sqcap
M_2}{1}$ and
$$f_{i,j}(L) \ =\ \newdn{\ (\,g(L) \pp M'_i\ \pp N'_j\,) \setminus (\{c,d\}
\cup \new{\Sigma_M}{1} \cup \new{\Sigma_N}{3})\ }{2}\textrm{ .}$$
Clearly $N'_0$ has no effect on the behaviour.
With $N'_0$, independently of what $\Sigma_N$ is, also the hiding with
$\new{\Sigma_N}{3}$ has no effect.

We show now that $L \CFFD f_{2,0}(L)$.
Before completing any minimal divergence trace, $f_{2,0}(L)$ behaves like $L$.
When $g(L)$ executes $c$, one of the two copies of $M_2$ in $M'_2$ is switched
on.
Then $g(L)$ tries to execute $\new{\sigma}{1}$.
If it fails because $M_2$ blocks it, then $f_{2,0}(L)$ diverges due to the
$\tau$-loops in Fig.~\ref{F:testdiv}.
That is still equivalent to $L$, because the trace that has been executed is a
minimal divergence trace.
For the same reason it is okay if $M_2$ diverges before completing $\sigma$.
The execution of $\sigma$ may also succeed, because $\sigma \in \Div(M_1)
\subseteq \Tr(M_1) = \Tr(M_2)$.
In that case, because $\sigma \notin \Div(M_2)$, $M_2$ is left in a
nondivergent state, having no effect on the further behaviour.
So $f_{2,0}(L)$ continues like $L$.

Because $M'_1$ has a copy of both $M_1$ and $M_2$, $f_{1,0}(L)$ behaves
otherwise like $f_{2,0}(L)$, but it has additional behaviour caused by $M_1$
starting in $M'_1$, executing $\sigma$ completely, and diverging.
In that case, every subsequent state of $f_{1,0}(L)$ is divergent.
Thus $L \CFFD f_{2,0}(L) \cgr f_{1,0}(L)$, $\Tr(f_{1,0}(L)) = \Tr(L)$,
$\Sf(f_{1,0}(L)) = \Sf(L)$, $\minD(f_{1,0}(L)) = \minD(L)$, $\Div(f_{1,0}(L))
= \Tr(L) \cap \extT(L)$, $\anI(f_{1,0}(L)) = \anI(L)$, and $\Inf(f_{1,0}(L)) =
\Inf(L)$.
As a consequence, $f_{1,0}$ qualifies as the $f$ of claim (a).

\begin{figure}[t]
\mbox{}\hfill\begin{pspicture}(259,89)
\arc(19,57)(19,50)
\aloopvd{19,48}\rput[r](10,52){$a_1$}
\GRAY{\aloopmr{19,48}\rput[t](18,37){$a_m$}}
\rput(7,47){$\ddots$}
\arc(20.2,49.6)(45.8,81.4)\rput(33,71){$c$}
\arc(20.2,46.4)(45.8,14.6)\rput(33,36){$c$}

\rput[lb](35,42){
\arc(14,41)(38,41)\rput[B](26,43){$\tau$}
\arc(42,41)(66,41)\rput[B](54,43){$\tau$}
\arc(40,39)(40,15)\rput[l](41,31){$\tau$}
\arc(70,41)(94,41)\rput[B](82,43){$\tau$}
\arc(68,39)(68,15)\rput[l](69,31){$\tau$}
\arc(98,41)(122,41)\rput[B](110,43){$\tau$}
\arc(96,39)(96,15)\rput[l](97,31){$\tau$}
\rput(132,41){$\cdots$}\rput(132,27){$\cdots$}
\aloopdl{12,13}\rput[r](6,7){$\tau$}
\acrud{40,13}{$a_1$}
\GRAY{\acrdu{40,13}{$a_m$}}
\rput(26,15){$\vdots$}
\acrud{68,13}{$a_1$}
\GRAY{\acrdu{68,13}{$a_m$}}
\rput(54,15){$\vdots$}
\acrud{96,13}{$a_1$}
\GRAY{\acrdu{96,13}{$a_m$}}
\rput(82,15){$\vdots$}
\acrud{124,13}{$a_1$}
\GRAY{\acrdu{124,13}{$a_m$}}
\rput(110,15){$\vdots$}
\rput(132,13){$\cdots$}
\psst(12,41){2}\psst(40,41){2}\psst(68,41){2}\psst(96,41){2}
\psst(12,13){2}\psst(40,13){2}\psst(68,13){2}\psst(96,13){2}
}

\aloopdr{47,13}\rput[l](54,7){$\tau$}
\arc(49,13)(73,13)\rput[B](61,16){$d$}
\aloopdr{75,13}\rput[l](82,7){$\tau$}
\arc(77,13)(101,13)\rput[B](89,16){$e_1$}
\aloopdl{103,13}\rput[r](97,7){$\tau$}
\arcud{103,13}{$a_1$}
\GRAY{\arcdu{103,13}{$a_m$}}
\rput(117,15){$\vdots$}
\aloopdr{131,13}\rput[l](138,7){$\tau$}
\arc(133,13)(157,13)\rput[B](145,16){$e_2$}
\aloopdl{159,13}\rput[r](153,7){$\tau$}
\arcud{159,13}{$a_1$}
\GRAY{\arcdu{159,13}{$a_m$}}
\rput(173,15){$\vdots$}
\aloopdr{187,13}\rput[l](194,7){$\tau$}
\arc(189,13)(213,13)\rput[B](201,16){$e_3$}
\aloopdl{215,13}\rput[r](209,7){$\tau$}
\arcud{215,13}{$a_1$}
\GRAY{\arcdu{215,13}{$a_m$}}
\rput(229,15){$\vdots$}
\rput(253,13){$\cdots$}
\psst(19,48){2}\psst(47,13){2}\psst(75,13){2}\psst(103,13){2}\psst(131,13){2}
\psst(159,13){2}\psst(187,13){2}\psst(215,13){2}
\end{pspicture}\hfill\mbox{}
\caption{A switchable LTS for detecting the infinite trace $e_1 e_2
\cdots$.}\label{F:testinftr2}
\end{figure}
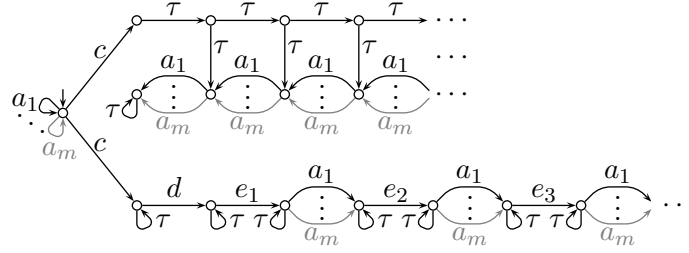

In the case of claim (b), there are $N_1$, $N_2$, and $\xi$ such that $N_1
\cgr N_2$ and $\xi \in \Inf(N_1) \setminus \Inf(N_2)$.
Let $e_1 e_2 \cdots = \new{\xi}{3}$, $\Sigma_N = \Sigma(N_1) = \Sigma(N_2)$,
and $\{a_1, a_2, \ldots{\gray, a_m}\} = \new{\Sigma_L}{2}$.
When $j \in \{1,2\}$, let $N'_j = T_\xi \pp d.\newup{N_j}{3}$, where $T_\xi$
is the LTS in Fig.~\ref{F:testinftr2} with the alphabet $\{c,d\} \cup
\new{\Sigma_L}{2} \cup \new{\Sigma_N}{3}$.

If $j \in \{1,2\}$, $c$ makes $T_\xi$ enter one of its two branches.
Its initial state and upper branch can parallel any finite execution of
$g(L)$.
Because $T_\xi$ never refuses any other subset of $\new{\Sigma_L}{2}$ than
$\emptyset$, and because of the stable states initially and in the upper
branch, $\Sf(f_{1,j}(L)) = \Sf(f_{1,0}(L))$.
Furthermore, $\Div(f_{1,j}(L)) = \Div(f_{1,0}(L))$, because $T_\xi$ cannot
diverge before executing $c$, and all traces that involve the execution of $c$
are in $\Div(f_{1,0}(L))$.

The upper branch of $T_\xi$ does not yield infinite traces.
In its lower branch $T_\xi$ switches $N_j$ on by executing $d$.
Thanks to the initial state of $T_\xi$ and because $N_2$ cannot execute $\xi$,
we have $\Inf(f_{1,2}(L)) = \anI(f_{1,0}(L))$.
Because $N_1$ can execute $\xi$, we have $\Inf(f_{1,1}(L)) =
\Inf(f_{1,0}(L))$.
We get $f_{1,0}(L) \CFFD f_{1,1}(L) \cgr f_{1,2}(L)$.
So $f_{1,2}$ qualifies as the $f$ of claim (b).\qed

\begin{theorem}\label{T:Tr-minD-anI}
If ``$\cgr$'' is a congruence, ``$\CFFD$'' implies ``$\cgr$'', and ``$\cgr$''
preserves $\Tr$, $\Sf$, and $\minD$ but neither $\Div$ nor $\Inf$, then
``$\cgr$'' is the equivalence induced by $\Sigma$, $\Tr$, $\Sf$, $\minD$, and
$\anI$.
\end{theorem}
\proof
By Lemma~\ref{L:minD=>anI}, ``$\cgr$'' preserves $\Sigma$ and $\anI$.
The $f$ of Lemma~\ref{L:noDiv}(b) qualifies as the $f$ of
Lemma~\ref{L:nocongr}.\qed

We have now two directions to go: one where $\Inf$ is preserved and another
where $\Div$ is preserved.
Given the work we have done already, the former is easy.

\begin{theorem}\label{T:Tr-minD-Inf}
If ``$\cgr$'' is a congruence, ``$\CFFD$'' implies ``$\cgr$'', and ``$\cgr$''
preserves $\Sf$, $\minD$, and $\Inf$ but not $\Div$, then ``$\cgr$'' is the
equivalence induced by $\Sigma$, $\Tr$, $\Sf$, $\minD$, and $\Inf$.
\end{theorem}
\proof
By Lemma~\ref{L:Inf=>Tr}, ``$\cgr$'' preserves $\Sigma$ and $\Tr$.
The $f$ of Lemma~\ref{L:noDiv}(a) qualifies as the $f$ of
Lemma~\ref{L:nocongr}.\qed

We still have the case where $\Div$ is preserved but $\Inf$ is not.

\begin{lemma}\label{L:noaenI}
If ``$\cgr$'' is a congruence, ``$\CFFD$'' implies ``$\cgr$'', and ``$\cgr$''
preserves $\Div$ but not $\aenI$, then for every LTS $L$ there is an LTS
$f(L)$ such that $L \cgr f(L)$, $\Sf(f(L)) = \Sf(L)$, $\Div(f(L)) = \Div(L)$,
and $\Inf(f(L)) = \eanI(L)$.
\end{lemma}
\proof
Let $M_1 \cgr M_2$ and $\xi \in \aenI(M_1) \setminus \aenI(M_2)$.
By Theorem~\ref{T:noSigma}, ``$\cgr$'' preserves $\Sigma$.
Let $\Sigma_M = \Sigma(M_1) = \Sigma(M_2)$, $c = \new{0}{0}$, $c_1 =
\new{1}{0}$, and $c_2 = \new{2}{0}$.
Because ``$\cgr$'' preserves $\Div$, $M_1$ and $M_2$ agree on which prefixes
of $\xi$ are divergence traces.
Infinitely many of them are not, by the definition of $\aenI$.
So non-empty $\sigma_1$, $\sigma_2$, $\sigma_3$, \ldots\ exist such that
$\sigma_1 \sigma_2 \sigma_3 \cdots = \new{\xi}{1}$ and $\sigma_1$, $\sigma_1
\sigma_2$, $\sigma_1 \sigma_2 \sigma_3$, \ldots\ are not divergence traces.
Let $T_\xi$ be the LTS whose alphabet is $\{c,c_1,c_2\} \cup
\new{\Sigma_M}{1}$ and whose graph is
\begin{center}
\newcommand{\azigzag}{\psline[linearc=3pt]{->}%
(0,0)(3,3)(6,0)(9,-3)(12,0)(15,3)(18,0)(21,-3)(24,0)(27,3)(30,0)(36,0)}
\begin{pspicture}(257,10)
\ainit{4,2}
\arc(6,2)(22,2)\rput[B](14,5){$c_1$}
\arc(26,2)(42,2)\rput[B](34,5){$c$}
\rput(46,2){\azigzag}\rput[B](64,5){$\sigma_1$}
\arc(86,2)(102,2)\rput[B](94,5){$c_2$}
\arc(106,2)(122,2)\rput[B](114,5){$c_1$}
\rput(126,2){\azigzag}\rput[B](144,5){$\sigma_2$}
\arc(166,2)(182,2)\rput[B](174,5){$c_2$}
\arc(186,2)(202,2)\rput[B](194,5){$c_1$}
\rput(206,2){\azigzag}\rput[B](224,5){$\sigma_3$}
\rput(252,2){$\cdots$}
\psst(4,2){2}\psst(24,2){2}\psst(44,2){2}\psst(84,2){2}\psst(104,2){2}
\psst(124,2){2}\psst(164,2){2}\psst(184,2){2}\psst(204,2){2}
\end{pspicture} .
\end{center}

For any LTS $L$, let $g(L)$ be the LTS that is obtained as follows.
First $L$ is replaced by $\Una(\newup{L}{2})$.
Then each transition whose label $a$ is visible and which ends in a
potentially divergent state is replaced by
\begin{pspicture}(68,12)(0,2)
\arc(0,2)(20,2)\rput[B](10,5){$a$}
\arc(24,2)(44,2)\rput[B](32,5){$c_1$}
\aloopur{46,2}\rput[r](45,8){$\tau$}
\arc(48,2)(68,2)\rput[B](60,5){$c_2$}
\psst(22,2){2}\psst(46,2){2}
\end{pspicture}.
The alphabet of the result is $\{c_1,c_2\} \cup \new{\Sigma_L}{2}$, where
$\Sigma_L = \Sigma(L)$.
When $i \in \{1,2\}$, let
$$f_i(L) \ =\ \newdn{\ (\,g(L) \pp T_\xi \pp c.\newup{M_i}{1}\,) \setminus
(\{c,c_1,c_2\} \cup \new{\Sigma_M}{1})\ }{2}\textrm{ .}$$
Each time when $g(L)$ is about to enter a potentially divergent state, it
executes $c_1$.
This makes $T_\xi$ move one step and then let $c.\newup{M_i}{1}$ try to
execute up to a nondivergent state.
If it succeeds, $T_\xi$ lets $g(L)$ continue by executing $c_2$.
In the opposite case, $g(L)$ is trapped in the $\tau$-loop between $c_1$ and
$c_2$.

The LTS $M_1$ has every prefix of $\xi$ as its trace.
By Lemma~\ref{L:Div=>Tr}, ``$\cgr$'' preserves $\Tr$.
So both $\newup{M_1}{1}$ and $\newup{M_2}{1}$ may succeed in executing
$\sigma_1 \sigma_2 \cdots \sigma_i$ for any $i$.
This implies $\Tr(f_1(L)) = \Tr(f_2(L)) = \Tr(L)$.
Clearly $g(L)$ mimics the divergence traces of $L$.
When $M_1$ or $M_2$ diverges, $g(L)$ is in a $\tau$-loop and the trace that
has been executed is a divergence trace.
Thus $\Div(f_1(L)) = \Div(f_2(L)) = \Div(L)$.

When $g(L)$ is in a stable state (other than the start states of $c_1$), then
$c.\newup{M_1}{1}$ and $c.\newup{M_2}{1}$ do not diverge, so $\Sf(f_1(L)) =
\Sf(f_2(L)) = \Sf(L)$.
Because $M_2$ does but $M_1$ does not necessarily prevent $g(L)$ from
infinitely many times continuing with $c_2$ after a divergence trace, we have
$\Inf(f_1(L)) = \Inf(L)$ but $\Inf(f_2(L)) = \eanI(L)$.
So $L \CFFD f_1(L) \cgr f_2(L)$.\qed

\begin{theorem}\label{T:Tr-Div-eanI}
If ``$\cgr$'' is a congruence, ``$\CFFD$'' implies ``$\cgr$'', and ``$\cgr$''
preserves $\Sf$ and $\Div$ but not $\aenI$, then ``$\cgr$'' is the equivalence
induced by $\Sigma$, $\Sf$, $\Div$, and $\eanI$.
\end{theorem}
\proof
By Lemma~\ref{L:Div=>eanI}, ``$\cgr$'' preserves $\Sigma$ and $\eanI$.
The $f$ of Lemma~\ref{L:noaenI} qualifies as the $f$ of
Lemma~\ref{L:nocongr}.\qed

\begin{lemma}\label{L:noInf}
If ``$\cgr$'' is a congruence, ``$\CFFD$'' implies ``$\cgr$'', and ``$\cgr$''
preserves $\Div$ and $\aenI$ but not $\Inf$, then for every LTS $L$ there is
an LTS $f(L)$ such that $L \cgr f(L)$, $\Sf(f(L)) = \Sf(L)$, $\Div(f(L)) =
\Div(L)$, and $\Inf(f(L)) = \aenI(L)$.
\end{lemma}
\proof
\newcommand{\eadI}{\mathit{eadI}}
For the purpose of this proof, we define \emph{eventually-always-divergent}
infinite traces as $\eadI(L) = \Inf(L) \setminus \aenI(L)$.
Let $M_1 \cgr M_2$ and $\xi \in \Inf(M_1) \setminus \Inf(M_2)$.
By Theorem~\ref{T:noSigma}, ``$\cgr$'' preserves $\Sigma$.
Let $\Sigma_M = \Sigma(M_1) = \Sigma(M_2)$, $c = \new{0}{0}$, $c_1 =
\new{1}{0}$, and $c_2 = \new{2}{0}$.
Because ``$\cgr$'' preserves $\aenI$, $\xi \in \eadI(M_1)$.
Because ``$\cgr$'' preserves $\Div$, $M_1$ and $M_2$ agree on which prefixes
of $\xi$ are divergence traces.
From some point on all of them are, because $\xi \in \eadI(M_1)$.

For any LTS $L$, let $\Sigma_L = \Sigma(L)$, and let $g(L)$ be obtained as
follows.
Each transition of $\Una(L)$ whose label $a$ is visible is replaced by
\begin{center}
\begin{tabular}{ll}
\begin{pspicture}(44,12)(0,2)
\arc(0,2)(20,2)\rput[b](10,3){$\new{a}{2}$}
\arc(24,2)(44,2)\rput[b](34,3){$c_1$}
\psst(22,2){2}
\end{pspicture}
&
, if it starts in a certainly nondiv.\ and ends in a potentially divergent
state;\\
\begin{pspicture}(44,12)(0,2)
\arc(0,2)(20,2)\rput[b](10,3){$\new{a}{2}$}
\arc(24,2)(44,2)\rput[b](34,3){$c_2$}
\psst(22,2){2}
\end{pspicture}
&
, if it starts and ends in a potentially divergent state;\\
\begin{pspicture}(20,12)(0,2)
\arc(0,2)(20,2)\rput[b](10,3){$\new{a}{3}$}
\end{pspicture}
&
, if it starts in a potentially divergent and ends in a certainly nondiv.\
state;\\
\begin{pspicture}(20,12)(0,2)
\arc(0,2)(20,2)\rput[b](10,3){$\new{a}{2}$}
\end{pspicture}
&
, if it starts and ends in a certainly nondivergent state.
\end{tabular}
\end{center}
If the initial state of $\Una(L)$ is potentially divergent, then a
$c_1$-transition is added to its front.
The alphabet of $g(L)$ is $\{c_1,c_2\} \cup \new{\Sigma_L}{2} \cup
\new{\Sigma_L}{3}$.

\begin{figure}[t]
\mbox{}\hfill\begin{pspicture}(129,70)(-6,0)
\ainit{4,62}
\arc(6,62)(30,62)\rput[B](18,65){$c_1$}
\arc(34,62)(58,62)\rput[B](46,65){$\tau$}
\arc(62,62)(86,62)\rput[B](74,65){$\tau$}
\arc(90,62)(114,62)\rput[B](102,65){$\tau$}
\rput(124,62){$\cdots$}
\arc(32,60)(32,34)\rput[l](33,53){$\tau$}
\arc(60,60)(60,34)\rput[l](61,53){$\tau$}
\arc(88,60)(88,34)\rput[l](89,53){$\tau$}
\rput(124,52){$\cdots$}
\psline[linearc=7pt,linewidth=2pt]{->}(114,43)(4,43)(4,60)
\psline[linearc=7pt,linewidth=2pt](30.6,33.4)(22,41)(12,43)
\psline[linearc=7pt,linewidth=2pt](58.6,33.4)(50,41)(40,43)
\psline[linearc=7pt,linewidth=2pt](86.6,33.4)(78,41)(68,43)
\rput[l](6,52.5){$\new{\Sigma_L}{3}$}
\arc(30,32)(6,32)\rput[B](18,35){$c_2$}
\arc(58,32)(34,32)\rput[B](46,35){$c_2$}
\arc(86,32)(62,32)\rput[B](74,35){$c_2$}
\arc(114,32)(90,32)\rput[B](102,35){$c_2$}
\rput(124,43){$\cdots$}
\rput(124,32){$\cdots$}
\arc(2,32)(-6,32)(-6,2)(2,2)\rput[lB](-3,35){$c$}
\aloopur{4,2}\rput[b](7,12.5){$\tau$}
\arc(6,2)(30,2)\rput[B](18,5){$b_1$}
\aloopur{32,2}\rput[b](35,12.5){$\tau$}
\arc(34,2)(58,2)\rput[B](46,5){$c_2$}
\aloopur{60,2}\rput[b](63,12.5){$\tau$}
\arc(62,2)(86,2)\rput[B](74,5){$b_2$}
\aloopur{88,2}\rput[b](91,12.5){$\tau$}
\arc(90,2)(114,2)\rput[B](102,5){$c_2$}
\rput(124,14){$\cdots$}
\rput(124,2){$\cdots$}
\psst(4,62){2}\psst(32,62){2}\psst(60,62){2}\psst(88,62){2}
\psst(4,32){2}\psst(32,32){2}\psst(60,32){2}\psst(88,32){2}
\psst(4,2){2}\psst(32,2){2}\psst(60,2){2}\psst(88,2){2}
\end{pspicture}\hfill\mbox{}
\caption{An LTS for detecting an infinite trace with only finitely many
nondivergent prefixes.
The thick arrows with $\new{\Sigma_L}{3}$ denote that there is a transition
from each start state of the thick arrows to their common end state for each
$a \in \new{\Sigma_L}{3}$.}\label{F:testinftr3}
\end{figure}
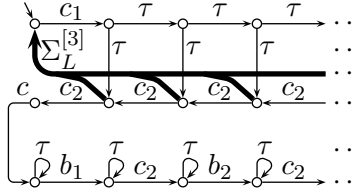

Let $b_1 b_2 \cdots = \new{\xi}{1}$.
Let $T_\xi$ be the LTS whose alphabet is $\{c,c_1,c_2\} \cup \new{\Sigma_M}{1}
\cup \new{\Sigma_L}{3}$ and whose graph is in Fig.~\ref{F:testinftr3}.
When $i \in \{1,2\}$, let
$$f_i(L) \ =\ \big((\,g(L) \pp T_\xi \pp c.\newup{M_i}{1}\,) \setminus
(\{c,c_1,c_2\} \cup \new{\Sigma_M}{1})\big)\Phi\textrm{ ,}$$
where $\Phi$ renames each $\new{a}{2}$ and each $\new{a}{3}$ to $a$.

While $g(L)$ traverses among certainly nondivergent states, $f_1(L)$ and
$f_2(L)$ behave like $L$.
When $g(L)$ enters a potentially divergent state, $T_\xi$ prepares for an
arbitrary finite number of transitions between potentially divergent states.
The divergence of $T_\xi$ is not a problem, because the trace that has been
executed is a divergence trace.
As long as $T_\xi$ is in its middle row excluding its leftmost state, $g(L)$
can execute transitions at will.
These states of $T_\xi$ are stable and offer all actions in $\Sigma(g(L)) \cap
\Sigma(T_\xi)$ except $c_1$ that also $g(L)$ refuses, so $\Sf$ is preserved.
If $g(L)$ enters a certainly nondivergent state, then $T_\xi$ goes back to its
initial state.
As a consequence, $f_1(L)$ and $f_2(L)$ have at least the same stable
failures, divergence traces, and always-eventually-nondivergent infinite
traces as $L$, and no extra stable failures, divergence traces, or infinite
traces have so far been found.

If $g(L)$ executes more transitions between potentially divergent states than
$T_\xi$ has been prepared for, $T_\xi$ reaches the leftmost state of its
middle row.
Then it executes $c$, switching $M_1$ or $M_2$ on.
From then on all states are divergent and $g(L)$ is prevented from leaving
potentially divergent states, so no new stable failures or divergence traces
are introduced.
$f_2(L)$ does not introduce any new infinite traces either, while $f_1(L)$ may
execute all the remaining infinite traces of $L$, that is, $\eadI(L)$.
So $\Sf(f_1(L)) = \Sf(f_2(L)) = \Sf(L)$, $\Div(f_1(L)) = \Div(f_2(L)) =
\Div(L)$, $\Inf(f_1(L)) = \Inf(L)$, and $\Inf(f_2(L)) = \aenI(L)$.
Clearly $L \CFFD f_1(L) \cgr f_2(L)$.\qed

\begin{theorem}\label{T:Tr-Div-aenI}
If ``$\cgr$'' is a congruence, ``$\CFFD$'' implies ``$\cgr$'', and ``$\cgr$''
preserves $\Sf$, $\Div$, and $\aenI$ but not $\Inf$, then ``$\cgr$'' is the
equivalence induced by $\Sigma$, $\Sf$, $\Div$, and $\aenI$.
\end{theorem}
\proof
By Theorem~\ref{T:noSigma}, ``$\cgr$'' preserves $\Sigma$.
The $f$ of Lemma~\ref{L:noInf} qualifies as the $f$ of
Lemma~\ref{L:nocongr}.\qed

Both branches of reasoning have now led to congruences that preserve both
$\Div$ and $\Inf$.
In this section also $\Sf$ is preserved.
The equivalence induced by $\Sigma$, $\Sf$, $\Div$, and $\Inf$ is ``$\CFFD$''.
So ``$\CFFD$'' is the last congruence in this section.

\begin{table}
\caption{All congruences when no two of deadlock, livelock, and
bothlock are equivalent}\label{B:all-different}
\smallskip
\begin{tabular}{l|l|l|c}
preserves & does not preserve & induced by & theorem\\
\hline
$\Sf$, $\minD$ & $\Tr$ & $\Sigma$, $\Sf$, $\minD$, $\anI$ & \ref{T:minDTr}\\
$\Tr$, $\Sf$, $\minD$ & $\Div$, $\Inf$ & $\Sigma$, $\Tr$, $\Sf$, $\minD$,
$\anI$ & \ref{T:Tr-minD-anI}\\
$\Sf$, $\minD$, $\Inf$ & $\Div$ & $\Sigma$, $\Tr$, $\Sf$, $\minD$, $\Inf$ &
\ref{T:Tr-minD-Inf}\\
$\Sf$, $\Div$ & $\aenI$ & $\Sigma$, $\Sf$, $\Div$, $\eanI$ &
\ref{T:Tr-Div-eanI}\\
$\Sf$, $\Div$, $\aenI$ & $\Inf$ & $\Sigma$, $\Sf$, $\Div$, $\aenI$ &
\ref{T:Tr-Div-aenI}\\
$\Sf$, $\Div$, $\Inf$ & & $\Sigma$, $\Sf$, $\Div$, $\Inf$\\
\end{tabular}
\end{table}

There are thus six congruences in this section.
They are summarized in Table~\ref{B:all-different}.
If a congruence is implied by ``$\CFFD$'' and preserves the sets in the first
column of the table but does not preserve the sets in the second column, then
it is the equivalence induced by the sets in the third column.
The sets in the third column that are not in the first column of the same row
must be added to meet the congruence requirement while preserving the sets in
the first column.
By Theorems~\ref{T:Sf} and~\ref{T:minD}, the congruence on the first row is
the weakest in this section.
By comparing the second column to the first column one may check that all
possibilities between the first row and ``$\CFFD$'' are covered.

\section{When Deadlock Is Not Livelock Is Bothlock}\label{S:BLisLL}

\newcommand{\nF}{\mi{nF}}
\newcommand{\snF}{\mi{snF}}
\newcommand{\anF}{\mi{anF}}
\newcommand{\sanF}{\mi{sanF}}
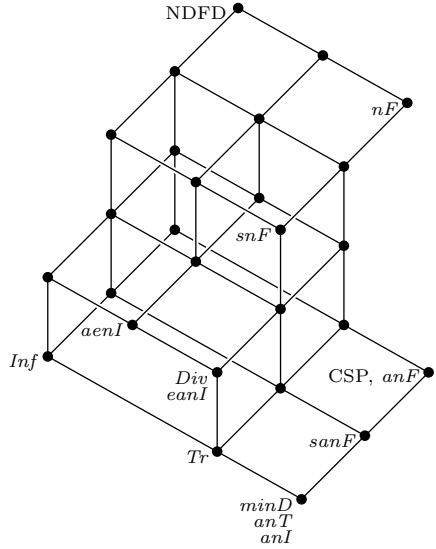
\begin{figure}[t]
\mbox{}\hfill\begin{pspicture}(-3,45)(158,250)
\newcommand{\Qx}{108}\newcommand{\Qy}{32}
\newcommand{\QT}[1]{\rput(-32,18){#1}}
\newcommand{\QA}[1]{\rput(-64,36){#1}}
\newcommand{\QI}[1]{\rput(-96,54){#1}}
\newcommand{\QH}[1]{\rput(24,24){#1}}
\newcommand{\QG}[1]{\rput(48,48){#1}}
\newcommand{\QN}[1]{\rput(72,72){#1}}
\newcommand{\QF}[1]{\rput(96,96){#1}}
\newcommand{\QM}[1]{\rput(0,30){#1}}
\newcommand{\QD}[1]{\rput(0,60){#1}}
\newcommand{\QE}[1]{\rput(0,90){#1}}
\newcommand{\Qc}[1]{\rput(\Qx,\Qy){\pscircle*(0,0){2}
\rput[r](-3,-2){\tiny{#1}}}}
\newcommand{\Qd}[1]{\rput(\Qx,\Qy){\rput[r](-3,-8){\tiny{#1}}}}
\newcommand{\Qe}[1]{\rput(\Qx,\Qy){\rput[r](-3,-14){\tiny{#1}}}}
\newcommand{\Qline}[2]{\psline[linewidth=2pt,linecolor=white](#1)(#2)
\psline(#1)(#2)}
\newcommand{\Qu}{\rput(\Qx,\Qy){\Qline{0,-30}{0,0}}}
\newcommand{\Quu}{\rput(\Qx,\Qy){\Qline{0,-60}{0,0}}}
\newcommand{\Ql}{\rput(\Qx,\Qy){\Qline{32,-18}{0,0}}}
\newcommand{\Qll}{\rput(\Qx,\Qy){\Qline{64,-36}{0,0}}}
\newcommand{\Qr}{\rput(\Qx,\Qy){\Qline{-24,-24}{0,0}}}
\newcommand{\Qrr}{\rput(\Qx,\Qy){\Qline{-48,-48}{0,0}}}
\newcommand{\Qrrrr}{\rput(\Qx,\Qy){\Qline{-96,-96}{0,0}}}
\newcommand{\ww}{\psframe*[linecolor=white]}

\QN{\QE{\QT{\Qr}}}\QN{\QE{\QA{\Qr\Ql}}}\QN{\QE{\QI{\Qr\Ql}}}

\QG{\QM{\Qr}}\QG{\QM{\QT{\Qr\Ql}}}\QG{\QM{\QI{\Qr\Qll}}}
\QG{\QD{\QT{\Qr\Qu}}}\QG{\QD{\QA{\Qr\Ql}}}\QG{\QD{\QI{\Qr\Qu\Ql}}}
\QG{\QE{\QT{\Qr\Qu}}}\QG{\QE{\QA{\Qr\Qu\Ql}}}\QG{\QE{\QI{\Qr\Qu\Ql}}}

\QH{\QM{\Qr}}\QH{\QM{\QT{\Qr\Ql}}}\QH{\QM{\QI{\Qr\Qll}}}
\QH{\QD{\QT{\Qr\Qu}}}\QH{\QD{\QA{\Qr\Ql}}}\QH{\QD{\QI{\Qr\Qu\Ql}}}
\QH{\QE{\QT{\Qu}}}\QH{\QE{\QA{\Qu\Ql}}}\QH{\QE{\QI{\Qu\Ql}}}

\QM{\QT{\Ql}}\QM{\QI{\Qll}}
\QD{\QT{\Qu}}\QD{\QA{\Ql}}\QD{\QI{\Qu\Ql}}

\QM{\Qc{$\minD$}\Qd{$\anT$}\Qe{$\anI$}}\QM{\QT{\Qc{$\Tr$}}}
\QM{\QI{\Qc{$\Inf$}}}
\QD{\QT{\Qc{$\Div$}\Qd{$\eanI$}}}\QD{\QA{\Qc{$\aenI$}}}\QD{\QI{\Qc{}}}

\QH{\QM{\Qc{\ww(-1,0)(3,4.5)$\sanF$}}}\QH{\QM{\QT{\Qc{}}}}\QH{\QM{\QI{\Qc{}}}}
\QH{\QD{\QT{\Qc{}}}}\QH{\QD{\QA{\Qc{}}}}\QH{\QD{\QI{\Qc{}}}}

\QG{\QM{\Qc{\ww(0,-0.5)(18,4.5)CSP, $\anF$}}}
\QG{\QM{\QT{\Qc{}}}}\QG{\QM{\QI{\Qc{}}}}
\QG{\QD{\QT{\Qc{}}}}\QG{\QD{\QA{\Qc{}}}}\QG{\QD{\QI{\Qc{}}}}

\QH{\QE{\QT{\Qc{\ww(0,-0.5)(9,4)$\snF$}}}}
\QH{\QE{\QA{\Qc{}}}}\QH{\QE{\QI{\Qc{}}}}

\QG{\QE{\QT{\Qc{}}}}\QG{\QE{\QA{\Qc{}}}}\QG{\QE{\QI{\Qc{}}}}

\QN{\QE{\QT{\Qc{$\nF$}}}}\QN{\QE{\QA{\Qc{}}}}\QN{\QE{\QI{\Qc{NDFD\,}}}}
\end{pspicture}\hfill\mbox{}
\caption{The congruences in Section~\ref{S:BLisLL} as a Hasse diagram.
Names in $\mi{italics}$ indicate the new preserved set(s).
Other names are the names of the congruences.}\label{F:BLisLL}
\end{figure}

In this section $\DLG \not\cgr \LLG \cgr \BLG$.
By Theorem~\ref{T:minD}, $\minD$ is preserved also in this section.
However, $\LLG \cgr \BLG$ implies that $\Sf$ is not preserved.
Subsection~\ref{S:new-fail} introduces the new kinds of failures that replace
$\Sf$.
The region is shown in Fig.~\ref{F:BLisLL}.
Its two lowest and the highest layer are surveyed in
Subsections~\ref{S:(s)anF} and~\ref{S:(s)nF}, respectively.

\subsection{New kinds of failures}\label{S:new-fail}

In this subsection we define four new kinds of failures and briefly analyse
their relation to divergence traces.

The next lemma reveals that the essence of $\LLG \cgr \BLG$ is that those
stable failures whose trace is a divergence trace do not matter.
The function $\nu$ in the lemma throws away all information on such failures,
by making $\nu(L)$ have the maximum possible set of them allowed by
$\Sigma(L)$, independently of what $L$ has.

\begin{lemma}\label{L:nF}
If ``$\cgr$'' is a congruence, ``$\CFFD$'' implies ``$\cgr$'', and $\LLG \cgr
\BLG$, then for every LTS $L$ there is an LTS $\nu(L)$ such that $\nu(L) \cgr
L$, $\Sigma(\nu(L)) = \Sigma(L)$, $\Sf(\nu(L)) = \Sf(L) \cup (\Div(L) \times
2^{\Sigma(L)})$, $\Div(\nu(L)) = \Div(L)$, and $\Inf(\nu(L)) = \Inf(L)$.
\end{lemma}
\proof
Let $M_1 = \LLG$ and $M_2 = \BLG$.
Let $c \notin \Sigma(L) \cup \{\tau\}$.
Let $g(L)$ be the LTS that is obtained by adding, from each divergent state of
$L$, a $c$-transition to a deadlock state.
When $i \in \{1,2\}$, let $f_i(L) = (g(L) \pp c.M_i) \setminus \{c\}$.
The only difference of $f_1(L)$ from $L$ is an additional divergence where $L$
already has a divergence, so $L \CFFD f_1(L)$.
On the other hand, $f_2(L)$ also has there a deadlock.
Thus $f_2(L)$ has the properties promised of $\nu(L)$.\qed

In this section we have to proceed in two dimensions.
On one hand, we have to start with no information on stable failures and add
it until we have all stable failures whose trace is not a divergence trace.
On the other hand, for each level of information on stable failures, we have
to investigate different kinds of divergence and infinite traces, like in the
previous section.

We will need four new kinds of failures: \emph{nondivergent}, \emph{strongly
nondivergent}, \emph{always nondivergent}, and \emph{strongly always
nondivergent}.
\begin{eqnarray}
\nF(L) & := & \{ (\sigma,A) \in \Sf(L) \mid \sigma \notin \Div(L)
\}\nonumber\\
\snF(L) & := & \{ (\sigma,A) \in \nF(L) \mid \forall a \in A: \sigma a \notin
\Div(L) \}\nonumber\\
\anF(L) & := & \{ (\sigma,A) \in \Sf(L) \mid \sigma \notin \extT(L)
\}\nonumber\\
\sanF(L) & := & \{ (\sigma,A) \in \anF(L) \mid \forall a \in A: \sigma a
\notin \minD(L) \}\nonumber
\end{eqnarray}
All these four sets $X(L)$ have the property that if $(\sigma, A) \in X(L)$,
then $(\sigma, \emptyset) \in X(L)$ and $\sigma \in \Tr(L)$.
Like before, with $X^\Tr(L)$ we denote the set $\{ \sigma \mid (\sigma,
\emptyset) \in X(L) \}$.
We have the following:
$$
\begin{array}{rcrcl}
\nF^\Tr(L)  & = & \snF^\Tr(L)  & = & \Tr(L) \setminus \Div(L)\\
\anF^\Tr(L) & = & \sanF^\Tr(L) & = & \anT(L)
\end{array}
$$
The $\nu$ of Lemma~\ref{L:nF} satisfies
$$\Sf(\nu(L)) \ =\ \Sf(L) \cup (\Div(L) \times 2^{\Sigma(L)}) \ =\ \nF(L) \cup
(\Div(L) \times 2^{\Sigma(L)})\textrm{ .}$$

The number of possible combinations of semantic sets is restricted a bit by
the next lemma.

\begin{lemma}\label{L:nF=>Div}
Any congruence that preserves $\nF$ or $\snF$ also preserves $\Div$.
\end{lemma}
\proof
By Theorem~\ref{T:noSigma}, it preserves $\Sigma$.
Let $\sigma = a_1 \cdots a_n$.
We have $\sigma \in \Div(L)$ if and only if $\sigma \notin \nF^\Tr(L \sqcap
\LTSsigma)$.
The same proof works for $\snF$.\qed

\subsection{(Strongly) always nondivergent failures}\label{S:(s)anF}

In this subsection, we essentially repeat the analysis in
Section~\ref{S:ALLnot} three times, with nothing, $\sanF$, or $\anF$ in the
place of $\Sf$.
Of course, we also prove that if any information on stable failures is
preserved then $\sanF$ is preserved, at the next level $\anF$ or $\snF$ is
preserved, and then both are preserved.

The next lemma is central in proving that if any information on stable
failures is preserved, then at least $\sanF$ must be preserved.

\begin{lemma}\label{L:nosanF}
If ``$\cgr$'' is a congruence, ``$\CFFD$'' implies ``$\cgr$'', ``$\cgr$''
preserves $\minD$ but not $\sanF$, and $\LLG \cgr \BLG$, then for every LTS
$L$ there is an LTS $h(L)$ such that $h(L) \cgr L$, $\Sf(h(L)) = \Tr(L) \times
2^{\Sigma(L)}$, $\Div(h(L)) = \Div(L)$, and $\Inf(h(L)) = \Inf(L)$.
\end{lemma}
\proof
By Theorem~\ref{T:noSigma}, ``$\cgr$'' preserves $\Sigma$.
Let $M_1 \cgr M_2$ and $(\sigma, A) \in \sanF(M_1) \setminus \sanF(M_2)$.
Let $\Sigma_M = \Sigma(M_1) = \Sigma(M_2)$, $b_1 \cdots b_n =
\new{\sigma}{1}$, and $\{a_1, \ldots{\gray, a_m}\} = \new{A}{1}$.
Let $L$ be any LTS and $\Sigma_L = \Sigma(L)$.
Let $T_{\sigma,A}$ be like in Fig.~\ref{F:testDl}, except that each
$\tau$-loop is replaced by an $\new{a}{2}$-loop for each $a \in \Sigma_L$, and
the alphabet is $\new{\Sigma_M}{1} \cup \new{\Sigma_L}{2}$.
When $i \in \{1,2\}$, let
$$g(M_i) \ =\ \newdn{\,(T_{\sigma,A} \pp \newup{M_i}{1}) \setminus
\new{\Sigma_M}{1}\,}{2}\textrm{ .}$$
By the definition of $\sanF$, $g(M_1)$ does not diverge.
Because ``$\cgr$'' preserves $\minD$, $g(M_2)$ does not diverge.
We have $\Run{\Sigma_L} \CFFD g(M_2) \cgr g(M_1) \CFFD \RD{\Sigma_L}$, where
$\RD{\Sigma_L}$ is obtained from $\Run{\Sigma_L}$ by adding a second state and
a $\tau$-transition to it from the original state (please see
Fig.~\ref{F:simpleLTSs}).

We have $L \bs L \pp \Run{\Sigma_L} \cgr L \pp \RD{\Sigma_L} \cgr \nu(L \pp
\RD{\Sigma_L})$, where $\nu$ is from Lemma~\ref{L:nF}.
The LTS $L \pp \RD{\Sigma_L}$ is otherwise like $L$, but its stable failures
are $\Sf^\Tr(L) \times 2^{\Sigma_L}$.
Therefore, and given~(\ref{E:Tr=DivSf}), $\nu(L \pp \RD{\Sigma_L})$ qualifies
as the $h(L)$.\qed

We can now list the first six congruences in this section, and prove that the
next ones must preserve $\sanF$.

\begin{table}[H]
\caption{The congruences of Theorem~\ref{T:nosanF}}\label{B:nosanF}
\smallskip
\begin{tabular}{l|l|l}
preserves & does not preserve & induced by\\
\hline
$\minD$ & $\Tr$, $\sanF$ & $\Sigma$, $\anT$, $\minD$, $\anI$\\
$\Tr$, $\minD$ & $\sanF$, $\Div$, $\Inf$ & $\Sigma$, $\Tr$, $\minD$, $\anI$\\
$\minD$, $\Inf$ & $\sanF$, $\Div$ & $\Sigma$, $\Tr$, $\minD$, $\Inf$\\
$\Div$ & $\sanF$, $\aenI$ & $\Sigma$, $\Tr$, $\Div$, $\eanI$\\
$\Div$, $\aenI$ & $\sanF$, $\Inf$ & $\Sigma$, $\Tr$, $\Div$, $\aenI$\\
$\Div$, $\Inf$ & $\sanF$ & $\Sigma$, $\Tr$, $\Div$, $\Inf$\\
\end{tabular}
\end{table}
\begin{theorem}\label{T:nosanF}
If ``$\cgr$'' is a congruence, ``$\CFFD$'' implies ``$\cgr$'', ``$\cgr$''
preserves the sets in the first column of Table~\ref{B:nosanF} but not the
sets in the second column, and $\LLG \cgr \BLG$, then it is the equivalence
induced by the sets in the third column.
\end{theorem}
\proof
Let [r1] to [r6] refer to the rows in the table.

Lemmas~\ref{L:Inf=>Tr} [r3], \ref{L:minD=>anT} [r1], \ref{L:minD=>anI}
[r1,2], \ref{L:Div=>Tr} [r4,5,6], and~\ref{L:Div=>eanI} [r4,5,6] imply that if
``$\cgr$'' preserves the sets in the first column, then ``$\cgr$'' also
preserves the additional sets in the third column.

To prove the first claim that ``$\cgr$'' can be no other equivalence, let
$f$ be the $f$ of Lemma~\ref{L:noTr} and $h$ be the $h$ of
Lemma~\ref{L:nosanF}.
We have $L \cgr f(L) \cgr h(f(L))$,
\begin{center}\begin{tabular}{rcccll}
$\Sf(h(f(L)))$ & $=$ & $\Tr(f(L)) \times 2^{\Sigma(f(L))}$ & $=$ &
$( \anT(L) \cup \extT(L) ) \times 2^{\Sigma(L)}$ & ,\\
$\Div(h(f(L)))$ & $=$ & $\Div(f(L))$ & $=$ & $\extT(L)$ & , and\\
$\Inf(h(f(L)))$ & $=$ & $\Inf(f(L))$ & $=$ & $\anI(L) \cup \extI(L)$ & .
\end{tabular}\end{center}
Because $\extT(L)$ and $\extI(L)$ are functions of $\Sigma(L)$ and $\minD(L)$,
Lemma~\ref{L:nocongr} applies and gives the claim.

The remaining five claims that ``$\cgr$'' can be no other equivalence are
proven in a similar way using the $f$ from Lemmas~\ref{L:noDiv}(b) [r2],
\ref{L:noDiv}(a) [r3], \ref{L:noaenI} [r4], and \ref{L:noInf} [r5], and the
function $f(L) = L$ [r6].
In all cases $\Sf(h(f(L))) = \Tr(f(L)) \times 2^{\Sigma(f(L))} = \Tr(L) \times
2^{\Sigma(L)}$.
Depending on the case, $\Div(h(f(L)))$ is $\Tr(L) \cap \extT(L)$ [r2,3] or
$\Div(L)$ [r4,5,6], and $\Inf(h(f(L)))$ is $\anI(L)$ [r2], $\eanI(L)$ [r4],
$\aenI(L)$ [r5], or $\Inf(L)$ [r3,6].\qed

The \emph{weakest livelock-preserving congruence} is the weakest congruence
that guarantees for every $L$ and $L'$ that if $\Div(L) = \emptyset \neq
\Div(L')$, then $L \not\cgr L'$.
In~\cite{PuV99} it was proven that the weakest livelock-preserving congruence
with respect to $L \setminus A$ and $L \pp L'$ is the equivalence induced by
$\Sigma$, $\anT$, $\minD$, and $\anI$.
Only equivalences that preserve $\Sigma$ were considered.
In the present publication, the apparently weaker starting point $\DLG
\not\cgr \BLG$ was used and the same result was obtained as
Theorem~\ref{T:minD} and Lemmas~\ref{L:minD=>anT} and~\ref{L:minD=>anI}.
When taking the preservation of $\Sigma$ as an assumption, their proofs only
use $L \setminus A$ and $L \pp L'$.
The equivalence induced by $\Sigma$, $\Tr$, $\Div$, and $\eanI$ is the weakest
congruence with respect to $L \setminus A$ and $L \pp L'$ that preserves
divergence traces~\cite{PuV99}.
This result corresponds to Lemmas~\ref{L:Div=>Tr} and~\ref{L:Div=>eanI}.

After adding $\sanF$, there is no unique next set of stable failures, but two.
So we need two different functions that throw out some information on stable
failures while preserving the congruence.

The function $h_1$ in the next lemma throws away all information on stable
failures at and after minimal divergence traces.
To facilitate the use of the lemma in two different situations, it has two
alternative assumptions on $\Div$.

\begin{lemma}\label{L:nosnF}
Assume that ``$\cgr$'' is a congruence, ``$\CFFD$'' implies ``$\cgr$'',
``$\cgr$'' preserves $\Sigma$ but not $\snF$, and $\LLG \cgr \BLG$.
For every LTS $L$ such that $\Div(L) = \Tr(L) \cap \extT(L)$ there is an LTS
$h_1(L)$ such that $h_1(L) \cgr L$, $\Sf(h_1(L)) = \anF(L) \cup (\,(\Tr(L)
\cap \extT(L)) \times 2^{\Sigma(L)}\,)$, $\Div(h_1(L)) = \Div(L)$, and
$\Inf(h_1(L)) = \Inf(L)$.
If ``$\cgr$'' preserves $\Div$, then the assumption $\Div(L) = \Tr(L) \cap
\extT(L)$ is not needed.

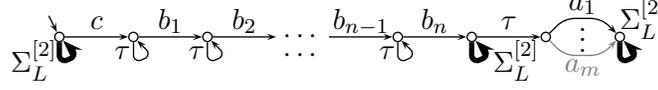
\begin{figure}[t]
\mbox{}\hfill\begin{pspicture}(244,32)
\ainit{17,13}
\arc(19,13)(43,13)\rput[B](31,16){$c$}
\aloopdr{45,13}\rput[r](44,7){$\tau$}
\arc(47,13)(71,13)\rput[B](59,16){$b_1$}
\aloopdr{73,13}\rput[r](72,7){$\tau$}
\arc(75,13)(99,13)\rput[B](87,16){$b_2$}
\rput(109,13){$\cdots$}
\rput(109,6){$\cdots$}
\arc(119,13)(143,13)\rput[B](131,16){$b_{n-1}$}
\aloopdr{145,13}\rput[r](144,7){$\tau$}
\arc(147,13)(171,13)\rput[B](159,16){$b_n$}
\arc(175,13)(199,13)\rput[B](187,16){$\tau$}
\arcud{201,13}{$a_1$}
\GRAY{\arcdu{201,13}{$a_m$}}
\rput(215,15){$\vdots$}
{
\psset{linewidth=2pt}
\aloopdr{17,13}\rput[r](16,4){$\new{\Sigma_L}{2}$}
\aloopdr{173,13}\rput[l](181,4){$\new{\Sigma_L}{2}$}
\aloopdr{229,13}\rput[l](229,20){$\new{\Sigma_L}{2}$}
}
\psst(17,13){2}\psst(45,13){2}\psst(73,13){2}\psst(145,13){2}\psst(173,13){2}
\psst(201,13){2}\psst(229,13){2}
\end{pspicture}\hfill\mbox{}
\caption{An LTS for detecting a strongly nondivergent failure.
The thick arrows denote that there is a transition for each $a \in
\new{\Sigma_L}{2}$.}\label{F:snF}
\end{figure}

\end{lemma}
\proof
Let $\nu$ be from Lemma~\ref{L:nF}.
If $\Div(L) = \Tr(L) \cap \extT(L)$, then $\nu$ qualifies as the $h_1$.

The case remains where ``$\cgr$'' preserves $\Div$.
Let $M_1 \cgr M_2$, $(\sigma, A) \in \snF(M_1) \setminus \snF(M_2)$, $\Sigma_M
= \Sigma(M_1) = \Sigma(M_2)$, $b_1 \cdots b_n = \new{\sigma}{1}$, $\{a_1,
\ldots{\gray, a_m}\} = \new{A}{1}$, $c = \new{1}{0}$, and $\Sigma_L =
\Sigma(L)$.
Let $T_{\sigma,A}$ be like in Fig.~\ref{F:snF} with the alphabet $\{c\} \cup
\new{\Sigma_M}{1} \cup \new{\Sigma_L}{2}$.
Let $g(L)$ be $\newup{\nu(L)}{2}$ with a $c$-transition added from each
divergent state to itself.
When $i \in \{1,2\}$, let $$f_i(L) \ =\ \newdn{\,(g(L) \pp T_{\sigma,A} \pp
c.\newup{M_i}{1}) \setminus (\{c\} \cup \new{\Sigma_M}{1})\,}{2}\textrm{ .}$$
By construction, $f_i(L)$ can do everything that $\nu(L)$ can do, but it
can also hiddenly execute $c$ from any divergent state.
After executing $c$, $f_i(L)$ tries to hiddenly execute $\new{\sigma}{1}$.
If that fails, then $f_i(L)$ is trapped in a divergence.
If that succeeds, then $g(L)$ can continue but $T_{\sigma, A}$ is in an
unstable state and $M_i$ is at an end state of $\sigma$.
We have $\Inf(f_i(L)) = \Inf(L)$.
By the definition of $\snF$, $M_1$ does not diverge when $T_{\sigma, A}$ is
in any of its last three states, but if $T_{\sigma, A}$ continues, then
$T_{\sigma, A} \pp c.\newup{M_1}{1}$ may deadlock.
Thanks to the use of $\nu$, also $g(L)$ may enter a stable state, resulting
in a total deadlock.
So $f_1(L)$ behaves otherwise like $L$, but has also the stable failures
$(\Tr(L) \cap \extT(L)) \times 2^{\Sigma(L)}$.

Because ``$\cgr$'' preserves $\Div$, $M_2$ cannot cause a divergence when
$T_{\sigma, A}$ is in any of its last three states.
It cannot cause a deadlock either, because $(\sigma, A) \notin \snF(M_2)$.
So $f_2(L) \CFFD \nu(L)$.
In conclusion, $L \cgr \nu(L) \CFFD f_2(L) \cgr f_1(L)$, and $f_1(L)$
qualifies as the $h_1$.\qed

The function $h_2$ in the next lemma throws away all information on stable
failures whose trace is or whose refused action would complete a divergence
trace.
Its construction requires that no state is the end state of both a divergence
trace and a nondivergent trace.
To cope with this problem, we use the function $\Una$ defined in the previous
section.

\begin{lemma}\label{L:noanF}
If ``$\cgr$'' is a congruence, ``$\CFFD$'' implies ``$\cgr$'', ``$\cgr$''
preserves $\minD$ but not $\anF$, and $\LLG \cgr \BLG$, then for every LTS $L$
there is an LTS $h_2(L)$ such that $h_2(L) \cgr L$, $\Sf(h_2(L)) = (\Div(L)
\times 2^{\Sigma(L)}) \cup \{ (\sigma, A_1 \cup A_2) \mid (\sigma, A_1) \in
\snF(L) \wedge \forall a \in A_2: \sigma a \in \Div(L) \}$, $\Div(h_2(L)) =
\Div(L)$, and $\Inf(h_2(L)) = \Inf(L)$.
\end{lemma}
\proof
\begin{figure}[t]
\mbox{}\hfill\begin{pspicture}(0,-4)(200,41)
\arc(29,22)(35.6,15.4)
{\psset{linewidth=2pt}
\aloopur{9,14}\rput[b](12,25){$\new{\Sigma_L}{2}$}
\aloopdl{9,14}\rput[l](10,4){$\new{\Sigma_L}{3}$}
\aloopur{37,14}\rput[b](40,25){$\new{\Sigma_L}{2}$}
\aloopdl{37,14}\rput[l](38,4){$\new{\Sigma_L}{3}$}
\aloopur{65,14}\rput[b](68,25){$\new{\Sigma_L}{2}$}
\aloopdl{65,14}\rput[l](66,4){$\new{\Sigma_L}{3}$}
\aloopur{137,14}\rput[b](140,25){$\new{\Sigma_L}{2}$}
\aloopdl{137,14}\rput[l](138,4){$\new{\Sigma_L}{3}$}
\aloopur{165,14}\rput[b](168,25){$\new{\Sigma_L}{2}$}
}
\arc(35,14)(11,14)\rput[B](23,17){$\tau$}
\arc(39,14)(63,14)\rput[B](51,17){$b_1$}
\arc(67,14)(91,14)\rput[B](79,17){$b_2$}
\aloopur{193,14}\rput[b](196,25){$\tau$}
\rput(101,14){$\cdots$}
\arc(111,14)(135,14)\rput[B](123,17){$b_{n-1}$}
\arc(139,14)(163,14)\rput[B](151,17){$b_n$}
\arc(167,14)(191,14)\rput[B](179,17){$d$}
\psst(9,14){2}\psst(37,14){2}\psst(65,14){2}\psst(137,14){2}\psst(165,14){2}
\psst(193,14){2}
\end{pspicture}\hfill\mbox{}
\caption{An LTS for detecting an always nondivergent failure.}\label{F:sanF}
\end{figure}
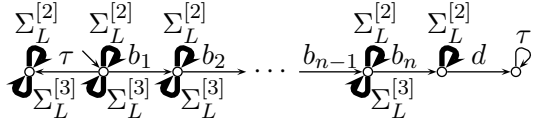
By Theorem~\ref{T:noSigma}, ``$\cgr$'' preserves $\Sigma$.
Let $M_1 \cgr M_2$, $(\sigma, A) \in \anF(M_1) \setminus \anF(M_2)$, $\Sigma_M
= \Sigma(M_1) = \Sigma(M_2)$, $b_1 \cdots b_n = \new{\sigma}{1}$, and $d =
\new{1}{0}$.
Let $L$ be any LTS and $\Sigma_L = \Sigma(L)$.
Let $T_\sigma^d$ be the LTS with the alphabet $\{d\} \cup \new{\Sigma_M}{1}
\cup \new{\Sigma_L}{2} \cup \new{\Sigma_L}{3}$ whose graph is in
Fig.~\ref{F:sanF}.
When $i \in \{1,2\}$, let
$$M'_i \ =\ (\,(T_\sigma^d \pp M_i\Phi^{[1],d}) \setminus
\new{\Sigma_M}{1}\,)\Phi_d^{[3]}\textrm{ ,}$$
where $\Phi_d^{[3]}$ renames $d$ to each $x \in \new{\Sigma_L}{3}$, and
$\Phi^{[1],d}$ renames each $x \in \Sigma_M$ to $\new{x}{1}$ and each $x \in
A$ also to $d$.
If $A = \emptyset$, we let $M_i \Phi^{[1],d} = \newup{M_i}{1} \pp
\Stop{\{d\}}$, so that $d$ is not accidentally left out from the alphabet
$\new{\Sigma_M}{1} \cup \{d\}$.
Clearly $M_i$ refuses $A$ if and only if $M_i \Phi^{[1],d}$ refuses $d$.
Let $\Xi = \new{\Sigma_L}{2} \cup \new{\Sigma_L}{3}$.
We have $\Sigma(M'_i) = \Xi$.

Clearly $\Inf(M'_1) = \Inf(M'_2) = \Xi^\omega$.
Because $(\sigma, A) \in \anF(M_1)$ and ``$\cgr$'' preserves $\minD$, $M'_i$
cannot diverge before executing $d$.
The leftmost state of $T_\sigma^d$ is stable, ensuring $(\sigma, \emptyset)
\in \Sf(M'_i)$ for every $\sigma \in \Xi^*$.
No other states of $T_\sigma^d$ can add to $\Sf(M'_i)$, except perhaps the
start state of the $d$-transition.
Because $M_2$ cannot execute $\sigma$ or refuse $A$ after it, $T_\sigma^d \pp
M_2\Phi^{[1],d}$ cannot refuse $d$ after $b_1 \cdots b_n$.
So $\Sf(M'_2) = \{ (\sigma, \emptyset) \mid \sigma \in \Xi^* \}$.
Since $M_1$ can, $\Sf(M'_1) = \{ (\sigma, B) \mid \sigma \in \Xi^* \wedge B
\subseteq \new{\Sigma_L}{3} \}$.

Let $g(L)$ be $\Una(L)$ with each visible label $x$ replaced by $\new{x}{3}$
if the transition ends in a potentially divergent state, and $\new{x}{2}$
otherwise.
Let $\Phi_{[2,3]}$ rename each $\new{x}{2}$ and $\new{x}{3}$ to $x$.
Consider $f_i(L) = (g(L) \pp M'_i)\Phi_{[2,3]}$.
When $M'_i$ diverges, $d$ has just been executed.
Thus $g(L)$ has just completed a divergence trace and $M'_i$ blocks the
visible transitions.
So $\Div(f_i(L)) = \Div(L)$.
$M'_2$ does not affect the behaviour of $g(L)$ in any other way, so $f_2(L)
\CFFD L$.
On the other hand, $M'_1$ can block all actions that would complete a nonempty
divergence trace of $L$.
So $$\Sf(f_1(L)) \ =\ \{ (\sigma, B_1 \cup B_2) \mid (\sigma, B_1) \in \Sf(L)
\wedge \forall a \in B_2: \sigma a \in \Div(L) \}\textrm{ .}$$
It implies
\begin{eqnarray}
\nF(f_1(L)) & = & \{ (\sigma, B_1 \cup B_2) \mid (\sigma, B_1) \in \nF(L)
\wedge \forall a \in B_2: \sigma a \in \Div(L) \}\nonumber\\
& = & \{ (\sigma, A_1 \cup A_2) \mid (\sigma, A_1) \in \snF(L) \wedge \forall
a \in A_2: \sigma a \in \Div(L) \}\textrm{ ,}\nonumber
\end{eqnarray}
where the last equality is obtained by letting $A_2 = \{ a \in B_1 \cup B_2
\mid \sigma a \in \Div(L) \}$ and $A_1 = (B_1 \cup B_2) \setminus A_2$.

As a consequence, $\nu(f_1(L))$ qualifies as the $h_2(L)$, where $\nu$ is from
Lemma~\ref{L:nF}.\qed

\noindent The following theorem lists the next six congruences and
points direction to the next nine.

\begin{table}[H]
\caption{The congruences of Theorem~\ref{T:noanF}}\label{B:noanF}
\smallskip
\begin{tabular}{l|l|l}
preserves & does not preserve & induced by\\
\hline
$\sanF$, $\minD$ & $\Tr$, $\anF$ & $\Sigma$, $\sanF$, $\minD$, $\anI$\\
$\Tr$, $\sanF$, $\minD$ & $\anF$, $\Div$, $\Inf$ & $\Sigma$, $\Tr$, $\sanF$,
$\minD$, $\anI$\\
$\sanF$, $\minD$, $\Inf$ & $\anF$, $\Div$ & $\Sigma$, $\Tr$, $\sanF$, $\minD$,
$\Inf$\\
$\sanF$, $\Div$ & $\anF$, $\snF$, $\aenI$ & $\Sigma$, $\Tr$, $\sanF$, $\Div$,
$\eanI$\\
$\sanF$, $\Div$, $\aenI$ & $\anF$, $\snF$, $\Inf$ & $\Sigma$, $\Tr$, $\sanF$,
$\Div$, $\aenI$\\
$\sanF$, $\Div$, $\Inf$ & $\anF$, $\snF$ & $\Sigma$, $\Tr$, $\sanF$, $\Div$,
$\Inf$\\
\end{tabular}
\end{table}
\begin{theorem}\label{T:noanF}
If ``$\cgr$'' is a congruence, ``$\CFFD$'' implies ``$\cgr$'', ``$\cgr$''
preserves the sets in the first column of Table~\ref{B:noanF} but not the sets
in the second column, and $\LLG \cgr \BLG$, then ``$\cgr$'' is the equivalence
induced by the sets in the third column.
\end{theorem}
\proof
Let [r1] to [r6] refer to the rows in the table.

Lemmas~\ref{L:Inf=>Tr} [r3], \ref{L:minD=>anI} [r1,2], \ref{L:Div=>Tr}
[r4,5,6], and~\ref{L:Div=>eanI} [r4,5,6] imply that if ``$\cgr$'' preserves
the sets in the first column, then it preserves also the additional sets in
the third column.

To prove the claims that ``$\cgr$'' can be no other equivalence, let $f'$ be
the $f$ in Lemma~\ref{L:noTr} [r1], \ref{L:noDiv}(b) [r2], \ref{L:noDiv}(a)
[r3], \ref{L:noaenI} [r4], or \ref{L:noInf} [r5], or the function $f'(L) = L$
[r6].
So $\Div(f'(L))$ is either $\extT(L)$ [r1], $\Tr(L) \cap \extT(L)$ [r2,3], or
$\Div(L)$ [r4,5,6]; and $\Inf(f'(L))$ is either $\anI(L) \cup \extI(L)$ [r1],
$\anI(L)$ [r2], $\eanI(L)$ [r4], $\aenI(L)$ [r5], or $\Inf(L)$ [r3,6].
Furthermore, $\Sf(f'(L)) = \Sf(L)$ and $f'(L) \cgr L$.
Because ``$\cgr$'' preserves $\minD$ or $\Div$, we have $\minD(f'(L)) =
\minD(L)$ and $\extT(f'(L)) = \extT(L)$.
Excluding [r1], we also have $\Tr(f'(L)) = \Tr(L)$.

Let $h_1$ and $h_2$ be like in Lemmas~\ref{L:nosnF} and~\ref{L:noanF}, and let
$f(L) = h_2(h_1(f'(L)))$.
The validity of some assumptions of Lemma~\ref{L:nosnF} is not immediately
obvious, so let us check them.
By Lemma~\ref{L:Div=>Tr}, $\Div$ is not preserved on [r1].
By Lemma~\ref{L:nF=>Div}, $\snF$ is not preserved on [r1,2,3].
It is explicitly given in the table that $\snF$ is not preserved on [r4,5,6].
On [r4,5,6], $\Div$ is preserved.
We show next that [r1,2,3] satisfy $\Div(f'(L)) = \Tr(f'(L)) \cap
\extT(f'(L))$.
By Lemma~\ref{L:noDiv}, [r2,3] have $\Div(f'(L)) = \Tr(L) \cap \extT(L) =
\Tr(f'(L)) \cap \extT(f'(L))$.
By Lemma~\ref{L:noTr}, [r1] has $\Div(f'(L)) = \extT(L) = \extT(f'(L)) =
\Tr(f'(L)) \cap \extT(f'(L))$, because $\Div(f'(L)) \subseteq \Tr(f'(L))$ by
the definition of $\Div$.
So Lemma~\ref{L:nosnF} can be used.

We have $L \cgr f'(L) \cgr h_1(f'(L)) \cgr f(L)$, $\Div(f(L)) = \Div(f'(L))$,
and $\Inf(f(L)) = \Inf(f'(L))$.
All assumptions of Lemma~\ref{L:nocongr} can now be checked except the
$\Sf(f(L))$ assumption.
To facilitate checking it, too, we show next that $\Sf(f(L)) = F(L)$, where
\begin{eqnarray}
F(L) & = & \big(\,\big(\Tr(f'(L)) \cap \extT(f'(L))\big) \times
2^{\Sigma(L)}\,\big) \cup \nonumber\\
&& \{ (\sigma, A_1 \cup A_2) \mid (\sigma, A_1) \in \sanF(f'(L)) \wedge
\forall a \in A_2: \sigma a \in \minD(f'(L)) \}\textrm{ .}\nonumber
\end{eqnarray}
Let $\sigma \in \Tr(f'(L))$, $A \subseteq \Sigma(L)$, $A_2 = \{ a \in A \mid
\sigma a \in \Div(f'(L)) \}$, and $A_1 = A \setminus A_2$.

Assume first that $\sigma \in \extT(f'(L))$.
Then clearly $(\sigma,A) \in F(L)$.
By Lemma~\ref{L:nosnF}, $(\sigma,A) \in \Sf(h_1(f'(L)))$.
If $\sigma \in \Div(h_1(f'(L)))$, then the first part and otherwise the second
part of the expression for $\Sf(h_2(\ldots))$ in Lemma~\ref{L:noanF} yields
$(\sigma,A) \in \Sf(f(L))$.

In the remaining case $\sigma \notin \extT(f'(L))$.
That implies $\sigma \in \anT(f'(L))$.
Then $\sigma a \in \minD(f'(L))$ if and only if $\sigma a \in \Div(f'(L))$ if
and only if $\sigma a \in \Div(h_1(f'(L)))$.
Furthermore, $(\sigma, A) \in F(L)$ if and only if $(\sigma, A_1) \in
\sanF(f'(L))$ if and only if $(\sigma, A_1) \in \sanF(h_1(f'(L)))$ if and only
if $(\sigma, A_1) \in \snF(h_1(f'(L)))$ if and only if $(\sigma, A) \in
\Sf(f(L))$.

We have shown $\Sf(f(L)) = F(L)$.

Because $f'$ preserves $\Sf$ and $\minD$, we have $\sanF(f'(L)) = \sanF(L)$.
On [r1],
$$
\Tr(f'(L)) \ =\ \Div(f'(L)) \cup \Sf^\Tr(f'(L)) \ =\ \extT(L) \cup
\Sf^\Tr(L) \ =\ \extT(L) \cup \sanF^\Tr(L)\textrm{ ,}
$$
because if $\sigma \notin \extT(L)$ and $(\sigma, \emptyset) \in \Sf(L)$, then
$(\sigma, \emptyset) \in \sanF(L)$.
In the remaining cases ``$\cgr$'' preserves $\Tr$, so $\Tr(f'(L)) = \Tr(L)$.
Thus Lemma~\ref{L:nocongr} applies in all cases.\qed

The equivalence induced by $\Sigma$, $\sanF$, $\minD$, and $\anI$ is the
weakest ``any-lock''-preserving congruence (that is, the weakest congruence
that distinguishes LTSs that can stop executing visible actions from those
that cannot) with respect to $L \setminus A$ and $L \pp L'$, as was proven
in~\cite{Puh01}.

Six more congruences follow.

\begin{table}[H]
\caption{The congruences of Theorem~\ref{T:nosnF}}\label{B:nosnF}
\smallskip
\begin{tabular}{l|l|l}
preserves & does not preserve & induced by\\
\hline
$\anF$, $\minD$ & $\Tr$ & $\Sigma$, $\anF$, $\minD$, $\anI$\\
$\Tr$, $\anF$, $\minD$ & $\Div$, $\Inf$ & $\Sigma$, $\Tr$, $\anF$, $\minD$,
$\anI$\\
$\anF$, $\minD$, $\Inf$ & $\Div$ & $\Sigma$, $\Tr$, $\anF$, $\minD$, $\Inf$\\
$\anF$, $\Div$ & $\snF$, $\aenI$ & $\Sigma$, $\Tr$, $\anF$, $\Div$, $\eanI$\\
$\anF$, $\Div$, $\aenI$ & $\snF$, $\Inf$ & $\Sigma$, $\Tr$, $\anF$, $\Div$,
$\aenI$\\
$\anF$, $\Div$, $\Inf$ & $\snF$ & $\Sigma$, $\Tr$, $\anF$, $\Div$, $\Inf$\\
\end{tabular}
\end{table}
\begin{theorem}\label{T:nosnF}
If ``$\cgr$'' is a congruence, ``$\CFFD$'' implies ``$\cgr$'', ``$\cgr$''
preserves the sets in the first column of Table~\ref{B:nosnF} but not the sets
in the second column, and $\LLG \cgr \BLG$, then ``$\cgr$'' is the equivalence
induced by the sets in the third column.
\end{theorem}
\proof
The proof is like the proof of Theorem~\ref{T:noanF} with the following
differences.
Now $h_2$ is not used, so $f(L) = h_1(f'(L))$.
By the definition of $h_1$, $$\Sf(f(L)) \ =\ \anF(f'(L)) \cup (\,(\Tr(f'(L))
\cap \extT(f'(L))) \times 2^{\Sigma(L)}\,)\textrm{ .}$$
We have $\anF(f'(L)) = \anF(L)$.
On [r1], $\Tr(f'(L)) = \extT(L) \cup \sanF^\Tr(L) = \extT(L) \cup
\anF^\Tr(L)$.\qed

\newcommand{\Cfail}{\mi{CFail}}
\newcommand{\Cdiv}{\mi{CDiv}}
The congruence induced by $\Sigma$, $\anF$, $\minD$, and $\anI$ is the same as
the well-known failures-divergences equivalence in the CSP
theory~\cite{Ros10}.
It is more often defined by requiring that $\Sigma$, $\Cfail$, and $\Cdiv$ are
preserved, where (in our terminology) $\Cdiv(L) = \extT(L)$ and $\Cfail(L) =
\Sf(L) \cup (\Cdiv(L) \times 2^{\Sigma(L)})$.
That $\anI$ is preserved is not required, because the LTSs are assumed to be
finitely branching, that is, for every $s$, the set $\{ s' \mid \exists a:
(s,a,s') \in \Delta \}$ is finite.
It makes $\anI$ a function of $\anT$.
Often other parallel composition operators than the one defined in this
publication are used, making it unnecessary to talk about $\Sigma$.

In CSP theory, the congruence was defined using a fixed-point method that
gives a meaning to recursively defined process expressions without appealing
to LTSs.
A natural consequence of this method is that the resulting congruence
preserves no information beyond minimal divergence traces.
With it, each divergence is equivalent to $\RDL{\Sigma_L}$ in
Fig.~\ref{F:simpleLTSs}.
This phenomenon is called \emph{catastrophic divergence} and
$\RDL{\Sigma_L}$ is called \emph{chaos}.
The phenomenon is harmful in many applications.
This motivated the development and name of CFFD-equivalence, that is,
chaos-free failures divergences equivalence.
Recently, a complicated fixed-point definition for the equivalence induced by
$\Sigma$, $\Tr$, $\Div$, and $\eanI$ has been found~\cite{Ros05}.
To this, $\Sf$ can be added.

\subsection{(Strongly) nondivergent failures}\label{S:(s)nF}

We still have to consider the congruences that preserve $\snF$ or more and
satisfy $\LLG \cgr \BLG$.
There are three groups of them.
Again, each group corresponds to Section~\ref{S:ALLnot}.
However, because of Lemma~\ref{L:nF=>Div}, each group only contains
congruences that preserve $\Div$, so it only contains three congruences.

\begin{table}[H]
\caption{The congruences of Theorem~\ref{T:snF}}\label{B:snF}
\smallskip
\begin{tabular}{l|l|l}
preserves & does not preserve & induced by\\
\hline
$\snF$ & $\anF$, $\aenI$ & $\Sigma$, $\snF$, $\Div$, $\eanI$\\
$\snF$, $\aenI$ & $\anF$, $\Inf$ & $\Sigma$, $\snF$, $\Div$, $\aenI$\\
$\snF$, $\Inf$ & $\anF$ & $\Sigma$, $\snF$, $\Div$, $\Inf$
\end{tabular}
\end{table}
\begin{theorem}\label{T:snF}
If ``$\cgr$'' is a congruence, ``$\CFFD$'' implies ``$\cgr$'', ``$\cgr$''
preserves the sets in the first column of Table~\ref{B:snF} but not the sets
in the second column, and $\LLG \cgr \BLG$, then ``$\cgr$'' is the equivalence
induced by the sets in the third column.
\end{theorem}
\proof
Lemmas~\ref{L:nF=>Div} and~\ref{L:Div=>eanI} imply that ``$\cgr$'' preserves
$\Div$, $\Sigma$, and $\eanI$.

To prove the claims that ``$\cgr$'' can be no other equivalence, let $f'$ be
the $f$ in Lemma~\ref{L:noaenI} or \ref{L:noInf}, or the function $f'(L) = L$.
Let $h_2$ be like in Lemma~\ref{L:noanF}, and let $f(L) = h_2(f'(L))$.
We have $L \cgr f(L)$, $\Div(f(L)) = \Div(L)$, $\Inf(f(L)) = \Inf(f'(L))$, and
$$\Sf(f(L)) = (\Div(L) \times 2^{\Sigma(L)}) \cup \{
(\sigma, A_1 \cup A_2) \mid (\sigma, A_1) \in \snF(L) \wedge \forall a \in
A_2: \sigma a \in \Div(L) \}\textrm{ .}$$
Furthermore, $\Inf(f'(L))$ is either $\eanI(L)$, $\aenI(L)$, or $\Inf(L)$.
So Lemma~\ref{L:nocongr} applies.\qed

\begin{lemma}\label{L:anFsnF}
If ``$\cgr$'' is a congruence, ``$\CFFD$'' implies ``$\cgr$'', ``$\cgr$''
preserves $\Div$ but not $\nF$, and $\LLG \cgr \BLG$, then for every LTS $L$
there is an LTS $h(L)$ such that $h(L) \cgr L$, $\Div(h(L)) = \Div(L)$,
$\Inf(h(L)) = \Inf(L)$, and
\begin{eqnarray}
\Sf(h(L)) & = & \anF(L) \cup (\Div(L) \times 2^{\Sigma(L)}) \cup\nonumber\\
&& \{(\sigma, A_1 \cup A_2) \mid (\sigma, A_1) \in \snF(L) \wedge \sigma \in
\extT(L) \wedge \forall a \in A_2: \sigma a \in \Div(L) \}\textrm{ .}\nonumber
\end{eqnarray}
\end{lemma}
\begin{figure}[t]
\mbox{}\hfill\begin{pspicture}(-14,-3)(301,35)
\arc(35,16)(11,16)\rput[B](23,19){$\tau$}
\ainit{37,16}
\arc(67,16)(91,16)\rput[B](79,19){$c$}
\aloopdr{93,16}\rput[r](92,10){$\tau$}
\arc(95,16)(119,16)\rput[B](107,19){$b_1$}
\aloopdr{121,16}\rput[r](120,10){$\tau$}
\arc(123,16)(147,16)\rput[B](135,19){$b_2$}
\rput(157,16){$\cdots$}
\rput(157,9){$\cdots$}
\arc(167,16)(191,16)\rput[B](179,19){$b_{n-1}$}
\aloopdr{193,16}\rput[r](192,10){$\tau$}
\arc(195,16)(219,16)\rput[B](207,19){$b_n$}
\arc(223,16)(259,16)\rput[B](247,19){$\tau$}
\arc(263,16)(287,16)\rput[B](275,19){$d$}
\aloopur{289,16}\rput[l](296,22){$\tau$}
{
\psset{linewidth=2pt}
\aloopul{9,16}\rput[r](3,26){$\new{\Sigma_L}{3}$}
\aloopdr{9,16}\rput[r](8,5){$\new{\Sigma_L}{2}$}
\aloopdr{37,16}\rput[l](43,5){$\new{\Sigma_L}{2}$}
\arc(39,16)(63,16)\rput[B](51,23){$\new{\Sigma_L}{3}$}
\aloopul{221,16}\rput[l](222,26){$\new{\Sigma_L}{3}$}
\aloopdr{221,16}\rput[l](227,5){$\new{\Sigma_L}{2}$}
\aloopdr{261,16}\rput[l](267,5){$\new{\Sigma_L}{2}$}
}
\psst(9,16){2}\psst(37,16){2}\psst(65,16){2}\psst(93,16){2}\psst(121,16){2}
\psst(193,16){2}\psst(221,16){2}\psst(261,16){2}\psst(289,16){2}
\end{pspicture}\hfill\mbox{}
\caption{An LTS for detecting a nondivergent failure.}\label{F:nF}
\end{figure}
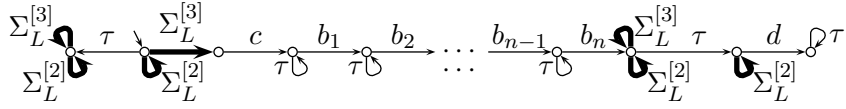
\proof
Theorem~\ref{T:noSigma} implies that ``$\cgr$'' preserves $\Sigma$.
Let $M_1 \cgr M_2$, $(\sigma, A) \in \nF(M_1) \setminus \nF(M_2)$, $\Sigma_M =
\Sigma(M_1) = \Sigma(M_2)$, $b_1 \cdots b_n = \new{\sigma}{1}$, $c =
\new{1}{0}$, and $d = \new{2}{0}$.
Let $L$ be any LTS and $\Sigma_L = \Sigma(L)$.
Let $T_\sigma^d$ be the LTS whose alphabet is $\{c,d\} \cup \new{\Sigma_M}{1}
\cup \new{\Sigma_L}{2} \cup \new{\Sigma_L}{3}$ and whose graph is in
Fig.~\ref{F:nF}.
When $i \in \{1,2\}$, let
$$M'_i \ =\ (\ (\,T_\sigma^d \pp c.(M_i\Phi^{[1],d})\,) \setminus (\{c\} \cup
\new{\Sigma_M}{1})\ )\Phi_d^{[3]}\textrm{ ,}$$
where $\Phi^{[1],d}$ renames each $x \in \Sigma_M$ to $\new{x}{1}$ and each
$x \in A$ also to $d$, and $\Phi_d^{[3]}$ renames $d$ to each $x \in
\new{\Sigma_L}{3}$.
We use the same trick as in the proof of Lemma~\ref{L:noanF} to ensure that
$\Sigma(M_i\Phi^{[1],d}) = \new{\Sigma_M}{1} \cup \{d\}$ even if $A =
\emptyset$.
Let $\Xi = \new{\Sigma_L}{2} \cup \new{\Sigma_L}{3}$.
We have $\Sigma(M'_i) = \Xi$.

Clearly $\Inf(M'_1) = \Inf(M'_2) = \Xi^\omega$.
Because $(\sigma, A) \in \nF(M_1)$ and ``$\cgr$'' preserves $\Div$, $\sigma
\notin \Div(M_1) = \Div(M_2)$.
Therefore, $\Div(M'_1) = \Div(M'_2) \subseteq \{ \sigma a \mid \sigma \in
\Xi^* \wedge a \in \new{\Sigma_L}{3} \}$.
The leftmost state of $T_\sigma^d$ is stable, ensuring $(\sigma, \emptyset)
\in \Sf(M'_i)$ for every $\sigma \in \Xi^*$.
No other states of $T_\sigma^d$ can affect $\Sf(M'_i)$, except perhaps the
start state of the $d$-transition.
Because $M_2$ cannot execute $\sigma$ or refuse $A$ after it, $T_\sigma^d \pp
c.(M_2\Phi^{[1],d})$ cannot refuse $d$ after $b_1 \cdots b_n$.
Therefore, $\Sf(M'_2) = \{ (\sigma, \emptyset) \mid \sigma \in \Xi^* \}$.
However, $M_1$ can, so we have $\Sf(M'_1) = \{ (\sigma, \emptyset) \mid \sigma
\in \Xi^* \}$ $\cup$ $\{ (\sigma a \rho, B) \mid \sigma\rho \in \Xi^* \wedge
a \in \new{\Sigma_L}{3} \wedge B \subseteq \new{\Sigma_L}{3} \}$.

Let $g(L)$ be $\Una(L)$ with each visible label $x$ replaced by $\new{x}{3}$
if the transition ends in a potentially divergent state of $g(L)$, and
$\new{x}{2}$ otherwise.
Let $\Phi_{[2,3]}$ rename each $\new{x}{2}$ and $\new{x}{3}$ to $x$.
Consider $f_i(L) = (g(L) \pp M'_i)\Phi_{[2,3]}$.
When $M'_i$ diverges, also $g(L)$ completes a divergence trace and $M'_i$
blocks the visible transitions.
$M'_2$ does not affect the behaviour of $g(L)$ in any other way, so $f_2(L)
\CFFD L$.
On the other hand, $M'_1$ can block all actions that would complete a
nonminimal divergence trace.

Let $\nu$ be like in Lemma~\ref{L:nF}.
Clearly $\nu(f_1(L)) \cgr L$, $\Div(\nu(f_1(L))) = \Div(L)$, and
$\Inf(\nu(f_1(L))) = \Inf(L)$.
By analysing in turn the stable failures whose trace is always-nondivergent,
divergent, or neither of them, we see that
\begin{center}
$\Sf(\nu(f_1(L))) \ =\ \anF(L) \cup (\Div(L) \times 2^{\Sigma(L)})$
$\cup$\hfill\mbox{}\\
\mbox{}\hfill
$\{ (\sigma, A_1 \cup A_2) \mid (\sigma, A_1) \in \snF(L) \wedge \sigma \in
\extT(L) \wedge \forall a \in A_2: \sigma a \in \Div(L) \}$.
\end{center}
Thus $\nu(f_1(L))$ qualifies as the $h(L)$ of the claim.\qed

\begin{table}[H]
\caption{The congruences of Theorem~\ref{T:anFsnF}}\label{B:anFsnF}
\smallskip
\begin{tabular}{l|l|l}
preserves & does not preserve & induced by\\
\hline
$\anF$, $\snF$ & $\nF$, $\aenI$ & $\Sigma$, $\anF$, $\snF$, $\Div$, $\eanI$\\
$\anF$, $\snF$, $\aenI$ & $\nF$, $\Inf$ & $\Sigma$, $\anF$, $\snF$, $\Div$,
$\aenI$\\
$\anF$, $\snF$, $\Inf$ & $\nF$ & $\Sigma$, $\anF$, $\snF$, $\Div$, $\Inf$
\end{tabular}
\end{table}
\begin{theorem}\label{T:anFsnF}
If ``$\cgr$'' is a congruence, ``$\CFFD$'' implies ``$\cgr$'', ``$\cgr$''
preserves the sets in the first column of Table~\ref{B:anFsnF} but not the
sets in the second column, and $\LLG \cgr \BLG$, then ``$\cgr$'' is the
equivalence induced by the sets in the third column.
\end{theorem}
\proof
The proof is like the proof of Theorem~\ref{T:snF}, but using the $h$ of
Lemma~\ref{L:anFsnF} instead of the $h_2$ of Lemma~\ref{L:noanF}.\qed

\begin{table}[H]
\caption{The congruences of Theorem~\ref{T:nF}}\label{B:nF}
\smallskip
\begin{tabular}{l|l|l}
preserves & does not preserve & induced by\\
\hline
$\nF$ & $\Sf$, $\aenI$ & $\Sigma$, $\nF$, $\Div$, $\eanI$\\
$\nF$, $\aenI$ & $\Sf$, $\Inf$ & $\Sigma$, $\nF$, $\Div$, $\aenI$\\
$\nF$, $\Inf$ & $\Sf$ & $\Sigma$, $\nF$, $\Div$, $\Inf$
\end{tabular}
\end{table}
\begin{theorem}\label{T:nF}
If ``$\cgr$'' is a congruence, ``$\CFFD$'' implies ``$\cgr$'', ``$\cgr$''
preserves the sets in the first column of Table~\ref{B:nF} but not the sets in
the second column, and $\LLG \cgr \BLG$, then ``$\cgr$'' is the equivalence
induced by the sets in the third column.
\end{theorem}
\proof
The proof is like the proof of Theorem~\ref{T:snF}, but using the $\nu$ of
Lemma~\ref{L:nF} instead of the $h_2$ of Lemma~\ref{L:noanF}.\qed

The equivalence induced by $\Sigma$, $\nF$, $\Div$, and $\eanI$ is the weakest
congruence that preserves all traces that can lead to an ``any-lock'' (that
is, deadlock or livelock) with respect to $L \setminus A$ and $L \pp L'$, as
was proven in~\cite{Puh01}.
The same (pre)congruence is the weakest that preserves so-called conditional
liveness properties~\cite{vGl10}.
The equivalence induced by $\Sigma$, $\nF$, $\Div$, and $\Inf$ is called
\emph{nondivergent failures divergences equivalence} or
\emph{NDFD-equivalence}.
In~\cite{KaV92} it was proven that it is the weakest congruence that preserves
all properties that can be formulated in the stuttering-insensitive linear
temporal logic of~\cite{MaP92}.
A variant of this result, where the logic is connected to LTSs in a more
intuitive way, was presented in~\cite{Val00}.

A comparison of the ``induced by'' and ``does not preserve'' colums of
Table~\ref{B:nosanF} to~\ref{B:nF} reveals that all possibilities with $\DLG
\not\cgr \LLG \cgr \BLG$ have been investigated.

\section{Conclusion}\label{S:conclusion}

\begin{figure}[t]
\mbox{}\hfill\begin{pspicture}(207,274)
\newcommand{\Qx}{108}\newcommand{\Qy}{32}
\newcommand{\QT}[1]{\rput(-32,18){#1}}
\newcommand{\QA}[1]{\rput(-64,36){#1}}
\newcommand{\QI}[1]{\rput(-96,54){#1}}
\newcommand{\QH}[1]{\rput(24,24){#1}}
\newcommand{\QG}[1]{\rput(48,48){#1}}
\newcommand{\QN}[1]{\rput(72,72){#1}}
\newcommand{\QF}[1]{\rput(96,96){#1}}
\newcommand{\QM}[1]{\rput(0,30){#1}}
\newcommand{\QD}[1]{\rput(0,60){#1}}
\newcommand{\QE}[1]{\rput(0,90){#1}}
\newcommand{\Qc}[1]{\rput(\Qx,\Qy){\pscircle*(0,0){2}
\rput[r](-3,-2){\tiny{#1}}}}
\newcommand{\Qd}[1]{\rput(\Qx,\Qy){\rput[r](-3,-8){\tiny{#1}}}}
\newcommand{\Qe}[1]{\rput(\Qx,\Qy){\rput[r](-3,-14){\tiny{#1}}}}
\newcommand{\Qline}[2]{\psline[linewidth=2pt,linecolor=white](#1)(#2)
\psline(#1)(#2)}
\newcommand{\Qu}{\rput(\Qx,\Qy){\Qline{0,-30}{0,0}}}
\newcommand{\Quu}{\rput(\Qx,\Qy){\Qline{0,-60}{0,0}}}
\newcommand{\Ql}{\rput(\Qx,\Qy){\Qline{32,-18}{0,0}}}
\newcommand{\Qll}{\rput(\Qx,\Qy){\Qline{64,-36}{0,0}}}
\newcommand{\Qr}{\rput(\Qx,\Qy){\Qline{-24,-24}{0,0}}}
\newcommand{\Qrr}{\rput(\Qx,\Qy){\Qline{-48,-48}{0,0}}}
\newcommand{\Qrrrr}{\rput(\Qx,\Qy){\Qline{-96,-96}{0,0}}}
\newcommand{\ww}{\psframe*[linecolor=white]}

\QF{\Qrrrr}\QF{\QT{\Qrrrr\Ql}}\QF{\QI{\Qrrrr\Qll}}
\QF{\QM{\Qu\Qrr}}\QF{\QM{\QT{\Qu\Ql\Qrr}}}\QF{\QM{\QI{\Qu\Qll\Qrr}}}
\QF{\QE{\QT{\Quu\Qr}}}\QF{\QE{\QA{\Ql\Qr}}}\QF{\QE{\QI{\Quu\Ql\Qr}}}

\QN{\QE{\QT{\Qr}}}\QN{\QE{\QA{\Qr\Ql}}}\QN{\QE{\QI{\Qr\Ql}}}

\QG{\QM{\Qr}}\QG{\QM{\QT{\Qr\Ql}}}\QG{\QM{\QI{\Qr\Qll}}}
\QG{\QD{\QT{\Qr\Qu}}}\QG{\QD{\QA{\Qr\Ql}}}\QG{\QD{\QI{\Qr\Qu\Ql}}}
\QG{\QE{\QT{\Qr\Qu}}}\QG{\QE{\QA{\Qr\Qu\Ql}}}\QG{\QE{\QI{\Qr\Qu\Ql}}}

\QH{\QM{\Qr}}\QH{\QM{\QT{\Qr\Ql}}}\QH{\QM{\QI{\Qr\Qll}}}
\QH{\QD{\QT{\Qr\Qu}}}\QH{\QD{\QA{\Qr\Ql}}}\QH{\QD{\QI{\Qr\Qu\Ql}}}
\QH{\QE{\QT{\Qu}}}\QH{\QE{\QA{\Qu\Ql}}}\QH{\QE{\QI{\Qu\Ql}}}

\Qu\QT{\Ql}\QI{\Qll}
\QM{\Qu}\QM{\QT{\Qu\Ql}}\QM{\QI{\Qu\Qll}}
\QD{\QT{\Qu}}\QD{\QA{\Ql}}\QD{\QI{\Qu\Ql}}

\rput(0,-30){\Qc{}}\Qc{$\Sigma$}\QT{\Qc{$\Tr$}}\QI{\Qc{$\Inf$}}

\QF{\Qc{$\Sf$\,}}\QF{\QT{\Qc{}}}\QF{\QI{\Qc{}}}

\QM{\Qc{\ww(0,0)(4,4)$\minD$}\Qd{$\anT$}\Qe{$\anI$}}\QM{\QT{\Qc{}}}
\QM{\QI{\Qc{}}}
\QD{\QT{\Qc{$\Div$}\Qd{$\eanI$}}}\QD{\QA{\Qc{$\aenI$}}}\QD{\QI{\Qc{}}}

\QH{\QM{\Qc{\ww(-1,0)(3,4.5)$\sanF$}}}\QH{\QM{\QT{\Qc{}}}}\QH{\QM{\QI{\Qc{}}}}
\QH{\QD{\QT{\Qc{}}}}\QH{\QD{\QA{\Qc{}}}}\QH{\QD{\QI{\Qc{}}}}

\QG{\QM{\Qc{\ww(0,-0.5)(18,4.5)CSP, $\anF$}}}
\QG{\QM{\QT{\Qc{}}}}\QG{\QM{\QI{\Qc{}}}}
\QG{\QD{\QT{\Qc{}}}}\QG{\QD{\QA{\Qc{}}}}\QG{\QD{\QI{\Qc{}}}}

\QH{\QE{\QT{\Qc{\ww(0,-0.5)(9,4)$\snF$}}}}
\QH{\QE{\QA{\Qc{}}}}\QH{\QE{\QI{\Qc{}}}}

\QG{\QE{\QT{\Qc{}}}}\QG{\QE{\QA{\Qc{}}}}\QG{\QE{\QI{\Qc{}}}}

\QN{\QE{\QT{\Qc{$\nF$}}}}\QN{\QE{\QA{\Qc{}}}}\QN{\QE{\QI{\Qc{NDFD\,}}}}

\QF{\QM{\Qc{}}}\QF{\QM{\QT{\Qc{}}}}\QF{\QM{\QI{\Qc{}}}}
\QF{\QE{\QT{\Qc{}}}}\QF{\QE{\QA{\Qc{}}}}\QF{\QE{\QI{\Qc{CFFD\,}}}}
\end{pspicture}\hfill\mbox{}
\caption{All abstract linear-time congruences with respect to $a.L$, $L
\setminus A$, $L\Phi$, and $L \pp L'$.
Names in $\mi{italics}$ indicate the new preserved set(s).
Other names are the names of the congruences.
There is a path from ``$\cgr_1$'' down to ``$\cgr_2$'' if and only if
``$\cgr_1$'' implies ``$\cgr_2$''.}\label{F:all}
\end{figure}
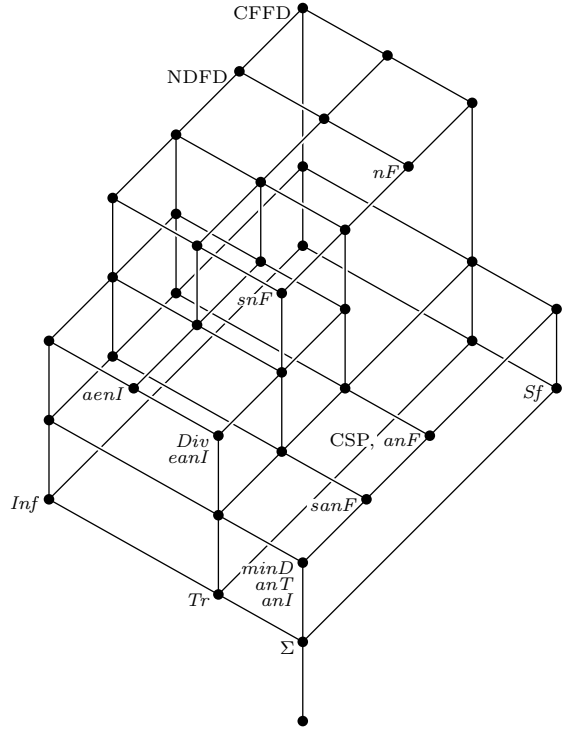

Fig.~\ref{F:all} shows the relations between the abstract linear-time
congruences discussed in this publication as a Hasse diagram.
There are altogether 40 of them.
If the set of considered operators is $a.L$, $L \setminus A$, $L\Phi$, and $L
\pp L'$, then for any stuttering-insensitive linear-time property, its optimal
congruence is among those in the figure.

For instance, what is the weakest linear-time congruence that distinguishes
$\LTSa{$a$}$ from $\LTSlr{$\tau$}{$a$}$?
Clearly the equivalence induced by $\Sigma$, $\Tr$, $\Div$, and $\Inf$ does
not separate them.
This also rules out the nine equivalences that are connected downstream to it
in the figure.
On the other hand, the equivalence induced by $\Sigma$ and $\Sf$ separates
them, and so does the equivalence induced by $\Sigma$, $\sanF$, $\minD$, and
$\anI$.
So there is no unique weakest linear-time congruence, but two.
It is worth mentioning that outside linear-time, also observation
equivalence~\cite{Mil89} separates them, although it is not strictly stronger
than the two linear-time congruences mentioned above.

With a smaller set of operators, there may be more congruences.
With a bigger set, there may be fewer.
However, it may also be that ``$\CFFD$'' is not a congruence with respect to
the bigger set.
Then it is necessary to strengthen ``$\CFFD$''.
This makes room for more congruences.
This happens if the ``choice'' operator of CCS is employed.
Then one must add one bit to the semantics that tells if the initial state is
stable~\cite{VaT91,VaT95}.
This splits some congruences in the figure to two, one with and another
without the initial stability bit.

If the LTSs are finite, then the distinction between $\Tr$, $\aenI$, and
$\Inf$ disappears, because then the infinite traces are determined by the
traces, as shown by~(\ref{E:Inf=}).
Then some congruences merge, leaving 20 distinct congruences.

\section*{Acknowledgement}
I thank Rob van Glabbeek, Bill Roscoe, and the anonymous reviewers for helpful
comments.
In particular, the anonymous reviewers found problems and suggested fixes in
Lemma~\ref{L:nosnF} and its proof.


\begin{thebibliography}{20}
\newcommand{\LNCS}[1]{Lecture Notes in Computer Science, vol.~#1,}

\bibitem{BoB87}
Bolognesi, T., Brinksma, E.:
Introduction to the ISO Specification Language LOTOS.
\emph{Computer Networks and ISDN Systems}, vol.~14, pp.~25--59 (1987)

\bibitem{DeV95}
De Nicola, R., Vaandrager, F.:
Three Logics for Branching Bisimulation.
\emph{Journal of the ACM} 42(2), 458--487 (1995)

\bibitem{GaF10}
Gazda, M., Fokkink, W.:
Congruence from the Operator's Point of View: Compositionality Requirements on
Process Semantics.
In: Aceto, L., Sobocinski, P.\ (eds.)
\emph{Proc.\ Seventh Workshop on Structural Operational Semantics}.
Electronic Proceedings in Theoretical Computer Science 32, 15--25 (2010)

\bibitem{vGl10}
van Glabbeek, R.:
The Coarsest Precongruences Respecting Safety and Liveness Properties.
In: Calude, C.S., Sassone, V.\ (eds.)
\emph{Proc.\ Theoretical Computer Science --- 6th IFIP TC 1/WG 2.2
Int.\ Conf., TCS 2010}.
IFIP AICT 323, Springer, 32--52 (2010)

\bibitem{vGl93}
van Glabbeek, R.:
The Linear Time --- Branching Time Spectrum II: The Semantics of Sequential
Systems with Silent Moves.
In: Best, E.\ (ed.)
\emph{Proc.\ CONCUR '93, Fourth International Conference on Concurrency
Theory},
\LNCS{715} 66--81 (1993)

\bibitem{GSL96}
Graf, S., Steffen, B., L\"uttgen, G.:
Compositional Minimisation of Finite State Systems Using Interface
Specifications.
\emph{Formal Aspects of Computing} 8(5), 607--616 (1996)

\bibitem{Hoa85}
Hoare, C.A.R.:
\emph{Communicating Sequential Processes}.
Prentice-Hall, Englewood Cliffs, NJ, 256~p.\ (1985)

\bibitem{KaV92}
Kaivola, R., Valmari, A.:
The Weakest Compositional Semantic Equivalence Preserving Nexttime-less Linear
Temporal Logic.
In: Cleaveland, R.\ (ed.)
\emph{Proc.\ CONCUR '92, Third International Conference on Concurrency
Theory},
\LNCS{630} 207--221 (1992)

\bibitem{MaV90}
Madelaine, E., Vergamini, D.:
AUTO: A Verification Tool for Distributed Systems Using Reduction of Finite
Automata Networks.
In: Vuong, S.T.\ (ed.)
\emph{Formal Description Techniques II (FORTE `89)},
North-Holland, 61--66 (1990)

\bibitem{MaP92}
Manna, Z., Pnueli, A.:
\emph{The Temporal Logic of Reactive and Concurrent Systems, Volume I:
Specification}.
Springer, Heidelberg, 427~p.\ (1992)

\bibitem{Mil89}
Milner, R.:
\emph{Communication and Concurrency}.
Prentice-Hall, Englewood Cliffs, NJ, 260~p.\ (1989)

\bibitem{Puh01}
Puhakka, A.:
Weakest Congruence Results Concerning ``Any-Lock''.
In: Kobayashi, N., Pierce, B.C.\ (eds.)
\emph{TACS 2001, Fourth International Symposium on Theoretical Aspects of
Computer Software},
\LNCS{2215} 400--419 (2001)

\bibitem{PuV99}
Puhakka, A., Valmari, A.:
Weakest-Congruence Results for Livelock-Preserving Equivalences.
In: Baeten, J.C.M., Mauw, S.\ (eds.)
\emph{CONCUR '99, 10th International Conference on Concurrency Theory},
\LNCS{1664} 510--524 (1999)

\bibitem{ReV07}
Rensink, A., Vogler, W.:
Fair Testing.
\emph{Information and Computation} 205(2), 125--198 (2007)

\bibitem{Ros05}
Roscoe, A.W.:
Seeing Beyond Divergence.
In: Abdallah, A.E., Jones, C.B., Sanders, J.W.\ (eds.)
\emph{Communicating Sequential Processes. The First 25 Years},
\LNCS{3525} 15--35 (2005)

\bibitem{Ros10}
Roscoe, A.W.:
\emph{Understanding Concurrent Systems}.
Springer, Heidelberg, 533~p.\ (2010)

\bibitem{Val95}
Valmari, A.:
The Weakest Deadlock-Preserving Congruence.
\emph{Information Processing Letters} 53(6), 341--346 (1995)

\bibitem{Val00}
Valmari, A.:
A Chaos-Free Failures Divergences Semantics with Applications to Verification.
\emph{Millennial Perspectives in Computer Science, Proceedings of the 1999
Oxford--Microsoft Symposium in Honour of sir Tony Hoare}, Palgrave, 365--382
(2000)

\bibitem{Val01}
Valmari, A.:
Composition and Abstraction.
In: Cassez, F., Jard, C., Rozoy, B., Ryan, M.D.\ (eds.)
\emph{Modeling and Verification of Parallel Processes},
LNCS Tutorials, \LNCS{2067} 58--98 (2001)

\bibitem{Val12a}
Valmari, A.:
All Linear-Time Congruences for Finite LTSs and Familiar Operators.
In: Brandt, J., Heljanko, K.\ (eds.)
\emph{Proc.\ Application of Concurrency to System Design, 12th Int.\ Conf.},
IEEE, 12--21 (2012)

\bibitem{Val12b}
Valmari, A.:
All Linear-Time Congruences for Familiar Operators Part 2: Infinite LTSs.
In: Koutny, M., Ulidowski, I.\ (eds.)
\emph{Proc.\ CONCUR 2012, 23rd International Conference on Concurrency
Theory},
\LNCS{7454} 162--176 (2012)

\bibitem{VaT91}
Valmari, A., Tienari, M.:
An Improved Failures Equivalence for Finite-State Systems with a Reduction
Algorithm.
In: Jonsson, B., Parrow, J., Pehrson, B.\ (eds.)
\emph{Proc.\ Protocol Specification, Testing and Verification XI},
North-Holland, 3--18 (1991)

\bibitem{VaT95}
Valmari, A., Tienari, M.:
Compositional Failure-Based Semantic Models for Basic LOTOS.
\emph{Formal Aspects of Computing} 7(4), 440--468 (1995)

\end{thebibliography}
\end{document}